\documentclass[aps,prx,preprint,superscriptaddress,showpacs]{revtex4-2}
\usepackage{amsmath,amssymb}
\usepackage{graphicx}
\usepackage{physics}
\usepackage{hyperref}
\usepackage{xcolor}
\usepackage{booktabs}
\usepackage{stfloats}
\usepackage{float} % pour le [H]
\usepackage{booktabs}
\usepackage{colortbl}
\usepackage{xcolor}
\usepackage{booktabs,makecell,array,xcolor,colortbl}
\definecolor{headergray}{RGB}{235,235,235}
\definecolor{emptygray}{RGB}{245,245,245}

\newcommand{\Z}{\mathbb{Z}}

\begin{document}

\title{Canonical Quantization of Cylindrical Waveguides:\\A Gauge-Based Approach}
%to Modal Dynamics
%on enlève aucun intéret

\author{Alexandre Delattre}
\affiliation{Institut Néel, CNRS/UGA, 38042 Grenoble, France}

\author{Eddy Collin}
\email{eddy.collin@neel.cnrs.fr}
\affiliation{Institut Néel, CNRS/UGA, 38042 Grenoble, France}

\date{\today}

\begin{abstract}
We present a canonical quantization of electromagnetic modes in cylindrical waveguides, extending a gauge-based formalism previously developed for Cartesian geometries \cite{Delattre2024}. 
%By introducing a systematic gauge-fixing strategy
By introducing the two field quadratures $X,Y$
%, we isolate the true dynamical degrees of freedom
 of %each 
%%% Tu avais oublié les TEM?
%%% j'explicite les acronymes, tjrs la 1ere fois.
TEM (transverse electric-magnetic), but also of 
 TM (transverse magnetic) and TE (transverse electric)  traveling modes, 
%  juste pour isister sur traveling... mode tout court est misleading, on pense a qqch de localisé
%revealing
% ca fait un peu guindé, je propose plutot le mot:
we identify for each
a 
%%%
characteristic 
%%% faut définir ce qu'il a de spécial?
one-dimensional scalar field 
(a {\em generalized flux} $\varphi$) %%% je le mets explicitement!
governed by a Klein-Gordon type equation. The associated Hamiltonian is derived explicitly from Maxwell’s equations, allowing the construction of bosonic 
% c'est un peu redondant, mais ca cadre car le langage peut différer d'un auteur a l'autre...
ladder operators. % and the quantification of zero-point fluctuations. 
% terriblement naif. Tu quantifies tout, pas que la zpf... ca veut assez rien dire, ca...
The generalized flux is directly deduced from the electromagnetic potentials $\mathbf A,V$ by a proper {\em gauge choice}, generalizing Devoret's approach \cite{DevoretQED}.
Our analysis unifies the treatment of cylindrical 
%%%
and Cartesian
%%% c'est bien eux que t'unifies aussi, à la fin!! tout est unifié!!
guided modes under a consistent and generic framework, ensuring both theoretical %rigor 
insight
%%  ca, c'est terriblement baif.... Evidemment, que c'est rigoureux, sinon tu publies pas!!! mets plutot "insight" qui veut dire que cela t'apporte une compréhension profonde du problème. Ce qui est bien plus pertinent ici!
and experimental relevance. 
%%% ADD REF 3!!!
Especially, we find out that TE$_{n=0}$ modes break a fundamental gauge symmetry that is observed by all other families.
%%%%%
We derive mode-specific capacitance and % (inverse) 
%il me semble que c'est la densité de L^-1 qui apparait... La parenthèse est juste pour etre rigoureux... A verifier que je dis pas une connerie (mais je crois pas).
%%% en fait si, c'est un peu une connerie... il y a visiblement un rafinement de nomenclature qu'on a zappé dans le précédent papier...
inductance %per unit length
 from the field profiles and express voltage and current 
 % operators
  in terms of the canonical field variables.
%%%% DONC: ca, il va falloir le discuter en APPENDIX ou en CONCLUSION...
%%%%%%
% This approach preserves the direct mapping between quantum observables and measurable quantities such as voltage noise, impedance, and power transport. 
%%%% c'est un peu nul comme phrase... le mapping? et l'impedance n'est pas un concept quantique...
%%%% je rephrase:
Measurable quantities are therefore properly defined from the mode quantum operators, especially for the non-trivial TM and TE ones.
%%%%
%%%%
The formalism shall %naturally 
extend in future works 
%%%% pour etre un peu moins catégorique.
 to 
 %coaxial guides 
 %Ca, t'es sensé le traiter, c'est le plus important expérimentalement...
 any other type of waveguides, 
 %and topologically non-trivial waveguides, 
 % sinon, que signifie non-trivial? Il faudrait le definir... Et là, tu t'embarques dans des emmerdes.
 %%% tu te rappelles, on en a parlé, il faut surement inclure un "twist" dans le guide, ce qui est exclu ici...
 % je te propose plus soft, et plus pertinent:
 especially on-chip coplanar geometries particularly relevant to quantum technologies.
 %%%
 % paving the way for rigorous modeling of hybrid quantum systems and advanced parametric amplification protocols in cylindrical geometries.
 %% là, c'est assez obscur... De quoi parles tu exactement? Je comprends que tu veux ouvrir le débat, mais tu peux pas le faire aussi naivemement en balancant un truc comme ca... Il faut définir qqchose de plus précis et physique. cf ma proposition ci-dessus.
 % aussi: ta phrase est vraiment bizarre... tu parles d'étendre à d'autre waveguides et à la fin tu reviens au cylindrique... ca va pas.
\end{abstract}

\maketitle

\newpage

\tableofcontents

\section{Introduction}

% The canonical quantization of electromagnetic fields in confined geometries is a cornerstone of modern quantum physics.
%%% Bin là t'en fait trop, ou bien t'es trop précis... Justement, jusque là ils s'en sont foutus, c'est pour ca que ton boulot est utile!
%%%% je te propose plutot:
The transfer of quantum information via light fields (let them be lasers or microwave signals) is at the heart of modern quantum technologies.
%%% tu noteras le technologies, et pas physics, et transfer plutot que canonical quantization... ca, c'est le truc auquel tu dois arriver plus tard pour dire ta nouveauté.
%%%%
%From the early 
%%% early c'est comme si c'était ultra vieux... 
Building on the original 
developments of quantum electrodynamics (QED)~\cite{CohenQED, LoudonQED, GardinerZoller,Clerk2010}, 
% to 
the circuit-based approach of Devoret and collaborators (cQED)~\cite{DevoretQED, VoolDevoretReview, DevoretSchoelkopf2013} %, identifying 
identified 
generalized variables (flux and charge) %has 
that enabled 
%%%% attention, les 1ers exemple sont pour la lumière générique, pas pour la cQED... Faut faire une distinction...
a rigorous Hamiltonian formulation of superconducting quantum circuits. This formalism has been instrumental in the rise of cQED, 
%%%% du coup je mets l'acronyme
where superconducting qubits interact with quantized microwave modes~%of microwave resonators
%%%% ils interagissent avec des modes résonants (localisés) et des modes traveling, ta phrase était limitante pour rien, je laisse général en retirant juste le mot.
\cite{Blais2004cQED, Blais2021cQED,Gely2017Multimode}, as well as in the design of 
modern
%%%% attention: un qubit est aussi un nonlinear quantum device!!! donc faut différencier les deux...
nonlinear quantum devices such as traveling-wave parametric amplifiers (TWPAs)~\cite{Macklin2015TWPA, White2023TWPA}.
%%%% C'est pas mal du tout comme entrée en matière!

%While the canonical treatment of lumped elements and Cartesian waveguides has been extensively 
%%% bin là t'y va fort, tu peux pas dire que c'est extensively developed! on vient juste de sortir un papier!!!... 
% je rephrase:
While the explicit %%% ajouté
quantization of all types of waves (TEM, TM and TE) in Cartesian waveguides has been recently 
developed~\cite{Delattre2024}, the extension to cylindrical geometries remained %comparatively underexplored.
%%%% bin personne l'a fait, donc c'est plus que underexplored...
to be performed.
%%%% mauvais réflexe de francais de fiare des phrases à rallonge: COUPE!
%despite their ubiquity in 
%% je te propose:
These are indeed ubiquitous in 
both classical microwave engineering~\cite{CollinFieldTheory, Jackson1999, PozarMW} and modern quantum platforms. 
%%%%%
Cylindrical cavities and waveguides are routinely used in particle accelerators~\cite{Padamsee1998SRF}, cavity optomechanics (in its microwave version)~\cite{Aspelmeyer2014CavityOpto,Regal2008}, quantum hybrid systems~\cite{Kurizki2015Hybrid}, and topological photonics~\cite{Ozawa2019TopoPhotonics}. 
%%%%
%Their modal structure, %governed 
%%% governed c'est un peu strong
%described by Bessel functions, naturally gives rise to rich features such as degeneracies and azimuthal polarization splitting \cite{polsplit}, % [REF XXX], %% WHAT? XXX
%on top of the universal TEM, TM and TE mode families \cite{PozarMW}.
% universal families of modes 
%TE$_{m,n}$, TM$_{m,n}$, and TEM in coaxial geometries. 
%%%% tu as déjà introduit les TEM, TE et TM en abstract! donc il faut que tu fasses le lien; pas la peine de mettre les index si tu ne les définit pas!
% (TEM, TM and TE as already mentioned).
%%%% 
%Recent efforts to generalize canonical quantization beyond Cartesian structures highlight the necessity of a transparent framework, where the electromagnetic fields are expressed with explicit dependence on measurable parameters: wavevectors, cutoff frequencies, permittivities, permeabilities, and geometry-dependent profiles~\cite{Gely2017Multimode, VoolDevoretReview}. Such an approach ensures continuity between theoretical predictions and experimental observables, and directly supports the design of quantum microwave devices with engineered dispersion and confinement.
%%%
%%% ce bout là n'apporte rien, en particulier à ce point ci de ton texte. J'ai bougé tes refs d'ici plus haut, là où elle fitte très bien
%%%
%%%
%%% Là tu peux dire explicitement directement que c'est ta motivation,... OK!

In this work, we extend the canonical quantization framework of Ref.~\cite{Delattre2024} 
%%%% tu t'aimes tellement que tu t'es défini 2 fois en biblio!!
%%% CollinDelattre2025
to cylindrical waveguides. Starting from Maxwell’s equations
for electric $\mathbf E$ and magnetic $\mathbf B$ fields %% ajouté
in cylindrical coordinates, we re-derive the 
%%%% modif REF 3
explicit modal profiles for 
%%% manque TEM
TEM, TM and TE modes, 
%%%% là je reprend ta phrase qui est lourde:
%and express generalized fluxes and charges in analogy with Devoret’s prescription, and construct the Hamiltonian. 
and express all the fields' characteristics in terms of a generalized flux variable $\varphi$, in analogy with Devoret’s prescription, 
%%%% t'as plus de ref à la jauge, là? c'est pourtant ca qui est super important!! Devoret = Jauge!
imposing a proper electromagnetic gauge 
to the potentials $\mathbf A$, $V$. %%%% ajouté
%%%%
Our %treatment
modeling emphasizes the universality of the canonical formalism, 
%while making manifest the distinctive role of cylindrical symmetry and radial confinement. 
%%% là je comprend pas ce que tu veux dire: c'est trivialement les fonctions de Bessel, pas d'intéret... par contre effectivement, comparer Cartesien et cylindrique pour voir que c'est out pareil, ca c'est profond! C'est ca qu'il faut surtout dire!!!!
leading to a one-to-one correspondence between the Cartesian treatment and the cylindrical one.
%This paves the way to a rigorous treatment of coaxial TEM modes, which will be the subject of forthcoming work.
% non c'est fait.
%%%
%%% pour finir, je pense qu'il faut que l'on insiste sur un truc qu'on a pris comme évident dans l'autre papier..;
Especially, the notion of {\em virtual electrodes} is invoked, and 
 the specificity of the corresponding gauge is discussed.
 %%% ADD REF 3
 It turns out that it breaks a fundamental gauge symmetry obeyed by all other types of traveling waves.
%%%%%
With the identification of proper mode-dependent voltage and current quantities, 
%%%%%%%%%%%
the mapping to transmission line theory (with the definition of effective capacitance and inductance per unit length) can be naturally obtained for {\em any} type of waves.
To date, this is performed rigorously only for the TEM ones \cite{PozarMW,Clerk2010}, 
%%% en appendix, ou en conclusion? a discuter...
and shall be developed in oncoming articles. %future works. %%% OK?

\section{Prelude on Bessel functions}
%%%%%%%%%%%%%% ok... pourrait etre en appendix...

%%%% definis de suite a quoi sert cette section... comme ca les gens peuvent la sauter...
We give here the basic mathematical tools specific to the cylindrical symmetry, for the interested reader willing to redo the calculations.
%%%
The analysis of electromagnetic modes in cylindrical geometries \cite{PozarMW} 
%%% plus que des bouquins de maths, la ref de PHYSIQUE sur le cylindrique est importante...
%%%% Achtung! a nouveau tu as des refs multiples...
%%%pozar2012microwave
naturally leads to the \emph{Bessel differential equation}~\cite{abramowitz1964handbook, nist2010handbook}:
%%%%%
%%%%% c'est ultra-chiant, mais je te change tout tes m en n; la littérature définit cet indice comme n, c'est mieux d'etre raccord!
%%%%%
\begin{equation}
x^2 \frac{d^2 y}{dx^2} + x \frac{dy}{dx} + (x^2 - n^2)\, y = 0,
\label{eq:bessel_eq}
\end{equation}
where $n \in \mathbb{Z}$ is the azimuthal index, arising from the separation of variables in the angular coordinate $\theta$
%%%%%%%%%%%% separation de quoi et quoi...
and radial coordinate $r \geq 0 $; $x=k_c \, r$ with $k_c$ a positive real number to be defined (see below). 
%%% definir les coordonnées...
%%% sinon c'est obscur... qui est x??? par rapport à r???
%%% d'ou sort ton k_c plus bas, il est pas défini???
%%% a et b ne sont définis que plus tard!!
%%% je pense il manque une figure pour comprendre ca...
The two linearly independent solutions of Eq.~\eqref{eq:bessel_eq} are the \emph{Bessel functions of the first kind} $J_n(x)$ and the \emph{Bessel functions of the second kind} $Y_n(x)$, also known as Neumann functions. 
Their mathematical properties are extensively documented in the %classical 
%%% mot malheureux... vs quantum?
literature~\cite{abramowitz1964handbook, nist2010handbook}, and their application 
%to electromagnetic theory 
%%% c'est très bizarre ta phrase! Effectivement, on retoruve ca partout, pas QUE en électromagnétisme!!!
is standard in many areas of physics~\cite{arfken2013mathematical}.

\begin{figure}[H]
\centering
\hspace*{-0.7cm}
\includegraphics[width=1.2\linewidth]{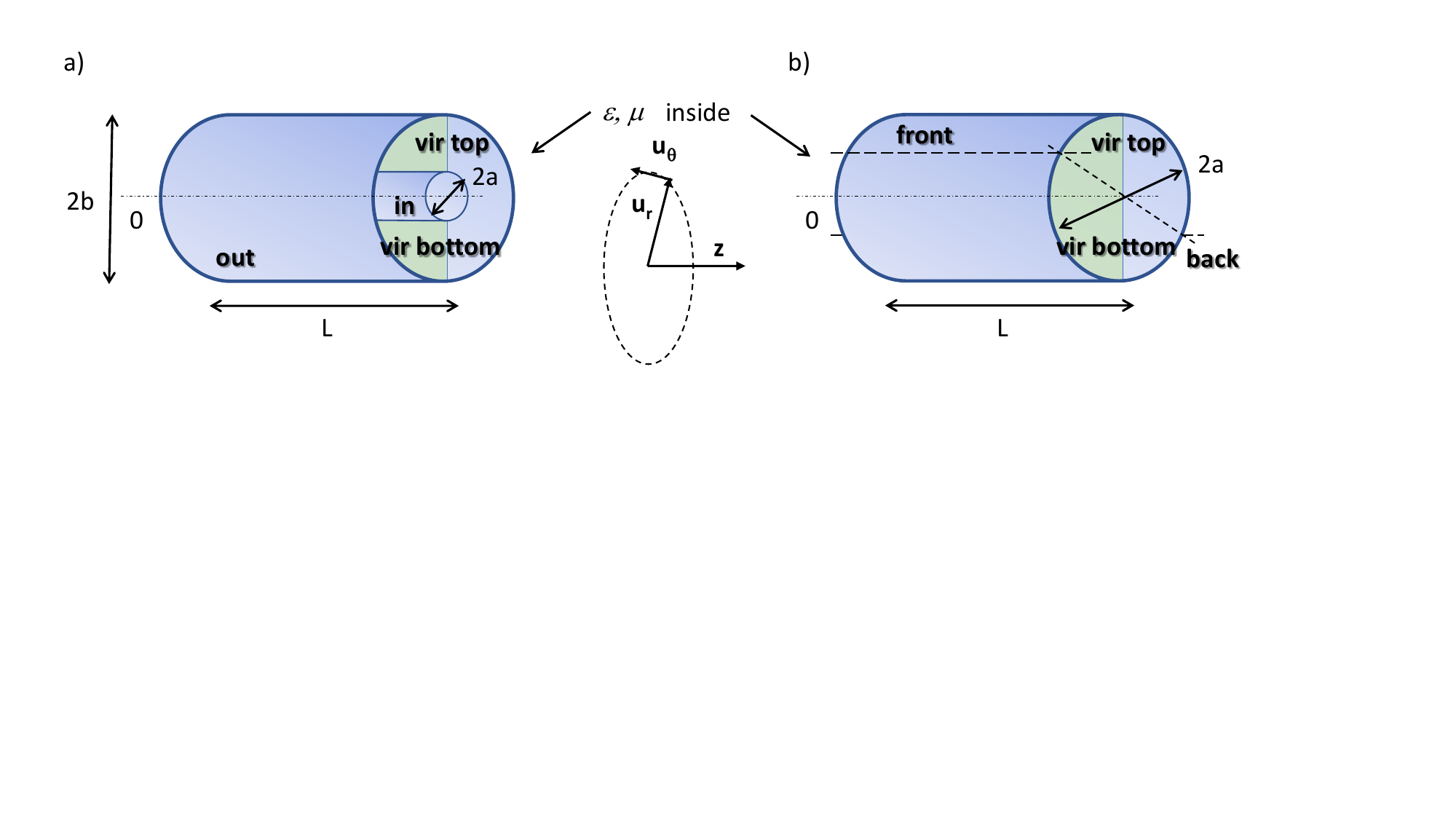} %{TM11_zpf_curve.png}
\vspace*{-6cm}
\caption{Geometries with cylindrical symmetry. 
 a) coaxial line. b) hollow cylinder.
 The axes corresponding to the coordinate system $r,\theta,z$ are displayed.
Real electrodes are shown in blue ($in$ and $out$ labels for coaxial, and $front$, $back$ for hollow pipe).
The {\em virtual electrodes} (labeled {\em vir top} and {\em vir bottom}) are shown in green. See text for details.
% Mettre $\mu$ and $\epsilon$ on graph! longueur $a, b, L$!
%%%% assez clair??????
}
\label{fig_1}
\end{figure}

The function $J_n(x)$ is finite at $x=0$, whereas $Y_n(x)$ diverges as $x \to 0$; 
%%% t'es un peu obligé de définir les n négatifs, tu t'en sers avec tes dérivés...
both verify $J_{-n}(x)=(-1)^n \, J_n(x)$ and $Y_{-n}(x)=(-1)^n \, Y_n(x)$ for integer $n \ge 0$.
%%%
For %domains
waveguides that include the axis $r=0$ (namely  
%% e.g.
%%% c'est pas, "par exemple", parce que... il n'y a QUE le cylindre!!...
% solid cylindrical waveguides 
%%% pas nécessaire?? Quel matériau??? 
hollow cylinders), regularity of the physical fields requires discarding $Y_n(x)$ in $\mathbf E, \mathbf B$ field expressions. 
In contrast, for %annular domains 
%such as 
%%% a nouveau, il n'y a qu'une possibilité!!!!!
coaxial waveguides the origin is excluded from the domain and both $J_n$ and $Y_n$ are admissible solutions. 
In such cases, the generic 
%%%% oui mon general!
radial profile takes the form:
\begin{equation}
R(r) = A\,J_n(k_c \, r) + B\,Y_n(k_c \, r),
\label{eq:bessel_general}
\end{equation}
with coefficients $A$ and $B$ determined (as well as $k_c$) by the electromagnetic boundary conditions on the conducting surfaces at $r=a$ and $r=b$, see Fig. \ref{fig_1}.
%%%%  image a faire pour etre clair!!! comme la fig 1 du papier cartesien!!!
%%% c'est necessaire!!!

Several identities involving $J_n$ and $Y_n$ are of practical importance in mode analysis. 
The \emph{Wronskian identity}:
\begin{equation}
J_n(x) Y_n'(x) - J_n'(x) Y_n(x) = \frac{2}{\pi x},
\label{eq:wronskian}
\end{equation}
is frequently used in normalization procedures. 
The {\em derivative recurrence} relations:
\begin{equation}
J_n'(x) = \frac{1}{2}\big[J_{n-1}(x) - J_{n+1}(x)\big], 
\quad
Y_n'(x) = \frac{1}{2}\big[Y_{n-1}(x) - Y_{n+1}(x)\big],
\label{eq:derivative_recurrence}
\end{equation}
allow one to express radial derivatives directly in terms of Bessel functions of adjacent orders. 
%%%%%
%The oscillatory behavior in the large-$x$ limit reflects the wave nature of the underlying modes, with $J_m$ and $Y_m$ playing roles analogous to standing-wave sine and cosine components in Cartesian geometries. 
%%%%%%%%%%%%%%%
%%% ce paragraphe est très misleading...
%%% d'abord, la "wave nature" c'est parce que tu as une propagation sinusoidale en z. Pas en radial.
%%% ensuite, à large x est misleading parce que les 2 fonctions tendent vers 0 à large x!!!
%%% Quel est vraiment le but de cette phrase? Je propose:
$J_n$ and $Y_n$ play here roles analogous to the transverse sine and cosine components found in Cartesian geometries \cite{Delattre2024}. 
%%%
In the following sections, these functions
will serve as the building blocks for the radial profiles of %TEM, 
TM and TE modes in coaxial waveguides and hollow cylinders, with the specific combination in Eq.~\eqref{eq:bessel_general} fixed by the mode type and boundary conditions \cite{PozarMW}.

%%%%%%
%%%%%%
%%%%%%
%%%%%%

\section{Basics: Maxwell’s Equations in Cylindrical Geometry}
%%%%% Dis de suite que tu as 2 types, et commence bien par celui que tu traiteras en premier.

%%%% je propose:
We consider two types of guides preserving the cylindrical symmetry with transverse coordinates $\{r,\theta\}$, see Fig. \ref{fig_1}. We assume that they are filled with a lossless,  homogeneous, and isotropic medium of permittivity $\epsilon$ and permeability $\mu$, over a length $L$.
%%% mu, epsilon pourquoi pas... ca correspond plus a un vrai coax...
%%%% ADD REF 3
Treating distributed losses in a quantum-mechanical way is not an easy task, and is far outside of our scope; see e.g. Ref. \cite{PhysRevApplied.12.034054} dealing with Traveling Wave Parametric Amplifiers. 
%%%%%%%% REF
%\bibitem{PhysRevApplied.12.034054} Loss Asymmetries in Quantum Traveling-Wave Parametric Amplifiers, Houde, M. and Govia, L.C.G. and Clerk, A.A., Phys. Rev. Appl. Vol. 12, 034054 (2019). 
%%%%

% We shall also consider 
The coaxial line
%cylindrical geometry, 
consists of {\it two} perfectly conducting concentric cylindrical surfaces of inner radius $r=a$ and outer radius $r=b$ (with $b>a$). The $z$-axis is aligned with the common axis of both conductors. %We label the inner conductor ($r=a$) as electrode $\zeta_1$ and the outer conductor ($r=b$) as electrode $\zeta_2$. 
%%% ca, c'est foireux. Appelle les inner et outer, non?
%%% et c'est pas tellement d'un nom pour l'électrode dont tu as besoin, c'est d'une désignation pour toutes les propriétés qui lui sont rattachées!!!!
%%% je propose:
The inner conductor properties are designated by a label $in$, while the outer ones take an $out$.
%%%% c'est tellement plus simple...
%%% t'as vu: on a défini tous les axes, r theta et z, et les deux electrodes  r=a et r=b. Et la longueur L! TOUT ca doit apparaitre en Fig 1, et du coup c'est clair.

The other type of guide possesses a {\it single} electrode: it consists in a cylinder
%We consider a lossless cylindrical waveguide 
of radius $a$, oriented along the $z$-axis.
%
%, filled with a homogeneous, isotropic medium of permittivity $\epsilon$ and permeability $\mu$. 
%%% Ca, ca s'applique aux 2 guides!!! donc je le mets plus haut.
As for the coaxial line, 
%%% pour faire le lien
the %inner 
%%% ce inner est misleading, parce qu'il y avait du inner/outer pour coax, et que la il y en a plus... j'enlève le mot 
surface at $r = a$ is assumed to be a perfect electric conductor. 
%, ensuring ideal confinement of the electromagnetic field. 
%%% useless
%%%
We attach to the properties related to this electrode the labels  $front$ and $back$, designating two radially opposite points of the surface, for any $\theta, \theta+\pi$.
%%% ca parait con, mais ca permettra d'avoir une systématique dans ton écriture je pense...
%%% c'est là que tu peut faire un lien au cartésien assez gratos:
This corresponds here to the two sets of equivalent electrodes defined in the Cartesian case ({\em top, bottom} and {\em left, right} respectively) \cite{Delattre2024}, but preserving the cylindrical symmetry. 

Treating %both 
Cartesian and cylindrical geometries within the same formalism is %thus 
%%% indeed marche mieux, avec ce que j'ai noté dessus.
indeed particularly 
%appealing:
enlightening: 
%%%%
%it allows one to capture the full modal spectrum (including TEM, TE$_{m,n}$, and TM$_{m,n}$ families) in a unified gauge-based framework, to make explicit the topological equivalence between parallel-plate and coaxial lines, and to exploit the concept of \emph{virtual electrodes} for modes whose profiles satisfy perfect conductor boundary conditions on non-physical surfaces. This unified approach is directly relevant to the modeling of hybrid quantum systems, precision metrology, and the design of low-noise parametric devices operating in cylindrical environments.
%%% t'en fais des caisses... on sait même plus c'est quoi le message. Soit plus sharp:
it enables to %demonstrate 
reveal the topological equivalence between these guided modes, and %enables
also to refine our understanding of 
\emph{virtual electrodes}, as introduced in Ref. \cite{Delattre2024}.
%%%% on introduit de suite... pour expliciter Fig. 1
They correspond to diameter planes, with {\em vir top} designating the top part of the diameter and {\em vir bottom} the other side 
(see Fig. \ref{fig_1}).
%%%%
 These are defined for specific TE modes only, see discussion below.
We demonstrate that %all
the concepts introduced in %this Reference 
the (simpler) Cartesian geometry
%%%% ok?
 do apply here, when properly adapted, especially regarding the gauge choice.
%%%% je pense que là, j'ai dit ce qu'il y avait à dire?

\subsection{Maxwell's equations % in vacuum
in an ideal material}

% In vacuum, in the linear regime (low intensity, microwave domain), 
%% tiens, plus haut tu as dit que tu avais un diélectrique???
%%% 
The electric and
magnetic fields $\mathbf{E}$ (V/m) and $\mathbf{B}$ (T) obey the well-known Maxwell equations:
\vspace*{-4mm}
\begin{align}
\nabla \cdot \mathbf{E}(r,\theta,z,t) &= 0, \label{eq:Maxwell1} \\
\nabla \cdot \mathbf{B}(r,\theta,z,t) &= 0, \label{eq:Maxwell2} 
\end{align}
\begin{align}
\mathbf{\nabla} \times \mathbf{E}(r,\theta,z,t) &= -\frac{\partial \mathbf{B}(r,\theta,z,t)}{\partial t}, \label{eq:Maxwell3} \\
\mathbf{\nabla} \times \mathbf{B}(r,\theta,z,t) &= \frac{1}{c^2}\,\frac{\partial \mathbf{E}(r,\theta,z,t)}{\partial t}, \label{eq:Maxwell4}
\end{align}
with $c=1/\sqrt{\mu \, \epsilon}$ the speed of light in the chosen dielectric ($\epsilon$ permittivity, $\mu$ permeability).
%vacuum. 
Since we aim at describing quantum information
transfer in the low-energy QED regime, these linear relations are all we need: 
we assume that the material is {\it ideal}, with no nonlinear properties. %: non-linear Euler–Heisenberg corrections due to light-by-light scattering can be safely ignored.
%%% dans ton cas, le diélectrique claquera bien avant!!!
Dielectric losses are also neglected, as already mentioned.
%%%% ptet lourd????

\subsection{Boundary conditions in cylindrical geometry}
\label{boundaryes}

In cylindrical coordinates $(r,\theta,z)$, the conductor walls of our configurations are % is 
located at $r=a$ and $r=b$, with a normal unit vector given by
$\mathbf{n} = \pm \mathbf{u}_r$ depending on the orientation of the surface ($+$ for the inner conductor of the coaxial guide, and $-$ for the outer one or for the cylinder).
%and its outward unit normal vector, pointing towards the inside of the guide, is $\mathbf{n} = -\mathbf{r}$. 
%% c'est maladroit parce que en coax tu as les 2 possibilités. Je rephrase avec un n général...
%%
The generic
% Mon general 
electromagnetic boundary conditions %at the conductor surface 
on the electrodes
%%% juste pour pas répéter surface ou conductor...
read:
\begin{align}
\mathbf{n} \cdot \mathbf{E}(r=\{a;b\},\theta,z,t) &= + \frac{\sigma_s(\theta,z,t)}{\varepsilon %_0
}, \label{eq:bc1} \\
\mathbf{n} \cdot \mathbf{B}(r=\{a;b\},\theta,z,t) &= 0, \label{eq:bc2} \\
\mathbf{n} \times \mathbf{E}(r=\{a;b\},\theta,z,t) &= 0, \label{eq:bc3} \\
\mathbf{n} \times \mathbf{B}(r=\{a;b\},\theta,z,t) &= +\mu %_0 
\, \mathbf{j}_s(\theta,z,t), \label{eq:bc4}
\end{align}
where $\sigma_s(\theta,z,t)$ is the surface charge density and $\mathbf{j}_s(\theta,z,t)$ is the surface current density on the wall.
%Equations~(\ref{eq:bc1}) and (\ref{eq:bc4}) show that the presence of a surface current or charge density is fully compatible with a perfect electric conductor boundary; these quantities are responsible for sustaining the tangential magnetic field and the normal electric field at the surface.
%%%%
%\begin{align}
%E_r(a,\theta,z,t) &= -\frac{\sigma_s(\theta,z,t)}{\varepsilon_0}, \label{eq:bc5} \\
%E_\theta(a,\theta,z,t) &= 0, \label{eq:bc6} \\
%E_z(a,\theta,z,t) &= 0, \label{eq:bc7} \\
%B_r(a,\theta,z,t) &= 0, \label{eq:bc8} \\
%B_\theta(a,\theta,z,t) &= - \mu_0 j_{s,z}(\theta,z,t), %\label{eq:bc9} \\
%B_z(a,\theta,z,t) &= \mu_0 j_{s,\theta}(\theta,z,t). \label{eq:bc10}
%\end{align}
%%% EUh.... la je suis désolé, mais c'est de la paraphrase: c'est construit comme ca!! On a compris, pas la peine d'insister aussi lourdement sur du textbook...
%%%%%%%
In the following sections, we will use these electromagnetic boundary conditions to determine the mode families supported by the guide. 
%: TEM, TE %$_{m,n}$  and TM %$_{m,n}$  families. 
%%%% ca suffit les familles!!! rien a foutre!!!
Our only physical assumption is that the %real 
%%%% je mets physique, parce que mathématiquement, tu separe les variables...
%%% trop vite... la notion d'électrode réelle ou virtuelle vient plus tard...
%%% on l'a simplement évoquer dans l'intro de Maxwell, ici, ca vient comme un cheveux sur la soupe...
electrodes are made of %good 
ideal conductors, so that losses can be neglected. % over the guide length $L$. 
%%% c'est plus haut qu'il faut définir L!! Là c'est un peu tard...
%%%%
But note that nothing is specified about the microscopic nature of the charges (free electrons or Cooper pairs): only their coupling to the electromagnetic field is relevant.
%%%%
%%%% là, je pense qu'il faut introduire en 1 mot la notion d'electrode virtuelle, sur laquelle on reviendra ensuite... Elles sont plus dures à introduire que dans le cas cartesien ;-)
The same boundary conditions Eqs. (\ref{eq:bc1}-\ref{eq:bc4}) shall apply to {\em virtual electrodes} (choosing $\mathbf{n}=\pm \mathbf{u_\theta}$ for normal vector, and $a<r<b$ or $0<r<a$ depending on the configuration), as discussed in the following.
%%%% on définit tout de suite!!!

\subsection{Factorized form of the fields}
\label{factor}

The factorized representation of the fields applies to all guided modal families supported 
by the geometry under consideration. 
%In a hollow cylindrical waveguide (single conducting 
%boundary at $r=a$ only), TE$_{m,n}$ and TM$_{m,n}$ modes exist; a TEM mode is not supported 
%because there is no second conductor to provide a uniform longitudinal potential. In a 
%coaxial cylindrical waveguide (two conducting boundaries at $r=a$ and $r=b$), TEM, TE$_{m,n}$ 
%and TM$_{m,n}$ modes are all possible. 
%%%
%%% je comprends pas pourquoi tu veux coller ca ici, alors que le but de la section c'est de donner la forme GENERIQUE des champs? T'as pas à discuter ca ici; ca fait au moins 2 ou 3 fois dans ton texte que tu y reveiens, alors que c'est pas le lieu!!!
Considering traveling wave solutions, we therefore 
%%% j'ajoute le mot
define, following conventions from Ref. \cite{Delattre2024} :
%%%% j'ajoute la ref!
\begin{align*}
f(z,t) &= X \cos(\omega t -\beta z +\phi_0) + Y\sin(\omega t -\beta z +\phi_0),\\
\tilde{f}(z,t) &= X \sin(\omega t -\beta z +\phi_0) - Y\cos(\omega t -\beta z +\phi_0),
\end{align*}
%%% pas de chance, theta_0 est misleading avec tes coordonnées; je change.
%%% dis tout de suite ce que tu as introduit!!
with $\beta= 2\pi\,l/L$ the wavevector (we use periodic boundaries, $l \in \Z^*$), $\omega = c \, k$ the angular frequency ($k>0$), and $\phi_0$ an arbitrary phase linked to the $t,z$ origins. The phase velocity reads $v_\phi=c\, k/|\beta|$.
%%%%%%%%%% OK? j'ai rajouté là la phase velocity...
%%%% Pitain!!! T'as même pas introduit X et Y !!!!!!
%%% la blague....
$X$ and $Y$ are the two field quadratures (no units), which quantify the state of the microwave field. 
%
%For any supported mode family,
%%% ca t'as deja utilisé l'expression ci-dessus
The electromagnetic field 
components can then be expressed as: %in the factorized form:
\begin{align}
E_r(r,\theta,z,t) &= E_m\, g_{E_r}(r,\theta)\, f(z,t), \label{eq:Er_fact} \\
E_\theta(r,\theta,z,t) &= E_m\, g_{E_\theta}(r,\theta)\, f(z,t), \label{eq:Ephi_fact} \\
E_z(r,\theta,z,t) &= E_m\, g_{E_z}(r,\theta)\, \tilde{f}(z,t), \label{eq:Ez_fact} \\
B_r(r,\theta,z,t) &= B_m\, g_{B_r}(r,\theta)\, f(z,t), \label{eq:Br_fact} \\
B_\theta(r,\theta,z,t) &= B_m\, g_{B_\theta}(r,\theta)\, f(z,t), \label{eq:Bphi_fact} \\
B_z(r,\theta,z,t) &= B_m\, g_{B_z}(r,\theta)\, \tilde{f}(z,t), \label{eq:Bz_fact}
\end{align}
where $E_m > 0$ sets the electric field amplitude, and $B_m = E_m/c$ the magnetic field amplitude. Note that $g_{E_i}$ and $g_{B_i}$ are \emph{dimensionless}:
%For now they can be left independent. 
%%% inutile
these functions %$g_i$
are the normalized transverse modal 
profiles, to be determined for each configuration by solving Maxwell’s equations with the boundary 
conditions Eqs. (\ref{eq:bc1})–(\ref{eq:bc4}). 
%%%%%%%%%% OK!
%%%%%%
In the next sections, %these modal structures families will be computed 
this is performed 
explicitly for the coaxial structure 
(leading to the well known TEM, TM, TE families), and then %successively 
for the hollow cylinder (TM and TE), 
%%%% et là je pense il faut mettre un lien a la litterature... qu'apportes-tu??
reformulating (for our purpose) results found in the literature \cite{PozarMW}.
%%% beaucoup plus sharp, non?
%%%%
%, showing their dependence on Bessel functions, as a direct analogue to the sinusoidal modes of the Cartesian case.
%%%%
%This structural choice tends to show the direct topological equivalence with the parallel plates and rectangular guides in Cartesian geometry. 
%%% ca, on l'a déjà dit plus haut!
The longitudinal envelopes verify, %are chosen 
%%% mieux comme mot
as in the Cartesian formulation:
\begin{equation}
\frac{\partial f}{\partial t}=-\omega\,\tilde f,\quad \frac{\partial f}{\partial z}=+\beta\,\tilde f,
\qquad
\frac{\partial \tilde f}{\partial t}=+\omega\,f,\quad \frac{\partial \tilde f}{\partial z}=-\beta\,f,
\label{eq:longitudinal_envelopes}
\end{equation}
which are identities used to simplify the calculations.
%%% c'est tout ce que ca porte comme info...
%%%%%
%%% non, utilise les memes notations partout, si tu as mis des  d/dt, utilise ca partout!!! Je corrige.
%%%%
%so that the transverse components are proportional to $\tilde f$ and the ``scalar'' components ($E_z$ for TM, $B_z$ for TE) are proportional to $f$.  
%%% là je vois pas le rapport avec la choucroute, et ton "scalar"????
%%%%%
%%%%%
%%%%%
%Unless otherwise stated, we adopt the following normalization for the modal profiles:
%\begin{equation}
%\text{TEM: } g_{E_r}(a)=1,\qquad
%\text{TE: } g_{B_z}(a)=1.
%\label{eq:normalizations}
%\end{equation}
%Azimuthal dependence follows the usual parity $\cos[m(\theta-\theta_0)]$ or $\sin[m(\theta-\theta_0)]$, indicated explicitly in each expression.
%%%%
%%%% alors là, j'ai rien compris... 
%%%% comment ca TE? c'est toujours vrai?
%%%% pourquoi parler de la dépendence azimuthal ici alors qu'on a pas encore les formules sous le nez????
%%%% tu ne parles pas de ta facon de normer les TM... pourquoi???
%%%%
%%%% Moi, je mettrais à ce niveau là une phrase "bateau" qui sera vrai tout le temps... C'est plus simple.
We take arbitrarily the sign convention for the transverse profiles such that the charge amplitude on the $r=a$ conductor (inner one for the coaxial line, or confining cylinder for the other type of guide) is positive,  
%%% J'ai conscience que c'est pas tellement plus clair que ce que tu avais écrit...
%%% j'ajoute de suite la normalisation du champ... plus subtile que dans le cas carré... est-ce bien compréhensible?
with a normalization   such that the electric field amplitude verifies  $|g_{E_r}(r=a)|=1$ on this electrode (regardless of %not considering
 the azimuthal $\cos$ or $\sin$ dependence when it exists; see below).
%%% et comme ca on n'y revient plus après!
%%% MAIS: cas particulier electrodes virtuelles... je le mentionne juste a cette etape, car elles sembler sortir du chapeau...
The case of  virtual electrodes shall be specifically discussed when necessary.
%%% OK? Hum...

%Sign conventions for $I$ and $\Delta V$ are specified below when introducing the electrodes.
%%%%
%%%% encore plus bizarre... I et delta V ne sont pas encore définis, d'où sortent-ils?

% =========================================================
% Modal profiles in cylindrical geometry — Final English version
% =========================================================

\subsection{Coaxial waveguide}
\label{sec:coax}

Three distinct families of propagating waves exist in a coaxial line \cite{PozarMW}. 
%%%%
They are in direct correspondence with the ones found for electromagnetic signals propagating between two plates \cite{Delattre2024}.
These are presented in the following.
%%%% ADD REF 3
All the mode shapes described below can be found in the literature, with schematics for the $\bf E, \bf B$ field profiles \cite{CollinFieldTheory, Jackson1999,PozarMW}. These can also be produced by numerical methods if required.
%%%
%%%% juste parce que je n'aime pas quand il y a pas de petite phrase de lien logique...

\subsubsection{TEM modes}
\label{sec:coax_TEM}

The TEM family is the simplest one, % pour justifier que tu commences là
and it satisfies:
\begin{equation}
k=|\beta|, %\qquad \omega_c=0, 
\qquad E_z=0,\quad B_z=0,
\end{equation}
%with no cutoff frequency and nondispersive propagation (\(v_\phi=v_g=c\)).  
%%% c'est pas comme ca qu'il faut l'écrire, tu n'as pas défini de cutoff avant, et tu n'as pas défini qui sont v_phi et v_g!! 
%%% je propose:
with no longitudinal components, and a linear dispersion relation (thus $v_\phi=c$).
The dimensionless profiles %, normalized by $g_{E_r}(a)=1$ according to \eqref{eq:normalizations}, 
%%% du coup je vire
are given in Tab. \ref{tab1} below.
%%% c'est comme ca qu'on fait.
\begin{table}[H]
\centering
\caption{Dimensionless modal profiles for the TEM modes in a coaxial waveguide.
%($m=0$), normalized by $g_{E_r}(a)=1$.
%%% pas de m, c'est pas défini, et la normalisation est donnée une fois pour toutes, on y revient pas
}
\label{tab:TEM_coax}
\begin{tabular}{@{}l l@{}}
\toprule
\textbf{Function} & \qquad \textbf{Expression} \\
\midrule
$g_{E_r}(r)$      & $=\;\dfrac{a}{r}$ \\[4pt]
$g_{E_\theta}(r)$ & $=\;0$ \\[4pt]
$g_{E_z}(r)$      & $=\;0$ \\[4pt]
$g_{B_r}(r)$      & $=\;0$ \\[4pt]
$g_{B_\theta}(r)$ & $=\;\dfrac{a}{r}\,\mathrm{sign}(\beta)$ \\[4pt]
$g_{B_z}(r)$      & $=\;0$ \\
\bottomrule
\end{tabular}
\label{tab1}
\end{table}
%%%% petit commentaire final:
The $\mathbf{E}$ field is radial, and the $\mathbf{B}$ field tangential \cite{PozarMW}. 
%%%
The wavevector $\beta$ characterizes each propagating mode.
%%% pour faire le lien aux autres qui ont besoin d'indices...

\subsubsection{TM %$_{m,n}$
modes}
\label{sec:coax_TM}

TM modes are defined by $B_z=0$ and $E_z\neq0$.  
%The perfect conductor boundary conditions impose the dispersion relation:
%%% c'est pas ca une relation de dispertion!!! c'est la formule ci-dessous... k^2=k_c^2+beta^2 !!!
%%% je rephrase: 
The boundary condition on the two electrodes impose ($n \geq 0$):
%%%% c'est malheureux... la convention standard c'est n,m, avec n pour la Bessel... je modifie...
\begin{equation}
J_n(k_c a)\,Y_n(k_c b)-J_n(k_c b)\,Y_n(k_c a)=0,
\label{eq:coax_TM_disp}
\end{equation}
from which one obtains the dispersion relation $k^2=k_c^2+\beta^2$, with $\omega_c=c\,k_c$ the cutoff frequency below which the TM waves cannot propagate.
% with $k_c$ fixed by \eqref{eq:coax_TM_disp}.
%%% là faut etre plus clair... qui est n? 
%%% je te propose:
Eq. (\ref{eq:coax_TM_disp}) leads to a discrete set of solutions for $k_c$, that we index by $m > 0$. % c'est visiblement la convention!!! Alors on va s'y tenir?
%%% j'exprime:
Each propagating mode is thus characterized by {\em two} indexes $n,m$ 
%%%% là, il y avait une remarque intéressante à faire...
%%%%   tu vois?
 (while in the parallel plate case, only one was enough \cite{Delattre2024}), plus a wavevector $\beta$.  %%% on en reparlera dans la suite...
The dimensionless %, derivative-free 
%%% what??? incompréhensible et inutile...
profiles are given in Tab. \ref{tab2}, with $\theta_0$ an arbitrary choice for the $\theta=0$ reference. 
%%%% je définis ce qu'est theta_0, sinon c'est bizarre...
% for $m\ge1$ are:
%%% pourquoi veux-tu distinguer ces cas??? Il n'y a rien de particulier là...
\begin{table}[H]
\centering
\caption{Dimensionless modal profiles for TM$_{n,m}$ in a coaxial waveguide ($n \geq 0, m>0$). We defined $A_{n,m}=(4 \beta)/[\pi \, k_c^2 a \, Y_n(k_c a)]$.}
\label{tab:TM_mn_coax}
\begin{tabular}{@{}l l@{}}
\toprule
\textbf{Function} & \qquad \textbf{Expression} \\
\midrule
$g_{E_r}(r,\theta)$     
& $=\;-\dfrac{\beta}{k_c \, A_{n,m}}\!\left[J_{n-1}(k_c r)-J_{n+1}(k_c r)-\dfrac{J_n(k_c a)}{Y_n(k_c a)}\!\left(Y_{n-1}(k_c r)-Y_{n+1}(k_c r)\right)\right]\!\cos[n(\theta-\theta_0)]$ \\[6pt]
$g_{E_\theta}(r,\theta)$
& $=\;+\dfrac{2\beta\,n}{k_c^2 r\,A_{n,m} }\!\left[J_n(k_c r)-\dfrac{J_n(k_c a)}{Y_n(k_c a)}\,Y_n(k_c r)\right]\!\sin[n(\theta-\theta_0)]$ \\[6pt]
%%%%% kc^2 au denom!!!!
$g_{E_z}(r,\theta)$     
& $=\;+\dfrac{2}{A_{n,m}} \left[J_n(k_c r)-\dfrac{J_n(k_c a)}{Y_n(k_c a)}\,Y_n(k_c r)\right]\!\cos[n(\theta-\theta_0)]$ \\[6pt]
$g_{B_r}(r,\theta)$     
& $=\;-\dfrac{2k\,n}{k_c^2 r\,A_{n,m} }\!\left[J_n(k_c r)-\dfrac{J_n(k_c a)}{Y_n(k_c a)}\,Y_n(k_c r)\right]\!\sin[n(\theta-\theta_0)]$ \\[6pt]
%%% pareil le k_c^2 !!!!!!
$g_{B_\theta}(r,\theta)$
& $=\;-\dfrac{k}{ k_c \,A_{n,m} }\!\left[J_{n-1}(k_c r)-J_{n+1}(k_c r)-\dfrac{J_n(k_c a)}{Y_n(k_c a)}\!\left(Y_{n-1}(k_c r)-Y_{n+1}(k_c r)\right)\right]\!\cos[n(\theta-\theta_0)]$ \\[4pt]
$g_{B_z}(r,\theta)$     
& $=\;0$ \\
\bottomrule
\end{tabular}
\label{tab2}
\end{table}
The $\mathbf{E}$  and  $\mathbf{B}$ fields have thus {\em both} a specific angular $\theta$ and radial $r$ dependence \cite{PozarMW}. They resemble pretty much the solutions found in Subsection \ref{sec:TM_hollow} for the hollow cylinder.

\subsubsection{TE modes}
\label{sec:coax_TE}

TE modes satisfy $E_z=0$ and $B_z\neq0$. 
% The dispersion relation enforced by perfect-conductor boundaries is
%%% c'est pas une relation de dispersion!!
The boundary condition writes this time, for $n \geq 0$:
\begin{equation}
J_n'(k_c a)\,Y_n'(k_c b)-J_n'(k_c b)\,Y_n'(k_c a)=0,
\label{eq:TE_coax_dispersion}
\end{equation}
%%% a nouveau, je remets n parce que c'est le standard
%with $k_c^2=(\omega/c)^2-\beta^2$. 
%% meme notation compacte:
with again $k^2=k_c^2+\beta^2$.
The cutoff frequency $\omega_c = c \, k_c $ is obtained through the zeroes of Eq. (\ref{eq:TE_coax_dispersion}), which solutions $k_c$ are indexed through $m >0$.
%%% a nouveau, il faut expliquer d'ou sort le 2eme indice!!!
%%%
As for TM waves, the coaxial case requires two indexes while the parallel plate needed only one \cite{Delattre2024}.
%Using the normalization $g_{B_z}(a)=1$ (cf. Eq.~\eqref{eq:normalizations}), the dimensionless profiles read:
%%
%%% je comprends pas pourquoi ti uas voulu changer de normalisation... C'est bien expliqué dans l'autre papier qu'on garde tjrs E comme reference!
%%
%% on introduit tableau et figures dans le texte avec leur nom.
%%% Mais ici, il y a une GROSSE SUBTILITE!!! c'est quand n==0!!!
%%% pourquoi t'as voulu détailler ce cas en TM, qui est impertinent, et PAS en TE là ou c'est essentiel??? Je comprends pas....
%%% Je fais donc la difference:
But there is a subtlety when $n=0$; we therefore give first   
the transverse profiles in Tab. \ref{tab3} for $n>0$, and treat then in Tab. \ref{tab4} the $n=0$ case separately.

%%%
%%% là, il faut mettre un Tab 4, et discuter le cas n==0.

%%% petit commentaire global come pour les autres modes, avant de discuter UN PEU le cas pathologique....

%%
\begin{table}[H]
\centering
\caption{Dimensionless modal profiles %$g_{E_i}$, $g_{B_i}$ 
for TE$_{n,m}$ % on mets toujours n,m dans la littérature?!
modes in a coaxial waveguide %(\(g_{B_z}(a)=1\))
($n>0,m>0$). We defined $A_{n,m}=-\frac{2 k\,n}{k_c^2 a}\dfrac{ J_n(k_c a)\,\bigl[\,Y_{n-1}(k_c a)-Y_{n+1}(k_c a)\,\bigr]
-\bigl[\,J_{n-1}(k_c a)-J_{n+1}(k_c a)\,\bigr]\,Y_n(k_c a) }{\bigl[\,Y_{n-1}(k_c a)-Y_{n+1}(k_c a)\,\bigr]}$.}
%%%% RQ: cette expression se simplifie!!! je l'ai po fait ici pour que tu vois mes changements; a faire ptete...
\label{tab:TE_mn_coax_profiles}
% (optionnel) compacter un peu l'espacement des colonnes
\setlength{\tabcolsep}{5pt}
\renewcommand{\arraystretch}{1.1}
% Redimensionne uniquement le tableau (pas la légende)
%\resizebox{\textwidth}{!}{%
\begin{tabular}{@{}l l@{}}
\toprule
\textbf{Function} & \textbf{Expression} \\
\midrule
$g_{E_r}(r,\theta)$
& $\!\!\! = -\dfrac{2 k\, n}{k_c^{2}\,r\, A_{n,m}}\;
\Biggl[ J_n(k_c r)\,
-\dfrac{\bigl[\,J_{n-1}(k_c a)-J_{n+1}(k_c a)\,\bigr]}{\bigl[\,Y_{n-1}(k_c a)-Y_{n+1}(k_c a)\,\bigr]}\,Y_n(k_c r)
\Biggr]
 \sin \bigl[n(\theta-\theta_0)\bigr]$
\\[12pt]

$g_{E_\theta}(r,\theta)$
& $\!\!\! = -\dfrac{k}{k_c\, A_{n,m} }\;
\Biggl( \bigl[\,J_{n-1}(k_c r)-J_{n+1}(k_c r)\,\bigr]  $ \\
& $
-\dfrac{\bigl[\,J_{n-1}(k_c a)-J_{n+1}(k_c a)\,\bigr]}{\bigl[\,Y_{n-1}(k_c a)-Y_{n+1}(k_c a)\,\bigr]}\bigl[\,Y_{n-1}(k_c r)-Y_{n+1}(k_c r)\,\bigr] \Biggr)
 \cos \bigl[n(\theta-\theta_0)\bigr]$
\\[12pt]

$g_{E_z}(r,\theta)$
& $=\;0$
\\[10pt]

$g_{B_r}(r,\theta)$
& $ \!\!\! = + \dfrac{\beta}{k_c\, A_{n,m}}\;
\Biggl(  \bigl[\,J_{n-1}(k_c r)-J_{n+1}(k_c r)\,\bigr]  $ \\
& $
-\dfrac{\bigl[\,J_{n-1}(k_c a)-J_{n+1}(k_c a)\,\bigr]}{\bigl[\,Y_{n-1}(k_c a)-Y_{n+1}(k_c a)\,\bigr]}\bigl[\,Y_{n-1}(k_c r)-Y_{n+1}(k_c r)\,\bigr] \Biggr)
 \cos \bigl[n(\theta-\theta_0)\bigr]$
\\[12pt]

$g_{B_\theta}(r,\theta)$
& $ \!\!\! = -\dfrac{2 \beta\, n}{k_c^{2}\,r\, A_{n,m}}\;
\Biggl[ \;J_n(k_c r)  
-\dfrac{\bigl[\,J_{n-1}(k_c a)-J_{n+1}(k_c a)\,\bigr]}{\bigl[\,Y_{n-1}(k_c a)-Y_{n+1}(k_c a)\,\bigr]}\,Y_n(k_c r) \Biggr]
 \sin \bigl[n(\theta-\theta_0)\bigr]$
\\[12pt]

$g_{B_z}(r,\theta)$
& $\!\!\! =-\dfrac{2}{  A_{n,m}}\; 
\Biggl[ \;J_n(k_c r) 
-\dfrac{\bigl[\,J_{n-1}(k_c a)-J_{n+1}(k_c a)\,\bigr]}{\bigl[\,Y_{n-1}(k_c a)-Y_{n+1}(k_c a)\,\bigr]}\,Y_n(k_c r) \Biggr]
 \cos \bigl[n(\theta-\theta_0)\bigr]$
\\[6pt]
\bottomrule
\end{tabular}
%} % end resizebox
\label{tab3}
\end{table}
%%%%
%%%%
%%% Bon... j'ai du réécrire tes formules.
%%% il y avait des fautes, et ta normalisation était débile (ne correspond pas au papier waveguide).
%%% regarde BIEN ce que j'ai fait, et compare à tes notes. 
%%% j'ai essayé de modifier AU MINIMUM pour que tu vois les différences. Mais c'est pas encore la facon la + propre de l'écrire. Ca se simplifie.
%%% remarque: quand ca passe pas, on fait pas plus petit, on écrit sur 2 lignes.

$\mathbf{E}$  and  $\mathbf{B}$ fields display again an angular $\theta$ and radial $r$ dependence \cite{PozarMW}.
%%% c'est comme ca que je les comprends:
These modes resemble the ones present in a hollow cylinder, see Section \ref{sec:TE_hollow}.
For $n>0$, the normalization has been chosen according to the 
%%%
electric field 
%%% c'est ca la convention!!!
convention discussed in Section \ref{factor}. 
%%%
However, for $n=0$ {\em the electric field is zero on the conductors}: this brings us to the concept of {\it virtual electrodes} introduced in Ref. \cite{Delattre2024}. 
%%%
%%% INTRODUIRE les elecs virtuelles, en citant la Fig. 1.
Looking at Tab. \ref{tab4}, one realizes that the metallic boundary conditions Eqs. (\ref{eq:bc1}-\ref{eq:bc4})
are met by all planes cutting the coaxial guide as a diameter, see Fig. \ref{fig_1}. Within any of these equivalent planes, one can then define {\em virtual charges and virtual currents} (see  Section \ref{sec:charges_currents_constants_cyl}) .
Following the same philosophy as for real electrodes, we chose to normalize the electric field to its maximal amplitude, namely  $|g_{E_\theta}(r=r_{max})|=1$.
$r_{max}$ is defined as the solution of:
\begin{equation}
    \frac{J_0(k_c r)-J_2(k_c r) }{Y_0(k_c r)-Y_2(k_c r)} - \frac{J_1(k_c a) }{ Y_1(k_c a)} =0,
\end{equation}
which is the closest to $r=a$. The normalization amplitude $A_m$ is then computed numerically. \\
%%% là, il faut que tu te creuses un peu les méninges sur la normalisation... je te l'ai écrit. Est-ce que c'est clair?...

%%% REMARQUE: j'ai réécrit pour une certaine cohérence entre les différents tableaux; il y a certainement encore des fioritures possibles pour rendre plus lisible.

%%%
%%% remarque générale sur les defs: tu n'as pas introduit le concept de "branche", et tu es resté sur "mode". Je laisse comme ca....
%%%

%%%% pitite intro de la section suivante pour faire un lien...
In all the preceding results, they are some differences, but more interestingly strong similarities with the Cartesian case \cite{Delattre2024}. We will discuss this explicitly in the following Subsection.

%%% tableau pour le cas n==0
%%
\begin{table}[H]
\centering
\caption{Dimensionless modal profiles 
for TE$_{n=0,m}$ 
modes in a coaxial waveguide 
($m>0$). See text for definition of $A_{n=0,m}=+ A_m \,\dfrac{ k \,a}{\bigl[\,Y_{-1}(k_c a)-Y_{+1}(k_c a)\,\bigr]} $.}
\label{tab:TE_mn_coax_profiles}
% (optionnel) compacter un peu l'espacement des colonnes
\setlength{\tabcolsep}{5pt}
\renewcommand{\arraystretch}{1.1}
% Redimensionne uniquement le tableau (pas la légende)
%\resizebox{\textwidth}{!}{%
\begin{tabular}{@{}l l@{}}
\toprule
\textbf{Function} & \textbf{Expression} \\
\midrule
$g_{E_r}(r,\theta)$
& $\!\!\! = 0$
\\[12pt]

$g_{E_\theta}(r,\theta)$
& $\!\!\! = -\dfrac{k}{k_c\, A_{0,m} }\;
\Biggl( \bigl[\,J_{-1}(k_c r)-J_{+1}(k_c r)\,\bigr] $ \\
& $
-\dfrac{\bigl[\,J_{-1}(k_c a)-J_{+1}(k_c a)\,\bigr]}{\bigl[\,Y_{-1}(k_c a)-Y_{+1}(k_c a)\,\bigr]}\bigl[\,Y_{-1}(k_c r)-Y_{+1}(k_c r)\,\bigr] \Biggr)$
\\[12pt]

$g_{E_z}(r,\theta)$
& $=\;0$
\\[10pt]

$g_{B_r}(r,\theta)$
& $ \!\!\! = + \dfrac{\beta}{k_c\, A_{0,m}}\;
\Biggl(  \bigl[\,J_{-1}(k_c r)-J_{+1}(k_c r)\,\bigr]  $ \\
& $
-\dfrac{\bigl[\,J_{-1}(k_c a)-J_{+1}(k_c a)\,\bigr]}{\bigl[\,Y_{-1}(k_c a)-Y_{+1}(k_c a)\,\bigr]}\bigl[\,Y_{-1}(k_c r)-Y_{+1}(k_c r)\,\bigr] \Biggr)$
\\[12pt]

$g_{B_\theta}(r,\theta)$
& $ \!\!\! = 0$
\\[12pt]

$g_{B_z}(r,\theta)$
& $\!\!\! =-\dfrac{2}{  A_{0,m}}\; 
\Biggl[ \;J_0(k_c r) 
-\dfrac{\bigl[\,J_{-1}(k_c a)-J_{+1}(k_c a)\,\bigr]}{\bigl[\,Y_{-1}(k_c a)-Y_{+1}(k_c a)\,\bigr]}\,Y_0(k_c r) \Biggr]
$
\\[6pt]
\bottomrule
\end{tabular}
%} % end resizebox
\label{tab4}
\end{table}
%%%%

%%% La on peut faire le lien à ce que ca veut dire vis-a-vis du cas cartesien...
%%% on fait une section exprès, OK? c'est ce qui me parait le plus didactique et en lien à ta structure?

\subsubsection{Comparison with Cartesian geometry}
\label{sec:compare}

%%%%
%%%% bon, c'est là qu'il y a 2-3 trucs intéressants à dire!

%%%% je te le phrase ici, et te mets mes commentaires en-dessous sur ton texte initial.
Let us define $b=a+d$ 
and $w = 2 \pi \, a$. 
%%%
$d$ is thus the gap between electrodes, and $w$ their transverse dimension (perimeter).
%%%% ADD REF 2
Considering $d \ll a$ guarantees that fringing fields  can be neglected, as well as any $x$-dependence  \cite{PozarMW}. 
%%%
Consider first the TEM wave: in Tab. \ref{tab1}, one obtains $g_{E_r} \approx 1$ and $g_{B_\theta} \approx \mathrm{sign}(\beta)$, which reproduce the Cartesian components of Ref. \cite{Delattre2024}.
Besides, Eqs. (\ref{eq:coax_TM_disp},\ref{eq:TE_coax_dispersion})
lead then to the simple solution:
\begin{equation}
    k_c = m \frac{\pi}{d}, \,\, m \in\mathbb{N}^* ,
\end{equation}
for any $n$. For $n=0$, the TM and TE solutions reproduce what is found in Ref. \cite{Delattre2024}. The interpretation of these modes becomes then clear: they correspond to the parallel plate guide {\em closed on itself}, with a large enough radius of curvature $\rho \gg a$. 
%%% ok? c'était ca que j'espérais...
The {\em virtual electrodes} of the TE parallel plate configuration, which are at the boundaries of the guide, transform naturally into the diameter planes discussed in the present manuscript (see Fig. \ref{fig_1}).
%%% ok?

But what about $n \neq 0$ solutions? These are actually reminiscent of the hollow cylinder configuration discussed in the following: these can be obtained by taking $a/b \rightarrow 0$, and modifying the normalization of the modal functions for $A_{n,m}$ (taking this time the outer guide as a reference). These modes cannot exist between the parallel plates because the symmetry imposes a strict translation invariance (compatible only with the $n=0$ solution of this Section), while the coaxial line provides periodic boundary conditions ($x \rightarrow x+m\,w$ for $x$ the transverse component, corresponding to the $\mathbf{u}_\theta$ direction here, $m \in \mathbb{N}^*$).

%%%% mettre appendix bifilar guide??? Hum...
The parallel plate configuration can also be continuously transformed into another type of guide, by rolling up each electrode on itself (creating a cylindrical bar). In the opposite limit where $d \gg a$, one obtains what is called a {\em bifilar waveguide} \cite{PozarMW}. %Its properties are briefly discussed in Appendix \ref{bifilar}:
In this open geometry, only TEM modes are allowed.

\subsection{Hollow % circular 
cylindrical waveguide}  %%% oups, hein?!
\label{sec:hollow}

In a single wall cylindrical waveguide, no TEM mode exists \cite{PozarMW};
%: sustaining a nontrivial transverse potential requires two conductors.  
%%% le mieux, c'est de pas s'épancher... ta phrase est pas tout a fait correcte, dans le vide tu peux définir du TEM mais y a pas du tout d'électrodes...
%%% ca t'emmène trop loin, donc on dit rien, et on cite un bon bouquin.
%
%The TEM solution ($E_z=B_z=0$) is incompatible with simple connectivity of the cross section (Laplace’s equation enforces a constant potential if $V$ is constant on the boundary).  
%%% a nouveau....
%%% ca t'emmène trop loin: t'as pas encore défini le potentiel!!!!! En plus, le but de ton papier c'est de redéfinir ce que l'on appelle Delta V communément! Donc là, tu vas trop vite.... Le mieux c'est de rien dire, ou de faire un commentaire plus tard si tu peux/veux.
the hollow %circular 
guide %therefore 
supports only TM ($E_z\neq0$ and $B_z=0$) and TE ($B_z\neq0$ and $E_z=0$) families, 
which will be described below.
%%% pour faire un petit lien...
%%% je t'ai rajouté le == 0, c'est quand même le sens du nom!!!

\subsubsection{TM %$_{m,n}$
modes}
\label{sec:TM_hollow}

Regularity at $r=0$ removes the Bessel $Y_n$ contribution, Eq. (\ref{eq:bessel_general}). %part. 
%%% tu l'as déjà dit!!!!! la mondre des choses, c'est de citer l'équation...
% With $E_z(a,\theta)=0$ on the perfect wall:
The boundary condition $E_z(r=a,\theta)=0$ leads to:
\begin{equation}
J_n(k_c a)=0 ,
%\;\Longrightarrow\;
%k_c=\frac{\xi_{m,n}}{a},\quad n\in\mathbb{N}^\ast,\;\; m\in\mathbb{N},
%\qquad
% k^2=\beta^2+k_c^2,\quad \omega_{c,mn}=c\,k_c.
\end{equation}
%%% bon... il faut etre un peu systématique; soit tu écris tout comme ca, soit jamais. Comme j'ai pas envie de réécrire ce que j'ai déjà corrigé, je te vire ca. Et je le mets au même format que ci-dessus pour le coax.
with similarly to the coaxial line $k^2=k_c^2+\beta^2$ and $\omega_c = c \, k_c$.
For each $n \geq 0$, the solutions to the above equation are indexed by $m > 0$.
%%%%  ok?
Modal profiles (properly normalized) are summarized in Tab. \ref{tab:TM_mn_hollow}. %bellow:

\begin{table}[H]
\centering
\caption{Dimensionless modal profiles %$g_{E_i}$, $g_{B_i}$ 
for TM$_{m,n}$ in a hollow %circular 
cylindrical waveguide
%(\textbf{family} $J_m(k_c a)=0$).
%%%% bin c'est quoi ca? on s'en fout?
($ n \geq 0, m>0$). We defined $A_{n,m}=+ \frac{\beta}{k_c} \Big[ J_{n-1}(k_c a)-J_{n+1}(k_c a)\Big]$.}
%%%% comme précédemment, je te mets la normalisation comme ca pour etre homogène. Tu t'étais encore gouré sur la normalisation, le E_rho faisait pas 1.... :-(
\label{tab:TM_mn_hollow}
\begin{tabular}{@{}l l@{}}
\toprule
\textbf{Function} & \qquad \textbf{Expression} \\
\midrule
$g_{E_r}(r,\theta)$     
& $=\;-\dfrac{\beta}{k_c \, A_{n,m}}\,\Big[J_{n-1}(k_c r)-J_{n+1}(k_c r) \Big] \cos[n(\theta-\theta_0)]$ \\[6pt]
$g_{E_\theta}(r,\theta)$
& $=\;+\dfrac{2\beta\,n}{k_c^2 r\, A_{n,m}}\, J_n(k_c r) \,\sin[n(\theta-\theta_0)]$ \\[6pt]
$g_{E_z}(r,\theta)$     
& $=\;+\dfrac{2}{A_{n,m}}\,J_n(k_c r)\,\cos[n(\theta-\theta_0)]$ \\[6pt]
$g_{B_r}(r,\theta)$     
& $=\;-\dfrac{2k\,n}{ k_c^2 r \, A_{n,m}}\, J_n(k_c r) \,\sin[n(\theta-\theta_0)]$ \\[6pt]
$g_{B_\theta}(r,\theta)$
& $=\;-\dfrac{k}{k_c\,A_{n,m}}\,\Big[ J_{n-1}(k_c r)-J_{n+1}(k_c r) \Big] \cos[n(\theta-\theta_0)]$ \\[4pt]
$g_{B_z}(r,\theta)$     
& $=\;0$ \\
\bottomrule
\end{tabular}
\end{table}
%%% je t'ai a nouveau swap m et n; sinon c'était OK.

%%% mais c'est quoi ca que tu me racontes là????? 
%Unlike the rectangular TM$_{n,m}$ case (where $n=0$ or $m=0$ is excluded because $E_z$ must vanish on \emph{all} walls), the hollow cylinder has only the real boundary $r=a$; regularity on axis removes $Y_m$ but does not impose extra azimuthal constraints. The family defined by $J_m(k_c a)=0$ therefore admits$ \,m\ge 0,\; n\ge 1\,$,
%so the axisymmetric TM$_{0n}$ modes are allowed and well defined.
%%%% c'est complètement n'imp, 
%%% la seule "fioriture" à noter ici est que dans le cas carré n==0 est exclut parce que sinon TOUS LES CHAMPS SONT NULS!!! Et pour le m non zero, c'est ici simplement un choix de numérotation!!!!!!!
%%% cf Pozar.... a citer....
%%% je te le réécris:
These modes share the same properties as their Cartesian counterpart. Note however that the index $n=0$ was excluded in this case \cite{Delattre2024}: even the lowest TM solutions have an oscillation pattern on {\em both axes} $\mathbf x, \mathbf y$. 
%A simple change $n \rightarrow n+1$ would then lead to the same numbering, but this is {\em not} what is used in the literature; we prefer to stick  here to conventional nomenclature \cite{PozarMW}, and will keep the $n=0$ writing.
%%%% tu vois... c'est trivial...
%%%% EH BIN NON!!! JE ME SUIS PLANTE....
%%% j'ai rephrasé ci-dessus correctement....
Here, the $n=0$ modes oscillate along $\mathbf u_r$, but not along $\mathbf u_\theta$. This is a peculiarity arising again from the higher symmetry of the cylindrical geometry.

%%%% bon, je te change a nouveau les m en n, et ta normalisation pourrie.
\begin{table}[H]
\centering
\caption{Dimensionless modal profiles %$g_{E_i}$, $g_{B_i}$ 
for TE$_{m,n}$ in a hollow %circular 
cylindrical waveguide 
($ n > 0, m>0$).
We defined $A_{n,m}=+ \frac{2 k \, n }{k_c^2 \, a}   J_{n}(k_c a)$.}
\label{tab:TE_mn_hollow}
\begin{tabular}{@{}l l@{}}
\toprule
\textbf{Function} & \qquad \textbf{Expression} \\
\midrule
$g_{E_r}(r,\theta)$     
& $=\;-\dfrac{2 k\, n}{k_c^2\, r \, A_{n,m} }\,
 J_n(k_c r) \,\sin[n(\theta-\theta_0)]$ \\[6pt]
$g_{E_\theta}(r,\theta)$
& $=\;-\dfrac{k}{k_c \, A_{n,m} }\,
\Big[ J_{n-1}(k_c r)-J_{n+1}(k_c r) \Big] 
\,\cos[n(\theta-\theta_0)]$ \\[6pt]
$g_{E_z}(r,\theta)$     & $=\;0$ \\[6pt]
$g_{B_r}(r,\theta)$     
& $=\;+\dfrac{\beta}{k_c \, A_{n,m}}\,
\Big[J_{n-1}(k_c r)-J_{n+1}(k_c r) \Big]   \,\cos[n(\theta-\theta_0)]$ \\[6pt]
$g_{B_\theta}(r,\theta)$
& $=\;-\dfrac{2\beta\, n}{k_c^2\, r \, A_{n,m}}\,
 J_n(k_c r) \,\sin[n(\theta-\theta_0)]$ \\[6pt]
$g_{B_z}(r,\theta)$     
& $=\;-\dfrac{2}{A_{n,m} }\,  J_n(k_c r) \,\cos[n(\theta-\theta_0)]$ \\
\bottomrule
\end{tabular}
\end{table}

\subsubsection{TE %$_{m,n}$ 
modes}
\label{sec:TE_hollow}

%For TE, regularity yields $B_z(r,\theta)=B_m J_m(k_c r)\cos[m(\theta-\theta_0)]$, and $E_\theta\propto\partial_r B_z$. %%%%%
%%%%% je comprends pas... la régularité impose à nouveau pas de Y_n!
%%%%% pourquoi tu veux discuter les autres points? C'est de la trivialité trouvée dans tous les bouquins! Ces parties ci du papier sont juste là pour donner les formules écrites dans nos conventions, pas pour recopier la littérature
%%% je rephrase:
Again regularity at $r=0$ imposes the absence of $Y_n$ in the solution Eq. (\ref{eq:bessel_general}).
%%%
The boundary 
condition  %%% explicite!
$E_\theta(r=a,\theta)=0$ gives $J_n'(k_c a)=0$, namely: %i.e. 
\begin{equation}
J_{n-1}(k_c a)=J_{n+1}(k_c a) ,
%\;\Longrightarrow\;
%k_c=\frac{\xi_{m,n}}{a},\quad n\in\mathbb{N}^\ast,\;\; m\in\mathbb{N},
\end{equation}
%leading to the vacuum dispersion:
%\begin{equation}
 %   k^2=\beta^2+k_c^2,\quad \omega_{c,mn}=c\,k_c.
%\end{equation}
%%% same remarque: pour les autres t'as pas explicité, tu as mis dans le texte. Soit tu mets ca explicite sur la 1ere, et ensuite tu t'y réfère, soit jamais.
%%% Je choisis jamais ;-)
with as previously $k^2=k_c^2+\beta^2$ and $\omega_c = c \, k_c$.
%%%%
The $k_c$ that solve the above equation are indexed by $m > 0$,
for $n \geq 0$. 
%%%%  ok?

%%%%%%% maintenant on donne les profiles, ET LA IL FAUT faire une distinction n not 0.
As for the coaxial line, we should distinguish the $n \neq 0$ from the $n=0$ cases. Modal profiles for $n >0$ are summarized in Tab. \ref{tab:TE_mn_hollow}. The normalization discussed in Section \ref{factor} has been applied.
%%% ok?
The $n=0$ modes are presented in Tab. \ref{tab:TE0n_hollow}.
Their understanding requires the introduction of {\em virtual electrodes}, as discussed in Subsection \ref{sec:coax_TE}.

% With the normalization $g_{B_z}(a)=1$, the profiles are:
%%% ouate de fouque? c'est pas du tout ce qu'on fait dans waveguide...
%%% zut....

%Because $k_c>0$ is required for any guided mode, the radial index satisfies $n\ge 1$.
%All azimuthal indices $m\ge 0$ are allowed; for $m=0$ the $E_r$ and $B_\theta$ components vanish identically.
%%% mais qu'est ce que tu me chantes là? C'est complètement non pertinent...

%\subsubsection{Axisymmetric TE$_{0n}$ branch}
%\label{sec:TE0n_hollow}
%%%
%%% bon, logiquement parlant, ca pose problème: la section d'avant traite des TE, donc cette section devrait formellement y etre incluse.
%%% aussi, tous ces modes sont axisymmetric!!! c'est pas ce qui distingue ce cas n=0 des autres!!!

%Here the only real boundary is $r=a$. For TE, $E_\theta(a)=0\Leftrightarrow J_0'(k_c a)=0$, and since
%$J_0'(x)=-J_1(x)$ this gives:
%\begin{equation}
%J_1(k_c a)=0,\qquad
%k_c=\frac{j_{1,n}}{a},\quad n=1,2,\ldots,\qquad
%k^2=\beta^2+k_c^2,\quad \omega_c=c\,k_c.
%\end{equation}
%With $g_{B_z}(a)=1$, the profiles are given in Table \ref{tab:TE0n_hollow}:
%%%%
%%%% on s'en fout, et en plus c'est pas la bonne normalisation...

\begin{table}[H]
\centering
\caption{Dimensionless 
modal profiles for %axisymmetric 
TE$_{n=0,m}$ in a hollow %circular PEC 
%%%% de quoi PEC???? 
cylindrical waveguide
%(family $J_1(k_c a)=0$). Normalization $g_{B_z}(a)=1$
($m>0$). See text for $A_{n=0,m}=+ A_m \, k/k_c$. }
\label{tab:TE0n_hollow}
\begin{tabular}{@{}l l@{}}
\toprule
\textbf{Function} & \qquad \textbf{Expression} \\
\midrule
$g_{E_r}(r)$      & $=\;0$ \\[4pt]
$g_{E_\theta}(r)$ & $=\;-\dfrac{ k}{k_c \, A_{0,m}}\,\Big[ J_{-1}(k_c r) - J_{+1}(k_c r) \Big]$ \\[4pt]
$g_{E_z}(r)$      & $=\;0$ \\[4pt]
$g_{B_r}(r)$      & $=\;+\dfrac{\beta}{k_c \, A_{0,m}}\,\Big[ J_{-1}(k_c r) - J_{+1}(k_c r) \Big]$ \\[4pt]
$g_{B_\theta}(r)$ & $=\;0$ \\[4pt]
$g_{B_z}(r)$      & $=\;-\dfrac{2}{A_{0,m} } \, J_0(k_c r)$ \\
\bottomrule
\end{tabular}
\end{table}
%%%% a nouveau, l'écriture peut etre simplifiée... je l'ai pas encore fait pour que tu vois bien...

Inspecting Tab. \ref{tab:TE0n_hollow}, we see that similarly to the TE$_{0,m}$ modes of the coaxial line, the planes cutting the hollow guide through a diameter verify metallic boundary conditions (see  Fig. \ref{fig_1}). These will enable to define the {\em virtual charges and currents} in the following. The required normalization condition reads again $|g_{E_\theta}(r=r_{max})|=1$, with this time $r_{max}$ solution of:
\begin{equation}
    J_0(k_c r)-J_2(k_c r)  =0,
\end{equation}
which is the closest to $r=0$. The $A_m$ parameter is then computed numerically.
%%%%% specif cylindre...
Note that these modes are {\em different} from their Cartesian counterpart \cite{Delattre2024}: in a square guide, both TE$_{0,m}$ and TE$_{n,0}$ are allowed, and simply correspond to equivalent patterns, rotated by 90$^\circ$ around the $\mathbf z$ axis. Here, we have only one family TE$_{0,m}$ with $m>0$. Besides, for the square guide the boundary conditions apply directly to the real electrodes, either the top/bottom pair or the left/right: {\em there is no virtual electrode}. The  TE$_{0,m}$ modes of the cylinder are in this sense quite peculiar, and do not have an equivalent in the Cartesian geometry, again because of their higher symmetry. This specificity will be further discussed in Subsection \ref{gaugePLUS}.

\section{Charges, currents, and constants of motion in cylindrical geometry}
\label{sec:charges_currents_constants_cyl}

Surface charge densities $\sigma_s$ and surface currents % $\vec{\jmath}_s$ 
$\mathbf{j}_s$,  %%% a nouveau, change pas de notations! T'as utilisé les vecteurs en gras, reste là dessus!
%follow from the perfect conductor boundary
as defined by the conditions Eqs. (\ref{eq:bc1}-\ref{eq:bc4}), 
%on each electrode $s$ (real or virtual).
when combined with Eq. \eqref{eq:Maxwell4} 
verify charge conservation:
%Since these relations are coordinate-invariant, only geometric factors change; in particular,
%charge conservation on any electrode surface reads similarly to the cartesian case:
%%% ca, on s'en fout... MAis je t'avouerai que moi, j'ai refait le calcul en cylinsrique pour m'assurer que c'est vrai...
\begin{equation}
\mathrm{div}_{\!s}\,\mathbf{j}_s + \frac{\partial \sigma_s}{\partial t} = 0,
\label{eq:surface_continuity_global}
\end{equation}
with $\mathrm{div}_{\!s}$ the divergence operator calculated on the $s$ surface.
%%% ca par contre, c'est ptet important de le dire explicitement.
%%%
%% on peut là dire un mot sur qui est s...
This surface can be part of a curved cylinder (real electrodes with {\em in,out} or {\em front,back} subscripts), or a flat diameter plane (virtual ones, with a {\em vir top,vir bottom} subscript), see Fig. \ref{fig_1}.  %%% OK? 

\subsection{Constants of motion of the electromagnetic field}
\label{subsec:constants_motion_cyl}

%%%
% As usual, 
% c'est un peu pauvre... cite qqch, par exemple toi!
Following Ref. \cite{Delattre2024}, the 
%conserved quantities 
constants of motion
are obtained by volume integration:
%in cylindrical coordinates
%($r\in[r_{\min},r_{\max}]$, $\theta\in[0,2\pi)$, $z\in[0,L]$) such that:
%%% les gens savent lire
\begin{align}
H &= \int_{0}^{L}\!\!\int_{r_{\min}}^{r_{\max}}\!\!\int_{0}^{2\pi}
\left[ \frac{\epsilon}{2}\,\mathbf{E}(\mathbf r,t)^2 + \frac{1}{2\mu}\,\mathbf{B}(\mathbf r,t)^2 \right]
\, r\, d\theta\, dr\, dz,
\qquad (\text{energy}).
\label{eq:H_cyl_full}\\[4pt]
\mathbf{P} &= \int_{0}^{L}\!\!\int_{r_{\min}}^{r_{\max}}\!\!\int_{0}^{2\pi}
\epsilon\,\big[\mathbf{E}(\mathbf r,t)\times \mathbf{B}(\mathbf r,t)\big] \, r\, d\theta\, dr\, dz,
\qquad (\text{momentum}),
\label{eq:P_cyl_full}\\[4pt]
\mathbf{J} &= \int_{0}^{L}\!\!\int_{r_{\min}}^{r_{\max}}\!\!\int_{0}^{2\pi}
\mathbf r \times \Big\{\epsilon\,\big[\mathbf{E}(\mathbf r,t)\times \mathbf{B}(\mathbf r,t)\big]\Big\}
\, r\, d\theta\, dr\, dz,
\qquad (\text{angular momentum}).
\label{eq:J_cyl_full}
\end{align}
%%% ADD REF 2
These quantities are $t$-independent, and represent all we must know about the confined fields \cite{CohenQED}.
%%%%
%%% a nouveau, tu melanges des notations!!! les vecteurs en flèches, puis en gras... NON! tiens toi a un type.
%%% bon, faut définir tout... les rmax et rmin sont nouveau...
For the coaxial line, $r_{\min}=a$ and $ r_{\max}=b$ while for the hollow cylinder we have $r_{\min}=0, r_{\max}=a$. %ok?
%%%
By symmetry about the guide axis, in $\mathbf{P}$ only the longitudinal momentum $P_z$ is nonzero; $\mathbf{J}$ is always identically zero.
Consequently, just as in the Cartesian analysis, the field configurations %studied here 
are fully characterized
by the pair $\{H,P_z \}$ (energy and longitudinal momentum).
%; we will not use $\mathbf J$ in the sequel.
%%%%
Thus, the light helicity does not play any role here.

\subsection{Generalized fluxes}
%%%%%%%
%\subsection{Capacitance and (inverse) inductance per unit surface %(electrode-wise)
\label{subsec:CdLd_defs_cyl}
%%%% le electrode-wise est de trop dans le titre; a expliquer dans le texte
%%%%% Mais... est ce que c'est pas plutot le flux généralisé le truc important de cette section???? C'est quand meme le coeur du papier?!

% We keep the factorized fields of Eqs.~\eqref{eq:longitudinal_envelopes}.
%%%
%%% on s'en fout! j'espère bien que tu les garde, c'est la base du medele introduit!

%%% c'est l'inverse inductance en fait... on en a parlé
At this stage, we must introduce surface capacitance densities (in F/m$^2$) and "inverse inductance densities" (actually, in H$^{-1}$, see Section \ref{sec:quantize}), which are defined for each electrode type. 
%, see Tab.  \ref{eq:surface_continuity_global}. 
%%%% c'est pas plus clair comme ca?
%%%% Bon, en fait, c'est pas à ce niveau qu'il faut un tableau... mais ptet plus tard...
%%%%
%\emph{electrode-wise} geometric coefficients: capacitances per unit surface
%and inverse inductances per unit surface as in the Cartesian paper, with the appropriate
%cylindrical effective gaps $h_{\mathrm{eff}}$. These coefficients, depending on the considered case developed in this part, are given in Table \ref{eq:surface_continuity_global}:
%%%%%%%
%%%%%%%
% Here $h_{\mathrm{eff}}^{(a)}$, $h_{\mathrm{eff}}^{(b)}$, $h_{\mathrm{eff}}^{[\theta]}$ are geometry-only
% lengthscales set by the \emph{transverse eigenproblem and boundary conditions} (Bessel-mode content);
%they will be computed family by family below.
They are obtained from a specific lengthscale $h_{\mathrm{eff}}$ which depends on the geometry, and will be given below for each specific case; 
%%%%%% ADD REF 3
A summary can be found at the end of the Section in Tab. \ref{tabADD3}.
 Explicitly, we write:
\begin{eqnarray}
    C_d & = & \frac{\epsilon }{h_{\mathrm{eff}}} , \\
    L_d^{-1} & = & \frac{1}{\mu\,h_{\mathrm{eff}}} .
\end{eqnarray}
%%%% c'est clair, c'est simple!?
%%%% Mais le truc clé, c'est d'EXPLIQUER la différence en électrodes comparé au cas cartésien: il n'y a qu'une famille d'électrodes (ton cas est plus symmétrique en ce sens), mais pour le coax, les électrodes ne sont pas équivalentes!!! Alors que c'était le cas en cartésien. C'est ca qu'il faut surtout expliquer...
%%%%
Let us point out two subtle differences with the Cartesian configuration. 
On one hand, the cylindrical geometry is more symmetric: the surfaces facing each other are of one kind only, while in a  rectangular guide with width $w$ different from height $d$ ($w \neq d$) {\em two sets} of electrodes must be considered \cite{Delattre2024}.
However, on the other hand in the parallel plate guide, the two electrodes confining the field are strictly identical, while in a coaxial line one is smaller than the other: {\em they are not equivalent}. 
%%%%
In this case, we will use a prime to designate quantities corresponding to the electrode opposite to the reference one.
%, similarly to a convention used in Ref. \cite{Delattre2024}.
%%% OK? c'est ca qu'il faut expliquer plutot que tout le reste! 
%%      et c'est assez fondamental...
%%%
%First of all, for the radial pair one can write:
%\begin{equation}
%\partial_t \phi \;=\; \Delta V^{(r)},\qquad
%\partial_z \phi \;=\; -\,\Delta A_z^{(r)}, 
%\label{eq:phi_gauge_devoret_cyl_r}
%\end{equation}
%where \(\Delta(\cdot)^{(r)}\) is between $r=a$ and $r=b$ in the coax (TEM/TM).
%In a hollow guide, the same definitions apply relative to a regular interior reference
%(e.g.\ $V(0)=0$ by gauge), and the entire radial drop is encoded in $h_{\mathrm{eff}}^{(a)}$.
%
% Mais non! T'as pas défini les potentiels!! D'où tu sors ces relations? Pour le moment, on ne définit que les phi. Les relations vont apparaitre avec les potentiels seulement...
%%% suis le déroulé du papier waveguide....

On the same model as what is performed in Ref. \cite{Delattre2024}, these lengthscales enable %also 
to define {\em generalized fluxes} with a magnitude given by:
\begin{equation}
\phi_m=\frac{E_m\,h_{\mathrm{eff}} }{\omega}. 
\end{equation}
These generalized fluxes live on the electrodes. When they correspond to real ones (the metallic guides), we have:
\begin{equation}
    \varphi (\theta,z,t) =\phi_m\,g_{real}(\theta)\,\tilde f(z,t), \label{varphi1}
\end{equation}
while when electrodes are virtual (diameter planes, Fig. \ref{fig_1}) we write:
\begin{equation}
    \varphi (r,z,t) =\phi_m\,g_{vir}(r)\,\tilde f(z,t). \label{varphi2}
\end{equation}
%One obtains:
%\begin{equation}
%    \phi(z,t)=\phi_m\,\tilde f(z,t),\qquad 
%\phi_m=\frac{E_m\,h_{\mathrm{eff}}^{(r)}}{\omega}.
%\end{equation}
%with $h_{\mathrm{eff}}^{(r)}\in\{h_{\mathrm{eff}}^{(a)},h_{\mathrm{eff}}^{(b)}\}$ depending on the considered case. 
%
% nan mais là t'es pas assez précis! cf ce que j'ai écrit.
%%% pitite phrase de transition et c'est fini!
The $g_{real}, g_{vir}$ functions represent the variations of the fluxes over the electrodes.
%%%% ok??? Tu les avais oubliés!!!
The point will be in the following to re-express the constants of motion as a function of these $\varphi$ quantities. We shall then introduce the canonically-conjugate variable $Q$, enabling the quantization of the $ \varphi,Q $ pair.
%%%
A proper gauge fixing must then link $\varphi$ to the electromagnetic potentials $\mathbf A,V$.
We demonstrate that the formalism developed for the Cartesian geometry \cite{Delattre2024} 
can be adapted to the specificities of the cylindrical symmetry.
%%%% vala!!!
%%% et j'insiste là dessus encore....

%Then, considering the azimuthal pair, we define the virtual-electrode angles:
%\begin{equation}
%    \theta_* \equiv \theta_0+\frac{p\pi}{m},\qquad p\in\mathbb Z,\ m\ge 1,
%\end{equation}
%
%as the nodal half-planes of $B_z\propto \cos[m(\theta-\theta_0)]$; the pair is
%\(\theta_* \to \theta_*+\tfrac{\pi}{m}\), with unit normal $\mathbf{n}^{[\theta]}=+\mathbf e_\theta$.
%Then:
%\begin{equation}
%\partial_t \phi' \;=\; \Delta V^{[\theta]},\qquad
%\partial_z \phi' \;=\; -\,\Delta A_z^{[\theta]},
%\label{eq:phi_gauge_devoret_cyl_theta}
%\end{equation}
%and
%\begin{equation}
%    \phi'(z,t)=\phi'_m\,\tilde f(z,t),\qquad 
%\phi'_m=\frac{E_m\,h_{\mathrm{eff}}^{[\theta]}}{\omega},
%\end{equation}
%with the radial interval on a virtual plane given by $(0,a)$ in a hollow guide and $(a,b)$ in a coaxial guide. This convention reproduces verbatim the Cartesian duality 
%$\dot\phi \leftrightarrow \Delta V$ (and $\partial_z\phi \leftrightarrow -\Delta A_z$) and will be used to build the one-dimensional Hamiltonian and the constants $(C_H,C_P,L_H^{-1})$ for each family.
%%%
%%% tout ca on s'en fout!
%%% les C_H C_P L_H sont pas définis ici!!! Ca sert a rien de les citer, on comprend pas de quoi tu parles!!

\subsection{Coaxial waveguide}
\label{subsec:coax_charges_currents}

\subsubsection{TEM modes}
%%% ta structure précédente créait des subsections pour les types d'onde... pourquoi ne pas la conserver ici????

As for the Cartesian case, the simplest wave propagation concerns TEM modes. We define:
\begin{eqnarray}
    h_{\rm eff} & = &  a \, \ln{\left[\frac{b}{a} \right]}, \\
    h_{\rm eff}' & = & \frac{b}{a} \, h_{\rm eff} , \\
    &\mbox{and:}& \\
    g_{real}(\theta) & = & 1 .
\end{eqnarray}
%, with $h_{\rm eff}^{(r)}(a)=a\ln\!\frac{b}{a}$ and $
%h_{\rm eff}^{(r)}(b)=b\ln\!\frac{b}{a}$. We treat this situation using the radial pair of electrodes $\phi$. We use the following convention as a wall selector:
%%
%%% ca mérite d'etre écrit en grand, a défaut de faire un tableau. On va définir un seul cas d'abord, et expliquer ensuite l'équivalence sur l'autre électrode, ce sera bien plus clair.
%%%
%%%
% in preamble or just before use
%\newcommand{\sgnalpha}{s_{\alpha}} % sign selector for chosen wall (a or b)
%\begin{equation}
%\sgnalpha \equiv 
%\begin{cases}
%+1, & \alpha=a,\\
%-1, & \alpha=b.
%\end{cases}
%\label{eq:sgnalpha_def}
%\end{equation}
%%%
%Boundary conditions at each wall give the local charge/current–phase laws,
%\begin{align}
%\sigma^{(\alpha)}(\theta,z,t) 
%&= \sgnalpha\, C_d^{(\alpha)}\,\partial_t\phi, \label{eq:sigma_alpha}\\[4pt]
%j^{(\alpha)}_{z}(\theta,z,t) 
%&= -\,\sgnalpha\,\big(L_d^{(\alpha)}\big)^{-1}\,\partial_z\phi, \label{eq:jz_alpha}
%\end{align}
%%%
%%%
%%% tu peux pas t'empecher de faire compliqué???? Tes notations sont à nouveau différentes des précédents (pour les dérivées), on mélange pas!
%%%
%%% je fais simple:
The boundary conditions on the inner electrode bring:
\begin{eqnarray}
    \sigma_{in} & =& + C_d \frac{\partial \varphi (\theta,z,t) }{\partial t} , \\
    \mathbf{j}_{in} & = & - L_d^{-1} \frac{\partial \varphi(\theta,z,t) }{\partial z} \, \mathbf{z} ,  
\end{eqnarray}
while on the outer one we have:
\begin{eqnarray}
    \sigma_{out} & =& - C_d\,\!\!' \frac{\partial \varphi (\theta,z,t) }{\partial t} , \\
    \mathbf{j}_{out} & = & + L_d'^{-1} \frac{\partial \varphi(\theta,z,t) }{\partial z} \, \mathbf{z} ,  
\end{eqnarray}
%%
%%
%No azimuthal dependence ($\partial_\theta\equiv0$) and $j_{s,\theta}=0$.
%%% on comprend rien... définis la fonction g, et pose la == qqch. Alors, c'est clair.
with reversed signs. 
Note that it is the same $\varphi$ function (calculated with $ h_{\rm eff}$) that appears in the above; only the capacitance and inductance densities are different from the reference ones (and are calculated with $ h_{\rm eff}'$).
%%% ok?
%As for the cartesian case, signs are reversed when switching electrodes, consequence of the electromagnetic field symmetry. We thus write considering the wall at $r=\alpha$:
%\begin{align}
%    H&=\!\int_{0}^{L}\!\!\int_{0}^{2\pi}\!
%\Big[\tfrac12\,C_d^{(\alpha)}(\partial_t\phi)^2+\tfrac12\,(L_d^{(\alpha)})^{-1}(\partial_z\phi)^2\Big]\;
%\alpha\, d\theta\,dz,
%\\
%P_z&=\!\int_{0}^{L}\!\!\int_{0}^{2\pi}\!
%C_d^{(\alpha)}\;[-\partial_t\phi\,\partial_z\phi]\;\alpha\, d\theta\,dz,
% \end{align}
%%%
%%% meme remarque... trop compliqué et notations pas compatibles!!
Energy and momentum write:
\begin{eqnarray}
H & = & \int_{0}^{L}\!\!\int_{0}^{2\pi}\! \left[ \frac{1}{2}   C_d \left( \frac{\partial \varphi(\theta,z,t) }{\partial t}\right)^2 +\frac{1}{2} L_d^{-1} \left( \frac{\partial \varphi(\theta,z,t) }{\partial z}\right)^2  \right] a\, d\theta \,  dz , \label{Htem} \\
\mathbf{P} & = &  \int_{0}^{L}\!\!\int_{0}^{2\pi}\! C_d \left[-\frac{\partial \varphi(\theta,z,t) }{\partial t}\frac{\partial \varphi(\theta,z,t) }{\partial z} \right] a\,   d\theta \,dz \, \mathbf{z} , 
\end{eqnarray}
and the charge conservation equation Eq. (\ref{eq:surface_continuity_global}) leads to:
%Surface continuity at a wall, $\partial_t\sigma^{(\alpha)}+\partial_z j_z^{(\alpha)}=0$, gives the $\phi$ wave propagation equation:
%%%
%%%
%%% surface continuity??? Mais ca veut rien dire!!
%
%
%\begin{equation}
%\partial_z^2\phi-\frac{1}{c^2}\,\partial_t^2\phi=0.
%\end{equation}
\begin{equation}
    \frac{\partial^2 \varphi(\theta,z,t)}{\partial z^2}-\frac{1}{c^2} \frac{\partial^2 \varphi(\theta,z,t)}{\partial t^2}=0 ,
    \label{eq:phi_wave_TEM_coax}
\end{equation}
which is nothing but the conventional (d'Alembert) wave propagation equation for $\varphi$, at the speed of light $c$ in the medium. 
%
%One can identify the \emph{field energy per unit surface} stored in the quadratic form:
% --- Paires radiales (mur réel r=α): TEM et TM_{m,n} ---
%\begin{equation}
%H_d^{(\alpha)}(\theta,z,t)
%= \frac{1}{2}\,\big(C_d^{(\alpha)}\big)^{-1}\,\big[\sigma^{(\alpha)}(\theta,z,t)\big]^2
%\;+\;
%\frac{1}{2}\,L_d^{(\alpha)}\,\big[j^{(\alpha)}_{z}(\theta,z,t)\big]^2.
%\label{eq:Hd_wall_alpha}\\
%\end{equation}
%
% c'est ultra trivial... faut le dire plus simplement, c'est surtout pour préparer aux différences avec les autres cas!
Finally, the bracket in 
Eq. (\ref{Htem}) 
can be identified with the surface charge/current energy density $H_d$:
\begin{equation}
    H_d = \frac{1}{2} C_d^{-1} \,\sigma_{in}^2 + \frac{1}{2} L_d \, \mathbf{j}_{in}^2 .
\end{equation}

{\em Note} - Obviously, a completely similar writing can be achieved while referencing the fields on the outer electrode, defining properly alternative $ h_{\rm eff}$ and $ h_{\rm eff}'$ quantities; the integrations run then onto an arc $b\,d\theta$ instead of $a\,d\theta$.
%%%%
This is also true for any of the other configurations considered below dealing with real electrodes.

\subsubsection{TM modes}
%%%% ok?
\label{subsec:coax_charges_currentsTM}

For TM modes, we define:
%Secondly considering TM$_{m,n}$ case, also associated to the radial pair $\phi$, the \emph{same} gauge relations hold ; only the wall laws pick up the usual TM factors:
%\begin{align}
%\sigma^{(\alpha)}(\theta,z,t) 
%&= s_\alpha\, C_d^{(\alpha)}\,\partial_t\phi,\\[4pt]
%j^{(\alpha)}_{z}(\theta,z,t) 
%&= -\,s_\alpha\,\big(L_d^{(\alpha)}\big)^{-1}\left(\frac{k}{\beta}\right)^{2}\,\partial_z\phi,\\[2pt]
%j^{(\alpha)}_{\theta}(\theta,z,t)&=0.
%\end{align}
%%%
%%% il faut DEFINIR les heff!!!
%%% C'EST TOUTE LA DIFFICULTE DU TRAVAIL AVEC BESSEL!!!!!!
%%%%%%%%%%%%%%%
\begin{eqnarray}
    h_{\rm eff} & = &   a \, \frac{ \pi^2}{2} \int_{a}^{b}\! \frac{k_c^2 r}{2} \left( J_n[k_c r] \, Y_n[k_c a] - 
  J_n[k_c a ] \, Y_n [ k_c r] \right)^2 \, dr , \\
    h_{\rm eff}' & = & \frac{b}{a} \left| \frac{J_n(k_c b)}{J_n(k_c a)} \right|\, h_{\rm eff} , \\
    &\mbox{and:}& \\
    g_{real}(\theta) & = & \cos [ n(\theta-\theta_0)] ,
\end{eqnarray}
%with $c(0)=1$ and $c(n \neq 0)=2$.
This time, the expression of $h_{\rm eff}$ is not analytic, and must be computed numerically. 
%%%%%% OK?
The metallic boundary condition on the inner electrode writes here:
\begin{eqnarray}
       \sigma_{in} & =& + C_d \frac{\partial \varphi (\theta,z,t) }{\partial t} , \\
    \mathbf{j}_{in} & = & - L_d^{-1} \left( \frac{k}{\beta} \right)^{\!\!2} \frac{\partial \varphi(\theta,z,t) }{\partial z} \, \mathbf{z} , 
\end{eqnarray}
and on the outer one we obtain:
\begin{eqnarray}
       \sigma_{out} & =& - (-1)^m \, C_d\,\!' \frac{\partial \varphi (\theta,z,t) }{\partial t} , \\
    \mathbf{j}_{out} & = & + (-1)^m \, L_d'^{-1} \left( \frac{k}{\beta} \right)^{\!\!2} \frac{\partial \varphi(\theta,z,t) }{\partial z} \, \mathbf{z} . 
\end{eqnarray}
As for TEM modes, the primed quantities are computed from $h_{\rm eff}'$; the $(-1)^m$ sign factor arises from the sign of $J_n(k_c b)/J_n(k_c a)$, characteristic of the symmetry of the mode profile.
Energy and momentum are obtained as:
\begin{eqnarray}
H & = & \int_{0}^{L}\!\!\int_{0}^{2\pi}\! \left[ \frac{1}{2}   C_d \left( \frac{\partial \varphi(\theta,z,t) }{\partial t}\right)^2 +\frac{1}{2}   C_d \left( \frac{k}{\beta}\right)^{\!\!2} (c\, k_c)^2  \varphi(\theta,z,t)^2  \right. \nonumber \\
& & \left. +\frac{1}{2} L_d^{-1} \left( \frac{k}{\beta}\right)^{\!\!4} \left( \frac{\partial \varphi(\theta,z,t) }{\partial z}\right)^2  \right] a\, d\theta \,  dz ,\label{eq:H} \\
\mathbf{P} & = &  \int_{0}^{L}\!\!\int_{0}^{2\pi}\! C_d  \left( \frac{k}{\beta}\right)^{\!\!2}\left[-\frac{\partial \varphi(\theta,z,t) }{\partial t}\frac{\partial \varphi(\theta,z,t) }{\partial z} \right] a\,   d\theta \,dz \, \mathbf{z} . 
\end{eqnarray}
%Once again, one can define energy and momentum: 
%\begin{align}
%    H&=\!\int_{0}^{L}\!\!\int_{0}^{2\pi}\!
%\Big[\tfrac12\,C_d^{(\alpha)}(\partial_t\phi)^2
%+\tfrac12\,(L_d^{(\alpha)})^{-1}\!\big(\tfrac{k}{\beta}\big)^{\!4}(\partial_z\phi)^2
%+\tfrac12\,C_d^{(\alpha)}\!\big(\tfrac{k}{\beta}\big)^{\!2}(c k_c)^2\phi^2\Big]\;
%\alpha\, d\theta\,dz,\label{eq:H} \\
%P_z&=\!\int_{0}^{L}\!\!\int_{0}^{2\pi}\!
%C_d^{(\alpha)}\big(\tfrac{k}{\beta}\big)^{\!2}\;[-\partial_t\phi\,\partial_z\phi]\;\alpha\, d\theta\,dz,
%\end{align}
%
% meme commentaire qu'avant...
The propagation equation derived from charge conservation is then:
\begin{equation}
    \frac{\partial^2 \varphi(\theta,z,t)}{\partial z^2}-\frac{1}{v_\phi^2} \frac{\partial^2 \varphi(\theta,z,t)}{\partial t^2}=0 ,
    \label{eq:phi_wave_TM_coax}
\end{equation}
where $v_\phi= c \, k/|\beta|$.
%with the propagation relation: 
%\begin{equation}
%    \partial_z^2\phi-\tfrac{1}{c^2}\big(\tfrac{\beta}{k}\big)^{\!2}\partial_t^2\phi=0.
%    \label{KG}
%\end{equation}
%%
%%% pareil.
The bracket in Eq. \eqref{eq:H} can be understood as the integral over a surface energy density $H_d$:
%is the wall energy density:
\begin{equation}
    H_d = \frac{1}{2} C_d^{-1} \,\sigma_{in}^2 + \frac{1}{2} L_d \, \mathbf{j}_{in}^2 + \Delta(\theta,z,t),
\end{equation}
%%%
%\begin{align}
%H_d^{(\alpha)}(\theta,z,t)
%&=\frac{1}{2}\,\big(C_d^{(\alpha)}\big)^{-1}\,[\sigma^{(\alpha)}(\theta,z,t)]^{2}
%+\frac{1}{2}\,L_d^{(\alpha)}\,\big(\tfrac{k}{\beta}\big)^{\!4}\,[j^{(\alpha)}_{z}(\theta,z,t)]^{2}
%+\Delta^{(\alpha)}(\theta,z,t),
%\\
%\Delta^{(\alpha)}(\theta,z,t)\;&=\;
%\frac{1}{2}\,C_d^{(\alpha)}\!\left(\frac{k}{\beta}\right)^{\!2}(c k_c)^2\,\phi(\theta,z,t)^2 .
%\end{align}
%%%%
%%%%
%%%% idem, je te le réécris. Et je splitte le Delta pour bien insister dessus.
but this time with an addendum term $\Delta$:
\begin{equation}
 \Delta(\theta,z,t)=\frac{1}{2}   C_d \left( \frac{k}{\beta}\right)^{\!\!2} (c\, k_c)^2  \varphi(\theta,z,t)^2  ,
\end{equation}
which  can be %$\Delta$ 
interpreted (as in the Cartesian case) as the \emph{potential energy cost} required
to confine a TM wave.
%photon. \\
%%% je préfère wave parce que a ce stade du papier, tu n'as pas encore quantifié, et tu n'as que des écritures classiques....

\subsubsection{TE modes}
%%%% ok?
\label{subsec:coax_charges_currentsTE}

Finally, the last situation to describe is %the 
TE %$_{m,n}$ 
modes. % family
Let us consider first $n>0$. 
%considering $m\ge 1,$ 
%%%%% Faisons simple...
In this case, we define:
\begin{eqnarray}
   && \!\!\!\!\!\!\!\!\!\!\!\!\!\!\!\!\!\! h_{\rm eff}  =   a \, \frac{2}{n^2} \!\! \int_{a}^{b}\! \frac{k_c^2 r}{2} \!\left(\frac{\left( J_{n+1}[k_c a]-J_{n-1}[k_c a]\right) \, Y_n[k_c r] - 
  J_n[k_c r ] \, \left( Y_{n+1} [ k_c a]-Y_{n-1} [ k_c a] \right) }{ \left( J_{n+1}[k_c a]-J_{n-1}[k_c a]\right) \, Y_n[k_c a] - 
  J_n[k_c a ] \, \left( Y_{n+1} [ k_c a]-Y_{n-1} [ k_c a] \right)} \right)^{\!\!2} \!  dr , \\
 && \!\!\!\!\!\!\!\!\!\!\!\!\!\!\!\!\!\!  h_{\rm eff}'  =  \frac{b}{a} \, \left|\frac{\pi}{4} (k_c b) \left[ \left( J_{n+1}[k_c b]-J_{n-1}[k_c b]\right) \, Y_n[k_c a] - 
  J_n[k_c a ] \, \left( Y_{n+1} [ k_c b]-Y_{n-1} [ k_c b] \right) \right] \right| \, h_{\rm eff} , \label{heffpTEc}\\
   & &\mbox{and:} \\
  && \!\!\!\!\!\!\!\!\!\!\!\! g_{real}(\theta)  =  \sin [ n(\theta-\theta_0)] .
\end{eqnarray}
As for TM modes, $ h_{\rm eff} $ must be computed numerically. 
%%%%%%%%%%%%%%%% OK?
From the boundary conditions of Section \ref{boundaryes}, we obtain:
\begin{eqnarray}
    \sigma_{in} & =& + C_d \frac{\partial \varphi (\theta,z,t) }{\partial t} , \\
    \mathbf{j}_{in} & = & - L_d^{-1} \frac{\partial \varphi(\theta,z,t) }{\partial z} \, \mathbf{z} - L_d^{-1} \left( \frac{ k_c^2 a^2}{n^2} \right) \frac{1}{a} \frac{\partial \varphi(\theta,z,t) }{\partial \theta}\, \mathbf{u_\theta}.  \label{eqcurrTEc}
\end{eqnarray}
A new term appeared in Eq. (\ref{eqcurrTEc}): a peripheral current flows in the electrode, due to the longitudinal magnetic field component.
Similarly, on the other electrode we have:
\begin{eqnarray}
    \sigma_{out} & =& - (-1)^{m+1} \, C_d' \frac{\partial \varphi (\theta,z,t) }{\partial t} , \\
    \mathbf{j}_{out} & = & + (-1)^{m+1}\, L_d'^{-1} \frac{\partial \varphi(\theta,z,t) }{\partial z} \, \mathbf{z} + (-1)^{m+1} \, L_d'^{-1} \left( \frac{ k_c^2 b^2}{n^2} \right) \frac{1}{b} \frac{\partial \varphi(\theta,z,t) }{\partial \theta}\, \mathbf{u_\theta} ,   
\end{eqnarray}
with primed quantities defined with  $ h_{\rm eff}' $.
The sign change $(-1)^{m+1}$ is due to the sign of the expression within the absolute value bars in Eq. (\ref{heffpTEc}). 
%%%% commentaire malin signes!!
At first sight, this $m+1$ (symmetry related) exponent seems different from the one found in Ref. \cite{Delattre2024}; but it is actually  a trivial consequence of the numbering. These TE $n \neq 0$ modes are reminiscent of the hollow cylinder configuration, which correspond for Cartesian geometries to the rectangular guide. Their numbering starts at $m = 0$ when $n>0$, while here it starts at $m=1$.
% sera repris plus loin... ou pas ;-)
%we shall comment this point further in Section \ref{subsec:hollow_charges_currents_en} below.
%%%%%
The expressions for energy and momentum are:
\begin{eqnarray}
H & = & \int_{0}^{L}\!\!\int_{0}^{2\pi}\! \left[ \frac{1}{2}   C_d \left( \frac{\partial \varphi(\theta,z,t) }{\partial t}\right)^2   \right. \nonumber \\
& & \left. +\frac{1}{2} L_d^{-1}  \left( \frac{\partial \varphi(\theta,z,t) }{\partial z}\right)^2  + \frac{1}{2}    L_d^{-1}  \frac{k_c^2 a^2}{n^2} \left( \frac{1}{a} \frac{\partial \varphi(\theta,z,t) }{\partial \theta}\right)^2 \right] a\, d\theta \,  dz ,\label{eq:coax_TE_energy} \\
\mathbf{P} & = &  \int_{0}^{L}\!\!\int_{0}^{2\pi}\! C_d    \left[-\frac{\partial \varphi(\theta,z,t) }{\partial t}\frac{\partial \varphi(\theta,z,t) }{\partial z} \right] a\,   d\theta \,dz \, \mathbf{z} . 
\end{eqnarray}
%We therefore define energy and momentum such as previously:
%\begin{align}
%H &=  \;=\; \int_{0}^{L}\!\!\int_{a}^{b}
%\left[
%\frac{1}{2}\,C_d^{[\theta]}(\partial_t\phi')^2
%+\frac{1}{2}\,(L_d^{[\theta]})^{-1}\,(\partial_z\phi')^2
%+\frac{1}{2}\,C_d^{[\theta]}\,(c\,k_c)^2\,(\phi')^2
%\right]\,dr\,dz,
%\label{eq:coax_TE_energy}\\[4pt]
%P_z &= \int_{0}^{L}\!\!\int_{a}^{b}
%C_d^{[\theta]}\,(-\partial_t\phi'\,\partial_z\phi')\,dr\,dz.
%\label{eq:coax_TE_momentum}
%\end{align}
%
%   pareil aue d'hab...
%%% je te l'ai réécrit....
The charge conservation equation brings us this time:
\begin{equation}
        \frac{\partial^2 \varphi(\theta,z,t)}{\partial z^2}-\frac{1}{c^2} \frac{\partial^2 \varphi(\theta,z,t)}{\partial t^2}=k_c^2 \, \varphi(\theta,z,t) , \label{KGTEcoax}
\end{equation}
which is a {\em Klein-Gordon} propagation equation. 
%%Using surface continuity on the plane
%%\begin{equation}
%\partial_z^2 \phi' \;-\; \frac{1}{c^2}\,\partial_t^2 \phi' \;=\; +\,k_c^2\,\phi',
%\end{equation}
%
%% pareil que d'hab.... 
%%%
%%% bon, c'est là qu'il faut expliciter ce que ca veut dire:
%%
%%%
%%%
%This confinement energy expression $\Delta^{[\theta]}$ is the
%standard “mass” form: it endows the 1D field $\phi'$ with an effective transverse
%inertia, quantized by $k_c$, and produces the Klein–Gordon propagation with the dispersion
%\(
%\omega^2=c^2(\beta^2+k_c^2).
%\)
%In quantum terms, this is identical to a relativistic particle with
%\emph{effective rest mass} \(m_{\rm eff}=\hbar k_c/c\): this term is the local potential energy that
%remains even when the longitudinal dynamics is frozen ($\partial_z\phi'=0$). Topologically, this is the exact analogue of the Cartesian TE case (virtual lateral
%plates): the role of the Cartesian transverse derivative is here played by $\partial_r$
%across the coaxial annulus. 
%%%%
%%%%
%%%% mais t'en fais des caisses!! Y a pas besoin, pourquoi tu remets la relation de dispertion??? 
%%%%
%%%% je fais:
Here, the term $\hbar\, k_c/c$ plays the role of a mass: there is a kinetic energy cost when confining TE waves (while there is a potential one for TM).
This will be discussed in more details in Section \ref{sec:quantize}.
%%%%%
Once again, the bracket in Eq. \eqref{eq:coax_TE_energy} can be recast into a surface energy density $H_d$: % on the plane:
%%%%%%%%%%%%%%%%%%%%%%%%%%%%%%%%%%%%%de quel plan tu parles??? Ici, on est sur une surface courbé de coax!!!
%%%
%\begin{align}
%H_d^{[\theta]}(r,z,t)
%&= \frac{1}{2}\,C_d^{[\theta]}[\partial_t\phi'(r,z,t)]^2
% + \frac{1}{2}\,\big(L_d^{[\theta]}\big)^{-1}[\partial_z\phi'(r,z,t)]^2
% + \Delta^{[\theta]}(r,z,t),\\
%  \Delta^{[\theta]}(r,z,t)
%&= \frac{1}{2}\,C_d^{[\theta]}\,(c\,k_c)^2\,\big[\phi'(r,z,t)\big]^2,
%\label{eq:coax_TE_density}
%\end{align}
%
%%%%%%%%%%% idem comments ci-dessus sur ton écriture. Dans ce cas-ci, je propose:
\begin{equation}
    H_d = \frac{1}{2} C_d^{-1} \,\sigma_{in}^2 + \frac{1}{2} L_d \, \left(\mathbf{j}_{in}.\mathbf{z}\right)^2 + \frac{1}{2} L_d \frac{n^2}{k_c^2 a^2}\, \left(\mathbf{j}_{in}.\mathbf{u_\theta}\right)^2 .
\end{equation}
The peripheral currents, to which we associate an inductance density $ L_d \frac{n^2}{k_c^2 a^2}$, are therefore associated to the mass appearance.
%%%% ADD REF 2
This transverse electronic motion is what we refer to by the wording "kinetic energy" here. \\
%%%% 

\begin{figure}[H]
\centering
%\vspace*{-2.5cm}
\hspace*{-1.2cm} 
\includegraphics[width=1.3\linewidth]{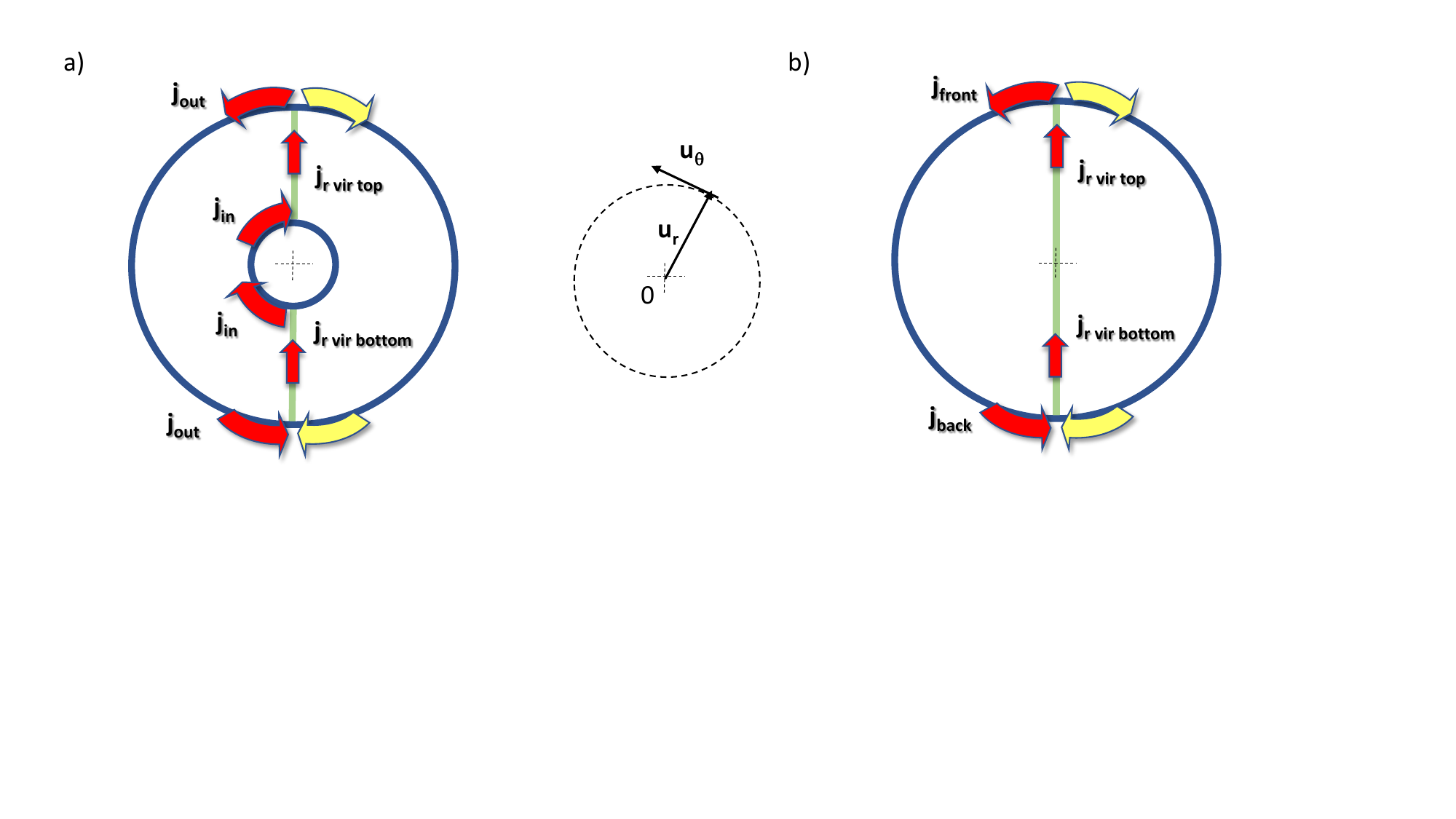}
\vspace*{-4cm}
\caption{Transverse currents of TE modes with $n=0$. a) coaxial case. b) hollow cylinder. The direction of the outer current circulation depends on the mode number $m$; for the coaxial line, it is also a function of the ratio $b/a$ (but the behavior becomes equivalent to the hollow guide at  $b/a \gg 1$). Considering this limit, we represent in red the $m$ even case; the outer current is reversed for $m$ odd (yellow). %%% Argh!? pas simple... mais c'est normal...
%%%%%%%%%%%%%%%%%%%  OK?
}
\label{fig_2}
\end{figure}

%For the axisymmetric family $m=0$, the boundary
%condition $E_\theta(a)=E_\theta(b)=0$ enforces
%\begin{equation}
%J_0'(k_c a)\,Y_0'(k_c b)\;-\;J_0'(k_c b)\,Y_0'(k_c a)\;=\;0
%\;\;\Longleftrightarrow\;\;
%J_1(k_c a)\,Y_1(k_c b)\;-\;J_1(k_c b)\,Y_1(k_c a)\;=\;0,
%\label{eq:coax_TE0n_disp}
%\end{equation}
%with $k^2=\beta^2+k_c^2$ and cutoff $\omega_c=c\,k_c$. 
%Since $k_c>0$ for any guided mode, the admissible indices are $m=0$ and $n=1,2,\ldots$ as there is no $\text{no TE}_{00}$. 
%%%%
%%%% Mais pourquoi tu répètes ça ici????? On l'a déjà fait AVANT!!!!!
Consider now the $n=0$ situation.
The peculiarity there is that the transverse fields are zero on the metallic guides, see Subsection \ref{sec:coax_TE}. We must thus introduce the concept of virtual electrodes, with:
\begin{eqnarray}
    h_{\rm eff} & = &   a \,\int_{a}^{b}\! \frac{4 r}{a^2} \!\left(\frac{J_1[k_c r] \, Y_1[k_c a] - J_1[k_c a] \, Y_1[k_c r] }{A_m  \, (k_c a) } \right)^{\!\!2} \!  dr , \\
    &\mbox{and:}& \\
    g_{vir}(r) & = & 2 \left(\frac{J_1[k_c a] \, Y_1[k_c r] - J_1[k_c r] \, Y_1[k_c a] }{A_m \, (k_c a)} \right), \label{gvirTEcoax}
\end{eqnarray}
and $A_m$ obtained numerically (Subsection \ref{sec:coax_TE}). $h_{\rm eff}$ is again non-analytic. Any diameter plane crossing the coaxial line can be taken as virtual electrode.
%%%
%%%
%There are only the \emph{real} coax electrodes at $r=a$ and $r=b$, and keeping in mine our wall selector variable definition:
%\begin{align}
%\sigma^{(\alpha)}&=j^{(\alpha)}_z=0,\\
%j^{(\alpha)}_\theta&=-s_{\alpha}\,\mu^{-1}B_z|_{r=\alpha}. 
%\end{align}
%
%  c'est bien joli, mais il faut donner EXPLICITEMENT les expressions!
Charges and currents are defined there as:
\begin{eqnarray}
    \sigma_{vir\,top} & =& + C_d \frac{\partial \varphi (r,z,t) }{\partial t} , \\
    \mathbf{j}_{vir\,top} & = & - L_d^{-1} \frac{\partial \varphi(r,z,t) }{\partial z} \, \mathbf{z}
    - L_d^{-1} \frac{1}{r} \frac{\partial \left(\, r \,\varphi [r,z,t] \,\right)}{\partial r} \, \mathbf{u_r},  \label{eqcurrTEcspec}
\end{eqnarray}
for the top part of the plane, for which we took $+ \mathbf u_\theta$ as normal (see Fig. \ref{fig_1}). The bottom part is oriented the other way, leading to:
\begin{eqnarray}
    \sigma_{vir\,bottom} & =& - C_d \frac{\partial \varphi (r,z,t) }{\partial t} , \\
    \mathbf{j}_{vir\,bottom} & = & + L_d^{-1} \frac{\partial \varphi(r,z,t) }{\partial z} \, \mathbf{z}
    + L_d^{-1} \frac{1}{r} \frac{\partial \left(\, r \,\varphi [r,z,t] \,\right)}{\partial r} \, \mathbf{u_r} .
\end{eqnarray}
%Thus the walls carry \emph{circulating} surface currents $j_{s,\theta}$ (opposed on $a$ and $b$), but no surface charge neither axial current. As $m=0$, there is no azimuthal virtual electrode: no  nodal for $B_z$ plan with $\theta=\text{cst}$ does exist.
%%%
%%% je fais maintenant le lien au "vrai" courant de surface, comme dans l'autre papier. La démarche est bien basé sur le phi, même s'il est d'abord défini sur une surface virtuelle...
On each real conductor, a peripheral current is flowing:
\begin{eqnarray}
\mathbf{j}_{in} & = &  
    + L_d^{-1} \, 2 k_c \,\left( \frac{J_1[k_c a] \, Y_0[k_c a] - J_0[k_c a] \, Y_1[k_c a] }{A_m (k_c a)} \right) \, \phi_m \,\tilde{f}(z,t)\, \mathbf{u_\theta} , \\
    \mathbf{j}_{out} & = &  
    - L_d^{-1} \,  2 k_c \,\left( \frac{ J_1[k_c a] \, Y_0[k_c b] - J_0[k_c b] \, Y_1[k_c a] }{A_m (k_c a)} \right) \, \phi_m \,\tilde{f}(z,t) \, \mathbf{u_\theta}.  \label{eqcurrTEcperiph}
\end{eqnarray}
These can be seen as connected by continuity in $r=a$ and $r=b$ by the virtual current component flowing in the direction $\pm \mathbf{u_r}$;
 this is depicted in Fig. \ref{fig_2}.
%%%%% ok?
Energy and momentum are then deduced with:
\begin{eqnarray} 
H & = & \int_{0}^{L}\!\!\int_{a}^{b}\! \left[ \frac{1}{2 }
C_d \left( \frac{\partial \varphi(r,z,t) }{\partial t}\right)^2   \right. \nonumber \\
& & \left. +\frac{1}{2} L_d^{-1}  \left( \frac{\partial \varphi(r,z,t) }{\partial z}\right)^2  + \frac{1}{2}    L_d^{-1}\left(\frac{1}{r} \frac{\partial \left(\, r \,\varphi [r,z,t] \,\right)}{\partial r} \right)^2  \right] \frac{2\pi \,r}{  h_{\rm eff}}\, dr \,  dz ,\label{eq:coax_TE_energyspe} \\
\mathbf{P} & = &  \int_{0}^{L}\!\!\int_{a}^{b}\!   
C_d \left[-\frac{\partial \varphi(r,z,t) }{\partial t}\frac{\partial \varphi(r,z,t) }{\partial z} \right]\frac{2\pi \, r}{  h_{\rm eff}}\, dr \,dz \, \mathbf{z} . 
\end{eqnarray}
The specific writing of these expressions will be commented below. 
%%%% bin oui, non?
%%%%
Charge conservation is applied in the virtual plane, treating $z,r$ as coordinates of a flat surface. One obtains: 
\begin{equation}
        \frac{\partial^2 \varphi(r,z,t)}{\partial z^2}-\frac{1}{c^2} \frac{\partial^2 \varphi(r,z,t)}{\partial t^2}=k_c^2 \, \varphi(r,z,t) ,
\end{equation}
which reads essentially the same as Eq. (\ref{KGTEcoax}).
There is indeed a geometrical subtlety here, arising from the mapping of the curved problem onto a plane, which leads to the normalization by $ h_{\rm eff}$ in the $H$ and $\mathbf{P}$ expressions.  The bracket in 
Eq. (\ref{eq:coax_TE_energyspe}) 
can still be interpreted as a charge/current energy density $H_d$, but with properly defined (and $r$-dependent) capacitance and inductance densities:
%%%%%%%%%   OK? c'était un peu tordu, non?
%%%%
\begin{equation}
    H_d = \frac{1}{2} \left[C_d^{-1}\frac{2\pi \, r}{  h_{\rm eff}} \right] \,\sigma_{vir\,top}^2 + \frac{1}{2} \left[L_d\frac{2\pi \, r}{  h_{\rm eff}} \right] \, \left(\mathbf{j}_{vir\,top}.\mathbf{z}\right)^2 + \frac{1}{2} \left[L_d\frac{2\pi \, r}{  h_{\rm eff}} \right]\, \left(\mathbf{j}_{vir\,top}.\mathbf{u_r}\right)^2 ,
\end{equation}
written in brackets in this formula.
The same expression holds obviously for the {\em vir bottom} terms.
As for the $n>0$ case, the light confinement in the guide comes with a kinetic cost, due to the peripheral  currents.

%%%%%% c'est pas si simple a voir, non? je propose une image....
%%%%%%
%%%%%%         Ca va aider, non????
%%%%%%
%%%%%%

\subsection{Hollow %circular
cylindrical waveguide}
\label{subsec:hollow_charges_currents_en}

\subsubsection{TM modes}
\label{subsec:hollow_charges_currents_enTM}

%As previously, we start considering the TM$_{m,n}$ family, keeping in mind for this hollow circular case that radial differences are taken between the regular interior and the wall: $ \Delta(\cdot)^{(r)}\equiv (\cdot)\big|_{r\to 0}- (\cdot)\big|_{r=a}$. As for the coaxial case, the radial pair $\phi$ is the relevant one for this situation. We thus set the local laws considering the single wall $r=a$, with the wall selector $s_a=-1$ ($\mathbf n^{(a)}=-\mathbf e_r$)
%%%
%%% j'ai rien compris a ce que tu racontes.
%%% En plus, normalement c'est déjà traité et donc ca devrait etre facile à introduire...
%%% je propose:
The TM modes of the hollow guide lead to:
\begin{eqnarray}
   && \!\!\!\!\!\!\!\!\!\!\!\! h_{\rm eff}  =   a \, \left[ 2\left(\frac{J_{n-1} [k_c a]}{J_{n-1} [k_c a]-J_{n+1} [k_c a]}\right)^2 \right], \\
   & &\mbox{and:} \\
  && \!\!\!\!\!\!\!\!\!\!\!\! g_{real}(\theta)  =  \cos [ n(\theta-\theta_0)] .
\end{eqnarray}
Charges and currents on the cylinder read:
\begin{eqnarray}
       \sigma_{front} & =& + C_d \frac{\partial \varphi (\theta,z,t) }{\partial t} , \\
    \mathbf{j}_{front} & = & - L_d^{-1} \left( \frac{k}{\beta} \right)^{\!\!2} \frac{\partial \varphi(\theta,z,t) }{\partial z} \, \mathbf{z} , 
\end{eqnarray}
on the "front" side, and we obtain on the "back" side $\theta \rightarrow \theta+\pi $: 
\begin{eqnarray}
       \sigma_{back} & =& - (-1)^{n+1} \, C_d \frac{\partial \varphi (\theta,z,t) }{\partial t} , \\
    \mathbf{j}_{back} & = & +(-1)^{n+1} \, L_d^{-1} \left( \frac{k}{\beta} \right)^{\!\!2} \frac{\partial \varphi(\theta,z,t) }{\partial z} \, \mathbf{z} . 
\end{eqnarray}
%\begin{align}
%\sigma^{(a)}(\theta,z,t) &= -\,C_d^{(a)}\,\partial_t\phi,\\[2pt]
%j^{(a)}_{z}(\theta,z,t)  &= +\,\big(L_d^{(a)}\big)^{-1}\!\left(\frac{k}{\beta}\right)^{\!2}\partial_z\phi, \\[2pt]
%j^{(a)}_{\theta}(\theta,z,t) &= 0.
%\end{align}
%%%
%%% a nouveau, notations... Et pourquoi définir partout le j_\theta ==0 comme ca? c'est lourd...
%%
%As usual, one has to refer to Table \ref{eq:surface_continuity_global} for details on capacitances and inductances.
%%%% mais... y avait rien dans ton tableau, t'avais rien calculé!!
This $(-1)^{n+1}$ sign change (due to the mode symmetry) seems in the first place different from what is found in Ref. \cite{Delattre2024}.
But it is a trivial consequence of the basis vector $\mathbf u_r$ being always oriented towards the outside of the guide.
% numbering choice chosen here, which starts from $n=0$ instead of $n=1$ % (see Subection \ref{sec:TM_hollow}). 
%%%%%%%%% on aura la meme discussion ci-dessous pour TE, qui a déjà été suggéré pour le coax...
Energy and momentum write:
\begin{eqnarray}
H & = & \int_{0}^{L}\!\!\int_{0}^{2\pi}\! \left[ \frac{1}{2}   C_d \left( \frac{\partial \varphi(\theta,z,t) }{\partial t}\right)^2 +\frac{1}{2}   C_d \left( \frac{k}{\beta}\right)^{\!\!2} (c\, k_c)^2  \varphi(\theta,z,t)^2  \right. \nonumber \\
& & \left. +\frac{1}{2} L_d^{-1} \left( \frac{k}{\beta}\right)^{\!\!4} \left( \frac{\partial \varphi(\theta,z,t) }{\partial z}\right)^2  \right] a\, d\theta \,  dz ,\label{eq:HTMhollow} \\
\mathbf{P} & = &  \int_{0}^{L}\!\!\int_{0}^{2\pi}\! C_d  \left( \frac{k}{\beta}\right)^{\!\!2}\left[-\frac{\partial \varphi(\theta,z,t) }{\partial t}\frac{\partial \varphi(\theta,z,t) }{\partial z} \right] a\,   d\theta \,dz \, \mathbf{z} , 
\end{eqnarray}
which are perfectly similar to the ones of the coaxial line (but with the integral $a\,d\theta$ running on the confining tube). As well, the propagation equation deduced from charge conservation is again:
%We then express energy and momentum defining surface integrals on $r=a$ such that:
%\begin{align}
%H &= \int_{0}^{L}\!\!\int_{0}^{2\pi}
%\Bigg[
%\frac{1}{2}\,C_d^{(a)}(\partial_t\phi)^2
%+\frac{1}{2}\,(L_d^{(a)})^{-1}\!\left(\frac{k}{\beta}\right)^{\!4}(\partial_z\phi)^2
%+\frac{1}{2}\,C_d^{(a)}\!\left(\frac{k}{\beta}\right)^{\!2}(c k_c)^2\,\phi^2
%\Bigg]\; a\,d\theta\,dz, \\[3pt]
%P_z &= \int_{0}^{L}\!\!\int_{0}^{2\pi}
%C_d^{(a)}\!\left(\frac{k}{\beta}\right)^{\!2}\,[-\partial_t\phi\,\partial_z\phi]\; a\,d\theta\,dz. 
%\end{align}
%
% comme d'hab....
%%%
\begin{equation}
    \frac{\partial^2 \varphi(\theta,z,t)}{\partial z^2}-\frac{1}{v_\phi^2} \frac{\partial^2 \varphi(\theta,z,t)}{\partial t^2}=0 ,
    \label{eq:phi_wave_TM_hollow}
\end{equation}
with $v_\phi$ the phase velocity. 
%%%%
%with the corresponding propagation equation:
%\begin{equation}
%\partial_z^2\phi-\frac{1}{c^2}\!\left(\frac{\beta}{k}\right)^{\!2}\partial_t^2\phi=0,
%\qquad k^2=\beta^2+k_c^2.
%\end{equation}
%
% comme d'hab. En quoi rappeler pour la nième fois la relation de dispersion apporte qqch????
%%
%%
The surface integral Eq. (\ref{eq:HTMhollow}) can be recast in terms of an energy density:
\begin{equation}
    H_d = \frac{1}{2} C_d^{-1} \,\sigma_{front}^2 + \frac{1}{2} L_d \, \mathbf{j}_{front}^2 + \Delta(\theta,z,t),
\end{equation}
following the same logic as for the coaxial line (the same expression can obviously be written with the {\em back} terms). The addendum term $\Delta$ is again:
\begin{equation}
 \Delta(\theta,z,t)=\frac{1}{2}   C_d \left( \frac{k}{\beta}\right)^{\!\!2} (c\, k_c)^2  \varphi(\theta,z,t)^2  ,
\end{equation}
and corresponds to a potential energy needed for the confinement of light in the guide.
%Following the usual procedure, one can define the wall energy density in the local form:
%\begin{align}
%H_d^{(a)}(\theta,z,t)
%&=\frac{1}{2}\,(C_d^{(a)})^{-1}\,[\sigma^{(a)}]^2
%+\frac{1}{2}\,L_d^{(a)}\,[j^{(a)}_{z}]^2
%+\Delta^{(a)}(\theta,z,t) ,\\
%\Delta^{(a)}(\theta,z,t) &=  \frac{1}{2}\,C_d^{(a)}\!\left(\frac{k}{\beta}\right)^{\!2}(c k_c)^2\,\phi^2
%\end{align}

\subsubsection{TE modes}
\label{TEmodesHollow}
%\label{subsec:hollow_charges_currents_enTE}

Let us present first the $n>0$ situation. 
%Concerning the TE$_{m,n}$ case, one shall start considering the case $m\ge 1$ with the azimuthal pair $\phi'$ defined as: one virtual plane $\theta=\theta_*\equiv\theta_0+\frac{p\pi}{m}$
%and the next one $\theta_*+\frac{\pi}{m}$ to close a single electromagnetic cell, the whole attached to a plane, $dS=dr\,dz$ with $r\in [0,a]$. Thus, on one virtual plane, one writes down the following local laws:
%\begin{align}
%\sigma^{[\theta]}(r,z,t) &= C_d^{[\theta]}\,\partial_t\phi'(r,z,t),\\
%j^{[\theta]}_{z}(r,z,t) &= -\,\big(L_d^{[\theta]}\big)^{-1}\,\partial_z\phi'(r,z,t), \\
%j^{[\theta]}_{r}(r,z,t) &= -\,\big(L_d^{[\theta]}\big)^{-1}\,\partial_r\phi'(r,z,t). 
%\end{align}
%
% comme d'hab... je réécris...
We define:
\begin{eqnarray}
   && \!\!\!\!\!\!\!\!\!\!\!\! h_{\rm eff}  =   a \, \left( \frac{k_c a}{n}\right)^{\!\!2}\!  \, \left( \frac{ (k_c a)\, J_n[k_c a]^2 - 2 n \, J_n[k_c a]J_{n+1}[k_c a]+  (k_c a)\, J_{n+1}[k_c a]^2 }{2\, (k_c a)\, J_n[k_c a]^2 } \right)  , \\
   & &\mbox{and:} \\
  && \!\!\!\!\!\!\!\!\!\!\!\! g_{real}(\theta)  =  \sin [ n(\theta-\theta_0)] .
\end{eqnarray}
Charges and currents on the "front" conductor read: 
\begin{eqnarray}
    \sigma_{front} & =& + C_d \frac{\partial \varphi (\theta,z,t) }{\partial t} , \\
    \mathbf{j}_{front} & = & - L_d^{-1} \frac{\partial \varphi(\theta,z,t) }{\partial z} \, \mathbf{z} - L_d^{-1} \left( \frac{ k_c^2 a^2}{n^2} \right) \frac{1}{a} \frac{\partial \varphi(\theta,z,t) }{\partial \theta}\, \mathbf{u_\theta},  \label{eqcurrTEhollow}
\end{eqnarray}
which reproduces the results found for the coaxial guide. 
The "back" position being defined by $\theta \rightarrow \theta +\pi$, 
we obtain:
\begin{eqnarray}
    \sigma_{back} & =& -(-1)^{n+1} \, C_d \frac{\partial \varphi (\theta,z,t) }{\partial t} , \\
    \mathbf{j}_{back} & = & +(-1)^{n+1} \, L_d^{-1} \frac{\partial \varphi(\theta,z,t) }{\partial z} \, \mathbf{z} +(-1)^{n+1} \, L_d^{-1} \left( \frac{ k_c^2 a^2}{n^2} \right) \frac{1}{a} \frac{\partial \varphi(\theta,z,t) }{\partial \theta}\, \mathbf{u_\theta}.
\end{eqnarray}
The symmetry sign change (which depends here on $n+1$) seems again different from the one reported in Cartesian geometries \cite{Delattre2024}. But as for TM waves, this is simply due to the orientation of the basis vector $\mathbf u_r$ which always points towards the outside of the guide.
%%%%
Energy $H$ and momentum $\mathbf P$ are written in the same form as for the coaxial guide:
\begin{eqnarray}
H & = & \int_{0}^{L}\!\!\int_{0}^{2\pi}\! \left[ \frac{1}{2}   C_d \left( \frac{\partial \varphi(\theta,z,t) }{\partial t}\right)^2   \right. \nonumber \\
& & \left. +\frac{1}{2} L_d^{-1}  \left( \frac{\partial \varphi(\theta,z,t) }{\partial z}\right)^2  + \frac{1}{2}    L_d^{-1}  \frac{k_c^2 a^2}{n^2} \left( \frac{1}{a} \frac{\partial \varphi(\theta,z,t) }{\partial \theta}\right)^2 \right] a\, d\theta \,  dz ,\label{eq:hollow_TE_energy} \\
\mathbf{P} & = &  \int_{0}^{L}\!\!\int_{0}^{2\pi}\! C_d    \left[-\frac{\partial \varphi(\theta,z,t) }{\partial t}\frac{\partial \varphi(\theta,z,t) }{\partial z} \right] a\,   d\theta \,dz \, \mathbf{z} . 
\end{eqnarray}
The propagation equation deduced from Eq. (\ref{eq:surface_continuity_global}) is also equivalent:
\begin{equation}
        \frac{\partial^2 \varphi(\theta,z,t)}{\partial z^2}-\frac{1}{c^2} \frac{\partial^2 \varphi(\theta,z,t)}{\partial t^2}=k_c^2 \, \varphi(\theta,z,t) , \label{KGTEhollow}
\end{equation}
displaying the {\em Klein-Gordon} form specific to TE waves. 
As well, Eq. (\ref{eq:hollow_TE_energy}) corresponds to the surface integral of the density $H_d$:
\begin{equation}
    H_d = \frac{1}{2} C_d^{-1} \,\sigma_{front}^2 + \frac{1}{2} L_d \, \left(\mathbf{j}_{front}.\mathbf{z}\right)^2 + \frac{1}{2} L_d \frac{n^2}{k_c^2 a^2}\, \left(\mathbf{j}_{front}.\mathbf{u_\theta}\right)^2 ,
\end{equation}
or equivalently with {\em back} subscripts. 
As for the coaxial guide, a mass term $\hbar \,k_c/c$ appeared, linked directly to the peripheral current. \\

Let us finally describe TE waves with $n=0$. As for the coaxial line, virtual electrodes must be used in this case. We define:
\begin{eqnarray}
    h_{\rm eff} & = &   a \,\left( 2 \frac{J_0[k_c a]^2+J_1[k_ca]^2}{A_m^2} \right), \\
    &\mbox{and:}& \\
    g_{vir}(r) & = &2 \frac{J_1[k_c r]}{A_m} ,
\end{eqnarray}
and $A_m$ is obtained numerically (Subsection \ref{sec:TE_hollow}). 
%%%%%%%%% ok?
Charges and currents on the top part of the virtual plane read:
\begin{eqnarray}
    \sigma_{vir\,top} & =& + C_d \frac{\partial \varphi (r,z,t) }{\partial t} , \\
    \mathbf{j}_{vir\,top} & = & - L_d^{-1} \frac{\partial \varphi(r,z,t) }{\partial z} \, \mathbf{z}
    - L_d^{-1} \frac{1}{r} \frac{\partial \left(\, r \,\varphi [r,z,t] \,\right)}{\partial r} \, \mathbf{u_r},  \label{eqcurrTEhollspec}
\end{eqnarray}
while for the bottom one we have:
\begin{eqnarray}
    \sigma_{vir\,bottom} & =& - C_d \frac{\partial \varphi (r,z,t) }{\partial t} , \\
    \mathbf{j}_{vir\,bottom} & = & + L_d^{-1} \frac{\partial \varphi(r,z,t) }{\partial z} \, \mathbf{z}
    + L_d^{-1} \frac{1}{r} \frac{\partial \left(\, r \,\varphi [r,z,t] \,\right)}{\partial r} \, \mathbf{u_r},  
\end{eqnarray}
the signs are reversed.
The peripheral current flowing in the real electrode is:
\begin{eqnarray}
\mathbf{j}_{front} =\mathbf{j}_{back} & = &  
    - L_d^{-1} \, 2 k_c  \left( \frac{  J_0[k_c a]   }{A_m  } \right) \, \phi_m \,\tilde{f}(z,t)\, \mathbf{u_\theta} , 
\end{eqnarray}
which connects to the virtual current by continuity in $r=a$ (see Fig. \ref{fig_2}).
The constants of motion are then obtained with the integrals:
\begin{eqnarray} 
H & = & \int_{0}^{L}\!\!\int_{0}^{a}\! \left[ \frac{1}{2 }
C_d \left( \frac{\partial \varphi(r,z,t) }{\partial t}\right)^2   \right. \nonumber \\
& & \left. +\frac{1}{2} L_d^{-1}  \left( \frac{\partial \varphi(r,z,t) }{\partial z}\right)^2  + \frac{1}{2}    L_d^{-1}\left(\frac{1}{r} \frac{\partial \left(\, r \,\varphi [r,z,t] \,\right)}{\partial r} \right)^2  \right] \frac{2\pi \,r}{  h_{\rm eff}}\, dr \,  dz ,\label{eq:hollow_TE_energyspe} \\
\mathbf{P} & = &  \int_{0}^{L}\!\!\int_{0}^{a}\!   
C_d \left[-\frac{\partial \varphi(r,z,t) }{\partial t}\frac{\partial \varphi(r,z,t) }{\partial z} \right]\frac{2\pi \, r}{  h_{\rm eff}}\, dr \,dz \, \mathbf{z} ,
\end{eqnarray}
exactly like in the coaxial case. We obtain again the Klein-Gordon propagation equation for the generalized flux $\varphi$:
\begin{equation}
        \frac{\partial^2 \varphi(r,z,t)}{\partial z^2}-\frac{1}{c^2} \frac{\partial^2 \varphi(r,z,t)}{\partial t^2}=k_c^2 \, \varphi(r,z,t) , \label{KGTEhollow2}
\end{equation}
from the charge conservation equation applied onto the virtual plane. 
The virtual surface charge/current density is again:
\begin{equation}
    H_d = \frac{1}{2} \left[C_d^{-1}\frac{2\pi \, r}{  h_{\rm eff}} \right] \,\sigma_{vir\,top}^2 + \frac{1}{2} \left[L_d\frac{2\pi \, r}{  h_{\rm eff}} \right] \, \left(\mathbf{j}_{vir\,top}.\mathbf{z}\right)^2 + \frac{1}{2} \left[L_d\frac{2\pi \, r}{  h_{\rm eff}} \right]\, \left(\mathbf{j}_{vir\,top}.\mathbf{u_r}\right)^2 ,
\end{equation}
and equivalently with a {\em vir bottom} subscript. 
The mass term and peripheral current terms, characteristic of a TE wave, appear here exactly like in the coaxial treatment.
%%
%%%
%% bon, je suis d'accord qu'il y a pas mal de formules qui se répètent, mot pour mot... Et c'est moche. M'enfin, elles sont pas dans les memes sous sections, c'est pour éviter au lecteur de faire des retours en arrière incessant...
%%%

%For $m=0$ there are no azimuthal virtual electrodes,
%hence no $\phi'$. The walls enforce $E_\theta(a)=0$ and the cutoff is set by:
%\begin{equation}
%J_1(k_c a)=0,\qquad n=1,2,\ldots\quad (m=0).
%\end{equation}
%Indeed, the same cutoff $ck_c$ exists for TE$_{0,n}$, but it is accessed via the \emph{volume}
%energy (or a circulation variable built from $A_\theta$), not through
%\eqref{eq:coax_TE_density}. Only circulating surface currents exist at the wall. The total energy and momentum are taken in volume form:
%\begin{align}
%H&=\!\int_0^L\!\!\int_0^a\!\!\int_0^{2\pi}
%\Big[\tfrac{\epsilon}{2}E_\theta^2+\tfrac{1}{2\mu}(B_r^2+B_z^2)\Big]\, r\,d\theta\,dr\,dz,
%\\
%P_z&=\!\int_0^L\!\!\int_0^a\!\!\int_0^{2\pi}
%\epsilon\,(-E_\theta B_r)\, r\,d\theta\,dr\,dz,  
%\end{align}
%and the envelope obeys the same TE Klein–Gordon law with $k_c$ from \eqref{KG}.
%
% no more comments...
%

%%% ADD REF 3
%%% XXXXXXXXX
\begin{table}[H]
\centering
\caption{Summary of the effective thicknesses $h_\mathrm{eff}$ obtained for the different configurations and wave families (see text for details).
}
\label{tab:TAB_ADD}
\begin{tabular}{@{}l l@{}}
\toprule
\textbf{Wave type} & \qquad \textbf{$h_\mathrm{eff}$ thickness} \\
\midrule
Coax. TEM       & $\,\,\,\,\,\, a \, \ln{\left[\frac{b}{a} \right]}$ \\[4pt]
Coax. TM        & $\,\,\,\,\,\, a \, \frac{ \pi^2}{2} \int_{a}^{b}\! \frac{k_c^2 r}{2} \left( J_n[k_c r] \, Y_n[k_c a] - 
  J_n[k_c a ] \, Y_n [ k_c r] \right)^2 \, dr$ \\[4pt]
Coax. TE $n>0$  & $\,\,\,\,\,\, a \, \frac{2}{n^2} \!\! \int_{a}^{b}\! \frac{k_c^2 r}{2} \!\left(\frac{\left( J_{n+1}[k_c a]-J_{n-1}[k_c a]\right) \, Y_n[k_c r] - 
  J_n[k_c r ] \, \left( Y_{n+1} [ k_c a]-Y_{n-1} [ k_c a] \right) }{ \left( J_{n+1}[k_c a]-J_{n-1}[k_c a]\right) \, Y_n[k_c a] - 
  J_n[k_c a ] \, \left( Y_{n+1} [ k_c a]-Y_{n-1} [ k_c a] \right)} \right)^{\!\!2} \!  dr$ \\[4pt]
Coax. TE $n=0$  & $\,\,\,\,\,\, a \,\int_{a}^{b}\! \frac{4 r}{a^2} \!\left(\frac{J_1[k_c r] \, Y_1[k_c a] - J_1[k_c a] \, Y_1[k_c r] }{A_m  \, (k_c a) } \right)^{\!\!2} \!  dr$ \\[4pt]
Cyl. TM         & $\,\,\,\,\,\, a \, \left[ 2\left(\frac{J_{n-1} [k_c a]}{J_{n-1} [k_c a]-J_{n+1} [k_c a]}\right)^2 \right]$ \\[4pt]
Cyl. TE  $n>0$  & $\,\,\,\,\,\, a \, \left( \frac{k_c a}{n}\right)^{\!\!2}\!  \, \left( \frac{ (k_c a)\, J_n[k_c a]^2 - 2 n \, J_n[k_c a]J_{n+1}[k_c a]+  (k_c a)\, J_{n+1}[k_c a]^2 }{2\, (k_c a)\, J_n[k_c a]^2 } \right)$ \\[4pt]
Cyl. TE $n=0$   & $\,\,\,\,\,\, a \,\left( 2 \frac{J_0[k_c a]^2+J_1[k_ca]^2}{A_m^2} \right)$ \\
\bottomrule
\end{tabular}
\label{tabADD3}
\end{table}

\section{Electromagnetic gauge}  %c'est plutot une section???
\label{sec:gauge_cyl}

%As precised in Ref\cite{Delattre2024}, 
%%% ouais bon... sur ce genre de chose, plutot que s'auto-citer tu ferais mieux de citer un bouquin... ou rien...
%%%% se donner du crédit pour un truc d'il y a 2 siècles, ça passe mal...

A key point of electromagnetic theory is the existence of  scalar and vector potentials $V,\mathbf{ A}$ 
%%%% pourquoi des parenthèses? t'en a pas mis pour E et B...
%%% je les vire.
from which Maxwell’s physical fields $\mathbf{ E},\mathbf{ B}$ can be derived \cite{CohenQED}:
\begin{align}
\mathbf{E}(\mathbf{r},t) &= -\,\frac{\partial \mathbf{A}(\mathbf{r},t)}{\partial t}\;-\;\mathbf{\nabla} V(\mathbf{r},t), \label{eq:E_from_pot}\\
\mathbf{B}(\mathbf{r},t) &= \mathbf{\nabla}\times \mathbf{A}(\mathbf{ r},t).\label{eq:B_from_pot}
\end{align}
%%%% encore une fois, tes notations... les dérivées,
%%%% et tu es parti sur du bold pour les vecteurs, et pas des flèches. Tiens toi au même notations tout du long!!!!!!!
%%%% 
These potentials are defined up to a transformation: 
\begin{equation}
\mathbf{A} (\mathbf{r},t)\;\to\; \mathbf{A}(\mathbf{r},t)+\mathbf{\nabla}\,\Pi(\mathbf{r},t), \qquad
V (\mathbf{r},t)\;\to\; V(\mathbf{r},t) -\frac{\partial \Pi(\mathbf{r},t)}{\partial t} ,
\label{eq:gauge_transform}
\end{equation}
with $\Pi(\mathbf{r},t)$ an arbitrary function called  gauge.
%%% ok?
%
%These relations are \emph{coordinate invariant}; nothing in
%\eqref{eq:E_from_pot}–\eqref{eq:gauge_transform} assumes Cartesian or cylindrical coordinates. This gauge invariance is a fundamental property of Nature: as showed for cartesian geometry, it can't be chosen arbitrarily, and we aim to show in this article that this feature is real for any geometry. 
%
% c'est horriblement naif...
This fact is known as {\em gauge invariance}, and is obviously valid for any coordinate system; we shall express it here in the cylindrical geometry.

%In cylindrical coordinates $\vec r=(r,\theta,z)$, only gauge functions involving the propagating degree of freedom along $(z,t)$ and sharing the transverse symmetries $(r,\theta)$ of the field are physically relevant. According to 
%%%
Following the reasoning of Ref. \cite{Delattre2024},
%we choose the separated form:
the relevant gauges should involve our $X,Y$ quadrature variables, and should share the  symmetry of the problem at hand. This enables to write: 
\begin{equation}
\Pi(\mathbf{r},t)\;=\;p(\mathbf{r},\theta)\,f(z,t)\;+\;\tilde p(\mathbf{r},\theta)\,\tilde f(z,t),
\label{eq:cyl82}
\end{equation}
with the potentials verifying the Lorenz gauge condition:
%The Lorenz gauge for the electromagnetic potentials reads:
\begin{equation}
\nabla\!\cdot\!\mathbf{A}(\mathbf{r},t)\;+\;\frac{1}{c^{2}}\,
\frac{\partial V(\mathbf{ r},t)}{\partial t}\;=\;0.
\label{eq:lorenz}
\end{equation}
%with $c=1/\sqrt{\mu\varepsilon}$ the wave speed in the medium.
% mais pourquoi le répéter?
Any other choice amounts to a trivial invariance with no physical content.
%Requiring the primed potentials to \emph{also} satisfy \eqref{eq:lorenz}
%imposes the standard constraint on the gauge function:
%\begin{equation}
%\Big(\nabla^{2}-\frac{1}{c^{2}}\partial_{t}^{2}\Big)\Pi(\vec r,t)\;=\;0.
%\label{eq:wavePi}
%\end{equation}
%
% bon ca c'est ptet un peu trivial non? on peut ptet passer au coeur du truc?
%
Injecting the %separated 
ansatz Eq. \eqref{eq:cyl82} into Eq. \eqref{eq:lorenz}, we obtain the pair of transverse
%Helmholtz 
equations for the gauge profiles:
\begin{equation}
\frac{1}{r}\,\frac{\partial }{\partial r} \left(r\,\frac{\partial  p(r,\theta)}{\partial r}\right)
\;+\; \frac{1}{r^{2}}\,\frac{ \partial^2  p(r,\theta)}{\partial \theta^2}
\;+\; \big(k^{2}-\beta^{2}\big)\,p(r,\theta) \;=\; 0,
\label{eq:gauge1}
\end{equation}
\begin{equation}
\frac{1}{r}\,\frac{\partial }{\partial r} \left(r\,\frac{\partial  \tilde{p}(r,\theta)}{\partial r}\right)
\;+\; \frac{1}{r^{2}}\,\frac{ \partial^2  \tilde{p}(r,\theta)}{\partial \theta^2}
\;+\; \big(k^{2}-\beta^{2}\big)\,\tilde{p}(r,\theta) \;=\; 0 .
\label{eq:gauge2}
\end{equation}
%Equations \eqref{eq:cyl84}–\eqref{eq:cyl85} are the cylindrical counterparts of
%the Cartesian Eqs.~(84)–(85) in Ref.\cite{Delattre2024}: they enforce that the admissible gauge functions
%share the modal cut-off $k_c^{2}\equiv k^{2}-\beta^{2}$ and the transverse
%symmetries of the fields. 
%
%% Bon, on peut faire plus sharp:
These equations must be solved for each case, taking into account the proper symmetries. 
%%%
Special care will be required by the TE $n=0$ modes, which involve virtual electrodes. 
%%% et surtout!:
We shall demonstrate below how a specific gauge choice can then be made, relating $\mathbf A$ and $V$ to the generalized flux $\varphi$ through Devoret-like expressions \cite{DevoretQED}. 
%%%%%%%%%%% OK?

\begin{table}[H]
\centering
\caption{Gauge description for TEM waves.}
\label{tab:coax_tem_gauge}
\begin{tabular}{@{}l l@{}}
\toprule
\textbf{Function} & \textbf{Expression} \\[5pt]
\midrule
$A_r(r,\theta,z,t)$ & $= \ 0$ \\ 
$A_\theta(r,\theta,z,t)$ & $= \ 0$ \\
$A_z(r,\theta,z,t)$ & $= \ +\phi_m\,\beta \, \left[\frac{1}{2}- \frac{ \ln(r/a)}{ \ln(b/a)}  \right]   f(z,t)$ \\
$V(r,\theta,z,t)$ & $= \ +\phi_m\,\omega \, \left[\frac{1}{2}- \frac{ \ln(r/a) }{\ln(b/a)} \right]   f(z,t)$ \\[5pt]
\midrule
\textbf{Gauge fixing} & $a_\pi = 0,\ \tilde a_\pi = 0$ \\
\textbf{Gauge invariance} & $b_\pi,\ \tilde b_\pi$ \\ [5pt] 
\bottomrule
\end{tabular} %%%% CHEEEECK OK, it was wrong...
%%%%%%%%%%%%%%%%%%%%%% t'avais une foooote de signe; aussi, t'avais pas la même definition symmétrique que dans waveguide
\end{table}
%%% les potentiels fonctionnent, on retrouve heff ici, et les signes sont opposés sur les 2 électrodes. OK? C'est VRAIMENT PAREIL que le Cartésien...

\subsection{Coaxial waveguide}
\label{sec:coaxPotentials}

\subsubsection{TEM modes}

For TEM waves, % in a two-conductor cylindrical geometry, 
the dispersion relation enforces
$k=\lvert\beta\rvert$. %, i.e.\ $k_c^2\equiv k^2-\beta^2=0$. Hence the transverse profiles obey the
%{\it Laplace} equation on $(r,\theta)$:
%\begin{equation}
%\frac{1}{r}\,\partial_r\!\big(r\,\partial_r p\big)
%\;+\;\frac{1}{r^2}\,\partial_\theta^2 p \;=\;0,
%\qquad
%\frac{1}{r}\,\partial_r\!\big(r\,\partial_r \tilde p\big)
%\;+\;\frac{1}{r^2}\,\partial_\theta^2 \tilde p \;=\;0.
%\label{eq:Laplace_cyl}
%\end{equation}
%
%In a coaxial guide (inner conductor at $r=a$, outer at $r=b$), the TEM solution
%is azimuthally invariant, $\partial_\theta p=\partial_\theta \tilde p=0$. 
%%% tout ca est inutile, on a bien compris avec le paragraphe d'avant!!! Va a l'essentiel:
Solving Eqs. 
(\ref{eq:gauge1},\ref{eq:gauge2}) with the azimuthal invariance characteristic of these fields %gives the harmonic (radial) primitives:
leads to:
\begin{equation}
p(r) \;=\; a_\pi \,\ln r \;+\; b_\pi,
\qquad
\tilde p(r) \;=\; \tilde a_\pi \,\ln r \;+\; \tilde b_\pi,
\label{eq:86cyl}
\end{equation}
with $a_\pi,b_\pi,\tilde a_\pi,\tilde b_\pi$ four real constants.
%%%
%%% c'est là que ca se corse, il faut faire le lien avec Devoret...
We define here:
\begin{eqnarray}
     \Delta V(z,t) & = & V(r=a,z,t)  -  V(r=b,z,t) , \\
     \Delta A_z(z,t) & = & A_z(r=a,z,t) - A_z(r=b,z,t) ,
\end{eqnarray}
the potential differences between the electrodes (with no $\theta$ dependence). These can be linked to the generalized flux through the relations:
\begin{eqnarray}
  \frac{\partial \varphi (\theta,z,t)}{\partial t}   & = &  \Delta V(z,t) ,  \\
   \frac{\partial \varphi(\theta,z,t)}{\partial z}     & = & - \Delta A_z(z,t) ,   
\end{eqnarray}
which express here the Devoret relationship \cite{Delattre2024,DevoretQED}.
%to be fixed by the
%electrode boundary data and by the gauge choice (see below). Let us now introduce the electrode labels $\zeta_1:\,r=a$ (inner) and
%$\zeta_2:\,r=b$ (outer), defining the \emph{voltage difference} such that $\Delta V(z,t) \;\equiv\; V(a,z,t) \;-\; V(b,z,t).$ Exactly as in the Cartesian treatment, the formalism based on the gauge scalar
%$\phi$ (generalized flux) matches the literature when one works with
%voltage differences $\Delta V$ and currents $I$, which then connect to the
%telegrapher's equation and lumped-element circuit theory. Upon quantization,
%these variables become the canonical ones of quantum circuits. Therefore, the key relationship used in that framework is:
%\begin{equation}
%\frac{\partial \phi}{\partial t} \;=\; \Delta V.
%\label{eq:87cyl}
%\end{equation}
%%%
%%%
%%%
%%%
Importantly, these formulas are \emph{not gauge invariant} (see Subection \ref{gaugePLUS}).
%it relies on fixing a particular gauge that renders \eqref{eq:87cyl} meaningful (we will make this choice explicit and consistent with Lorenz invariance).
%%%
%%% ce point est le SEUl vraiment important ici! 
%%%
%%%
%Precisely, starting from the initial gauge freedom and requiring Lorenz invariance
%for the transformed potentials, one must (partly or fully) fix the gauge in order
%to obtain relations of the type \eqref{eq:87cyl}. This is the gauge fixing
%mechanism in the present context. A direct consequence of the Lorenz condition
%\eqref{eq:lorenz}, evaluated on the two electrodes and subtracted, is the axial
%counterpart:
%\begin{equation}
%\frac{\partial \phi}{\partial z} \;=\; -\,\Delta A,
%\qquad
%\Delta A(z,t)\;\equiv\;A_z(a,z,t)\;-\;A_z(b,z,t),
%\label{eq:88cyl}
%\end{equation}
%where $\Delta A$ is the difference of the \emph{longitudinal} vector potential
%between inner and outer conductors. Equations \eqref{eq:87cyl}–\eqref{eq:88cyl}
%are thus the cylindrical (coaxial) counterparts of the Cartesian identities,
%obtained under the same logical steps: separation of transverse/longitudinal
%dependences, TEM condition $k=\lvert\beta\rvert$ (hence Laplace problems for the
%transverse gauge profiles), and Lorenz-invariant gauge fixing tying $(\phi,V,A_z)$
%to electrode differences.
%%%
%%%
%The TEM wave results for the coaxial line are summarized 
%The components of $\mathbf A, V$ are listed,
%with the gauge coefficients $a_\pi, b_\pi, \tilde a_\pi, \tilde b_\pi$.
The potentials corresponding to this gauge are given
in
Tab.~\ref{tab:coax_tem_gauge}, written such that $a_\pi, \tilde{a}_\pi=0$: this is the gauge fixing corresponding to the "Devoret gauge" in the coaxial geometry. 
We recognize the usual
\textit{longitudinal gauge}. %%% ca, je garde!

Similarly to the Cartesian case, the $b_\pi, \tilde{b}_\pi$ constants are undefined: this is the remaining gauge invariance  \cite{Delattre2024}.
%% et là, petite explique sur le choix fait dans le tableau...
Note that the choice made in the tabular is such that potentials have opposite signs on the two electrodes, matching Ref. \cite{Delattre2024}.
%%
%Exactly following our cartesian Reference \cite{Delattre2024}, we fix the gauge to set all coefficient to zero, leaving the remaining ones undetermined. Again, 
%, here in the cylindrical geometry. The same
%structure in the formula typesetting will be used for the other modal
%families, for reasons of readability.
This is a pure commodity, and is not required by the modeling, especially since the two electrodes are {\em different} in size (which breaks the symmetry). 
% The gauge fixing is analyzed below for every specific configuration.
%We follow below the same gauge fixing logic for all other configurations, as originally presented in this reference.

\subsubsection{TM modes}
\label{TMPotcoax}
%%% pourquoi tu faisais TE avant TM???
%% On a toujours fait le contraire, je remets dans l'ordre.

% --- TM waves (two-conductor cylindrical geometry) ---
For TM waves, % in a two-conductor cylindrical geometry, the dispersion likewise
the dispersion relation reads $ k^2-\beta^2 = k_c^2 > 0$. 
Eqs. (\ref{eq:gauge1},\ref{eq:gauge2}) therefore lead to the solutions:
\begin{eqnarray}
  && \!\!\!\!\!\!\!\!\!\!\!\!\!\!\!\!\!\!\!\!\!\!\!\!  p(r,\theta)   =  \Big( a_\pi J_n[k_c r] + b_\pi Y_n[k_c r]  \Big) \cos [n(\theta-\theta_0)] +  \Big( c_\pi J_n[k_c r] + d_\pi Y_n[k_c r]\Big) \sin [n(\theta-\theta_0)], \label{petgauge1} \\
  && \!\!\!\!\!\!\!\!\!\!\!\!\!\!\!\!\!\!\!\!\!\!\!\!  \tilde{p}(r,\theta)   =   \left( \tilde{a}_\pi J_n[k_c r] + \tilde{b}_\pi Y_n[k_c r]  \right) \cos [n(\theta-\theta_0)] +  \left( \tilde{c}_\pi J_n[k_c r] + \tilde{d}_\pi Y_n[k_c r]\right) \sin [n(\theta-\theta_0)], \label{petgauge2}
\end{eqnarray}
%Hence the transverse primitives obey the
%same Helmholtz equation on $(r,\theta)$:
%\[
%\frac{1}{r}\,\partial_r\!\big(r\,\partial_r p\big)
%+ \frac{1}{r^2}\,\partial_\theta^2 p
%+ k_c^2\,p = 0, 
%\qquad
%\frac{1}{r}\,\partial_r\!\big(r\,\partial_r \tilde p\big)
%+ \frac{1}{r^2}\,\partial_\theta^2 \tilde p
%+ k_c^2\,\tilde p = 0 .
%\]
%In a coaxial guide (inner at $r=a$, outer at $r=b$), the TM solution also
%carries the azimuthal order $m$ with the same real basis. Solving yields the
%harmonic (radial) primitives:
%\[
%p(r)=a_\pi\,J_m(k_c r)+b_\pi\,Y_m(k_c r), 
%\qquad
%\tilde p(r)=\tilde a_\pi\,J_m(k_c r)+\tilde b_\pi\,Y_m(k_c r),
%\]
%with $a_\pi,b_\pi,\tilde a_\pi,\tilde b_\pi$ fixed by the electrode boundary
%data and by the gauge choice (see below).
%%
%%% t'as pas besoin de ré"péter les équations 100 fois! On donne juste la solution, c'est du trivial ca.
%%% et du coup... t'as raté le fait qu'il y a du sinus et du consinus!!!!!!!!
which are here characterized by eight real coefficients when $n>0$; for $n=0$, the $c_\pi, d_\pi, \tilde{c}_\pi, \tilde{d}_\pi$ are irrelevant and we are again left with four parameters.  
%%%%
%%%% Où est la condition de jauge, avec l'expression de Devoret????
%%%% C'est pourtant CA, et QUE CA le point central de cette section!!!! DEFINIR LA bonne jauge pour chaque cas particulier!!!
As in Ref. \cite{Delattre2024}, we must define adapted "potential differences" that take into account the transverse symmetry of the mode profiles (Subsection \ref{subsec:coax_charges_currentsTM}). We pose:
\begin{eqnarray}
     \Delta V(\theta,z,t) & = & V(r=a,\theta,z,t)  +(-1)^{m} \,  V(r=b,\theta,z,t) , \label{eq:TMVcyl1} \\
     \Delta A_z(\theta,z,t) & = & A_z(r=a,\theta,z,t)  +(-1)^{m} \,  A_z(r=b,\theta,z,t) , \label{eq:TMVcyl2}
\end{eqnarray}
in which we either subtract or add up the potentials of the facing electrodes. 
With these, the Devoret expressions still apply with the proper angular dependence: 
\begin{eqnarray}
  \frac{\partial \varphi (\theta,z,t)}{\partial t}   & = &  \Delta V(\theta,z,t) , \label{eq:phidot} \\
   \frac{\partial \varphi(\theta,z,t)}{\partial z}     & = & - \Delta A_z(\theta,z,t) . \label{eq:phiprime} 
\end{eqnarray}
The gauge fixing underlying this choice is given in Tab. \ref{tabTMpotcyl}.
% tu noteras que je suis moi-meme assez lourd... il y a quand même pas mal de choses qui se répètent, mais techniquement, les formules sont différentes. Pour TEM, les Delta V, Delta A n'ont pad de dépendence en theta. Et pour TE n==0, la dépendence sera en r.... dans ces cas là, je réécrit les formules quand meme, pour bien expliciter les différences. C'est pas simple, et pour pas perdre le lecteur et bien montrer de quoi on parle, je pense que c'est quand même nécessaire. Y a bien 1 ou 2 formules qui sont rigouresement identiques et qui pourrait sauter, et qui sont là que pour la fluidité du texte (pour pa sauter en arriere...). Elles pourrait etre retirer. On verra ce que disent les referree...
%
But there is a subtlety here: because of the high symmetry of the problem at hand,  $J_n$ and $Y_n$ functions are both eligible in the potentials. This leads to an extra degeneracy represented by the real number $\alpha$ in the tabular. 
%%%%
With our conventions, we always have $V(r=a,\theta,z,t)>0$ while $V(r=b,\theta,z,t)$ displays the sign of $(-1)^m$. However, $|V(r=a,\theta,z,t)| \neq |V(r=b,\theta,z,t)|$ which is {\em different} from the TEM case discussed above, or the parallel plate result \cite{Delattre2024}. Only when $b/a \rightarrow 1$ do we recover the perfect symmetry/anti-symmetry of the electrode potentials (the same conclusions obviously apply to the $A_z$ component).
%%%
Actually, the parameter $\alpha$ vanishes from the expressions of the potentials on the electrodes, which is a consequence of the boundary condition.
%%%%
%%%%
As well, the gauge invariance is relaxed as compared to the Cartesian case. As soon as:
\begin{eqnarray}
    a_\pi & = & - \frac{Y_n[k_c a]+ (-1)^m \,Y_n[k_c b] }{J_n[k_c a]+ (-1)^m \,J_n[k_c b]} \ b_\pi ,  \label{eqTMGaugec1} \\
    c_\pi & = & - \frac{Y_n[k_c a]+ (-1)^m \,Y_n[k_c b] }{J_n[k_c a]+ (-1)^m \,J_n[k_c b]} \ d_\pi ,  \label{eqTMGaugec2}
\end{eqnarray}
and similarly for the tilded parameters, the gauge expressions Eqs. (\ref{eq:TMVcyl1},\ref{eq:TMVcyl2}) verify the Devoret laws Eqs. (\ref{eq:phidot},\ref{eq:phiprime}).
%%%
We can thus fix one constant of the pair (say $a_\pi$) as a function of the other one, which is kept unknown.
%%%
In this sense, the gauge invariance is %much 
higher than in the corresponding parallel plate case \cite{Delattre2024}.

\begin{table}[H]
\centering
\caption{Gauge description for $\mathrm{TM}_{n,m}$ waves in a coaxial waveguide with $n\geq 0, m>0$. %  ($a<r<b$). 
%The normalization constant $A_{m,n}$ is fixed by the electric-field normalization 
%$\epsilon\!\!\iint_{\mathcal S}\!(g_{E_r}^2+g_{E_\theta}^2+g_{E_z}^2)\,r\,dr\,d\theta=1$. 
%Here $R_m(r)=J_m(k_c r)\,Y_m(k_c a)-Y_m(k_c r)\,J_m(k_c a)$ and the TM dispersion relation 
%is $J_m(k_c a)Y_m(k_c b)-Y_m(k_c a)J_m(k_c b)=0$
%%%% mais pourquoi tu colles ca là??? on s'en fout, la normalisation c'est fait dans la partie sur le champ, ce qui importe ici c'est bien COMMENT tu fixes la jauge!!!
}
\label{tab:gauge_TM_mn_coax}
\setlength{\tabcolsep}{5pt}
\renewcommand{\arraystretch}{1.1}
% \resizebox{\textwidth}{!}{%
\hspace*{-1cm}
\begin{tabular}{@{}l l@{}}
\toprule
\textbf{Function} & \textbf{Expression} \\
\midrule
$A_r(r,\theta,z,t)$
& $=\;\phi_m \frac{k_c}{2} \left( \left[  \frac{(1-\alpha) \, [J_{n+1}(k_c r)-J_{n-1}(k_c r)]}{J_n(k_c a)+(-1)^m \, J_n(k_c b)}+\frac{\alpha \, [Y_{n+1}(k_c r)-Y_{n-1}(k_c r)]}{Y_n(k_c a)+(-1)^m \, Y_n(k_c b)} \right] \right. $  \\[10pt]
& $\!\!\!\!\!\!\!\!\!\!\!\! \left. + \frac{a \,  \pi}{2 \, h_{\rm eff}}  \left[ J_n(k_c a) [ Y_{n+1}(k_c r)-Y_{n-1}(k_c r)] - [J_{n+1}(k_c r)-J_{n-1}(k_c r)] Y_n(k_c a) \right]  \right)\cos[n(\theta-\theta_0)]\ \tilde{f}(z,t) $  \\[10pt]

$A_\theta(r,\theta,z,t)$
& $=\;\phi_m \frac{n}{r} \left( \left[ \frac{(1-\alpha) \, J_n(k_c r)}{J_n(k_c a)+(-1)^m \, J_n(k_c b)}+\frac{\alpha \, Y_n(k_c r)}{Y_n(k_c a)+(-1)^m \, Y_n(k_c b)} \right] \right. $  \\[10pt]
& $\left. + \frac{a \,  \pi}{2 \, h_{\rm eff}}  \left[J_n(k_c r) Y_n(k_c a)-J_n(k_c a)Y_n(k_c r)  \right]  \right)\sin[n(\theta-\theta_0)]\ \tilde{f}(z,t) $ \\[10pt]

$A_z(r,\theta,z,t)$
& $=\; \phi_m \, \beta \left( \left[ \frac{(1-\alpha) \, J_n(k_c r)}{J_n(k_c a)+(-1)^m \, J_n(k_c b)}+\frac{\alpha \, Y_n(k_c r)}{Y_n(k_c a)+(-1)^m \, Y_n(k_c b)} \right] \right. $\\[10pt]
& $\left. + \frac{a \,  \pi}{2 \, h_{\rm eff}} \frac{k_c^2}{\beta^2}  \left[ J_n(k_c r) Y_n(k_c a)-J_n(k_c a)Y_n(k_c r) \right]  \right)\cos[n(\theta-\theta_0)]\ f(z,t) $ \\[10pt]

$V(r,\theta,z,t)$
 & $=\; \phi_m \, \omega \left( \frac{(1-\alpha) \, J_n(k_c r)}{J_n(k_c a)+(-1)^m \, J_n(k_c b)}+\frac{\alpha \, Y_n(k_c r)}{Y_n(k_c a)+(-1)^m \, Y_n(k_c b)} \right)\cos[n(\theta-\theta_0)]\ f(z,t)$
\\[10pt] 

\midrule
\textbf{Gauge fixing} & $a_\pi  ,\ \ \tilde a_\pi   , \  \  c_\pi  ,\ \ \tilde c_\pi $ from Eqs. (\ref{eqTMGaugec1},\ref{eqTMGaugec2}) \\
\textbf{Gauge invariance} & $b_\pi,\ \ \tilde b_\pi,\ \ d_\pi,\ \ \tilde d_\pi$ and $\alpha$   \\
\bottomrule
\end{tabular}
% } % end resizebox  
\label{tabTMpotcyl}
\end{table}
%%%
%%% Et là, y a des choses à expliquer....

\subsubsection{TE modes}

TE waves verify the same dispersion relation as the previous TM ones, and share the same generic symmetry. Therefore $p(r,\theta)$, $\tilde{p}(r,\theta)$ have the same expressions.

\begin{table}[H]
\centering
\caption{Gauge description for $\mathrm{TE}_{n,m}$ modes in a coaxial waveguide, with $n>0,m>0$. % ($a<r<b$). 
%Gauge fixing and normalization identical to the Cartesian case. 
%The normalization constant $A_{n,m}$ is the same as in Table~\ref{tab:TE_mn_coax_profiles}, 
%and the dispersion relation is $J_n'(k_c a)Y_n'(k_c b)-Y_n'(k_c a)J_n'(k_c b)=0$.
%%%%%%%%%%%%%% c'est TOUT!!!!
}
%\label{tab:gauge_TE_mn_coax}
\setlength{\tabcolsep}{5pt}
\renewcommand{\arraystretch}{1.1}
%\resizebox{\textwidth}{!}{%
\hspace*{-1cm}
\begin{tabular}{@{}l l@{}}
\toprule
\textbf{Function} & \textbf{Expression} \\
\midrule

$A_r(r,\theta,z,t)$
& $=\;\phi_m  \left( \frac{k_c}{2}\left[  \frac{(1-\alpha) \, [J_{n+1}(k_c r)-J_{n-1}(k_c r)]}{J_n(k_c a)+(-1)^{m+1} \, J_n(k_c b)}+\frac{\alpha \, [Y_{n+1}(k_c r)-Y_{n-1}(k_c r)]}{Y_n(k_c a)+(-1)^{m+1} \, Y_n(k_c b)} \right] \right. $   \\[10pt]
& $  \left. + \frac{ a  }{ r \, h_{\rm eff}} 
\left[ \frac{J_n[k_c r] (Y_{n+1}[k_c a]-Y_{n-1}[k_ca])-Y_n[k_c r](J_{n+1}[k_c a]-J_{n-1}[k_c a] )}{(J_{n+1}[k_c a]-J_{n-1}[k_c a])Y_n[k_c a] -J_n[k_c a] (Y_{n+1}[k_c a]-Y_{n-1}[k_ca])} \right]  \right)\sin[n(\theta-\theta_0)]\ \tilde{f}(z,t) $  \\[10pt]

$A_\theta(r,\theta,z,t)$
& $=\;\phi_m  \left( - \frac{n}{r} \left[ \frac{(1-\alpha) \, J_n(k_c r)}{J_n(k_c a)+(-1)^{m+1} \, J_n(k_c b)}+\frac{\alpha \, Y_n(k_c r)}{Y_n(k_c a)+(-1)^{m+1} \, Y_n(k_c b)} \right] \right. $  \\[10pt]
& $\left. + \frac{k_c \, a}{2 \, n \, h_{\rm eff}}  \left[ -\frac{(J_{n+1}[k_c r]-J_{n-1}[k_c r])\,(Y_{n+1}[k_c a]-Y_{n-1}[k_ca])}{(J_{n+1}[k_c a]-J_{n-1}[k_c a])Y_n[k_c a] -J_n[k_c a] (Y_{n+1}[k_c a]-Y_{n-1}[k_ca])} \right. \right.$ \\[10pt]  
& $\left. \left. + \frac{(J_{n+1}[k_c a]-J_{n-1}[k_c a])\,(Y_{n+1}[k_c r]-Y_{n-1}[k_c r]) }{(J_{n+1}[k_c a]-J_{n-1}[k_c a])Y_n[k_c a] -J_n[k_c a] (Y_{n+1}[k_c a]-Y_{n-1}[k_ca])} \right]  \right)\cos[n(\theta-\theta_0)]\ \tilde{f}(z,t) $ \\[10pt]

$A_z(r,\theta,z,t)$
& $=\; \phi_m \, \beta \left(   \frac{(1-\alpha) \, J_n(k_c r)}{J_n(k_c a)+(-1)^{m+1} \, J_n(k_c b)}+\frac{\alpha \, Y_n(k_c r)}{Y_n(k_c a)+(-1)^{m+1} \, Y_n(k_c b)}    \right)\sin[n(\theta-\theta_0)]\ f(z,t) $ \\[10pt]

$V(r,\theta,z,t)$
 & $=\; \phi_m \, \omega \left( \frac{(1-\alpha) \, J_n(k_c r)}{J_n(k_c a)+(-1)^{m+1} \, J_n(k_c b)}+\frac{\alpha \, Y_n(k_c r)}{Y_n(k_c a)+(-1)^{m+1} \, Y_n(k_c b)} \right)\sin[n(\theta-\theta_0)]\ f(z,t)$
\\[10pt] 

\midrule
\textbf{Gauge fixing} & $a_\pi  ,\ \ \tilde a_\pi   , \  \  c_\pi  ,\ \ \tilde c_\pi $ from Eqs. (\ref{eqTEGaugec1},\ref{eqTEGaugec2})\\
\textbf{Gauge invariance} & $b_\pi,\ \ \tilde b_\pi,\ \ d_\pi,\ \ \tilde d_\pi$ and $\alpha$   \\
\bottomrule
\end{tabular}
%} % end resizebox  
\label{tabTEpotcyl}
\end{table}
%%%%%%%%%%%

Consider first the $n>0$ situation. 
From the transverse symmetry of these modes discussed in
Subsection \ref{subsec:coax_charges_currentsTE}, we define this time:
\begin{eqnarray}
     \Delta V(\theta,z,t) & = & V(r=a,\theta,z,t)  +(-1)^{m+1} \,  V(r=b,\theta,z,t) , \label{eq:TEVcyl1} \\
     \Delta A_z(\theta,z,t) & = & A_z(r=a,\theta,z,t)  +(-1)^{m+1} \,  A_z(r=b,\theta,z,t) . \label{eq:TEVcyl2}
\end{eqnarray}
%%
%For TE waves in a two-conductor cylindrical geometry, the dispersion enforces
%$k_c^2 = k^2-\beta^2 > 0$. Hence the transverse primitives obey the Helmholtz
%equation on $(r,\theta)$:
%\[
%\frac{1}{r}\,\partial_r\!\big(r\,\partial_r p\big)
%+ \frac{1}{r^2}\,\partial_\theta^2 p
%+ k_c^2\,p = 0, 
%\qquad
%\frac{1}{r}\,\partial_r\!\big(r\,\partial_r \tilde p\big)
%+ \frac{1}{r^2}\,\partial_\theta^2 \tilde p
%+ k_c^2\,\tilde p = 0 .
%\]
%In a coaxial guide (inner conductor at $r=a$, outer at $r=b$), the TE solution
%carries an azimuthal structure of integer order $m$; we use the real basis
%$\cos[m(\theta-\theta_0)]$ and $\sin[m(\theta-\theta_0)]$.
%Solving the equations above gives the harmonic (radial) primitives:
%\[
%p(r)=a_\pi\,J_m(k_c r)+b_\pi\,Y_m(k_c r), 
%\qquad
%\tilde p(r)=\tilde a_\pi\,J_m(k_c r)+\tilde b_\pi\,Y_m(k_c r),
%\]
%with $a_\pi,b_\pi,\tilde a_\pi,\tilde b_\pi$ four constants to be fixed by the
%electrode boundary data and by the gauge choice (see below). 
%
% fait dans TM. On se répète pas.
%%%
The same arguments as in the case of TM waves apply, with Devoret relations Eqs. (\ref{eq:phidot},\ref{eq:phiprime}) leading to the results given in 
Tab. \ref{tabTEpotcyl}. The role of the $\cos$ and $\sin$ terms in Eqs. (\ref{petgauge1},\ref{petgauge2}) are simply inverted. 
Again, we have a degeneracy linked to the high symmetry of the problem represented by the real number $\alpha$. This time:
\begin{eqnarray}
    a_\pi & = & - \frac{Y_n[k_c a]+ (-1)^{m+1} \,Y_n[k_c b] }{J_n[k_c a]+ (-1)^{m+1} \,J_n[k_c b]} \ b_\pi ,  \label{eqTEGaugec1} \\
    c_\pi & = & - \frac{Y_n[k_c a]+ (-1)^{m+1} \,Y_n[k_c b] }{J_n[k_c a]+ (-1)^{m+1} \,J_n[k_c b]} \ d_\pi ,  \label{eqTEGaugec2}
\end{eqnarray}
leads to the remaining gauge invariance.
%%%%
But as opposed to the TM situation described before, it is possible for TE waves to impose a perfect symmetry/antisymmetry  $|V(r=a,\theta,z,t)| = |V(r=b,\theta,z,t)|$ and $|A_z(r=a,\theta,z,t)| = |A_z(r=b,\theta,z,t)|$ of the electrodes' potentials.
This is due to the different boundary conditions applying for TE modes, and we have then:
\begin{equation}
\alpha = \frac{1}{2} \, \frac{\left(\, J_n[k_c a]- (-1)^{m+1} \,J_n[k_c b] \, \right) \left( \, Y_n[k_c b]+ (-1)^{m+1} \,Y_n[k_c a] \, \right) }{J_n(k_ca)Y_n(k_cb)-J_n(k_cb)Y_n(k_ca)} ,
\end{equation}
which indeed would diverge for TM waves.
%%%%%%
Note that imposing this condition {\em is not} required by our modeling, which is why we still keep $\alpha$ as a gauge undefined choice in Tab. \ref{tabTEpotcyl}. 
%%%
The gauge invariance is again larger than in the Cartesian case, see Ref. \cite{Delattre2024}.
\\

%%%%%%%%%%%%%%% example imperfect...
\begin{figure}[H]
\centering
\hspace*{-2.2cm}
\includegraphics[width=1.2\linewidth]{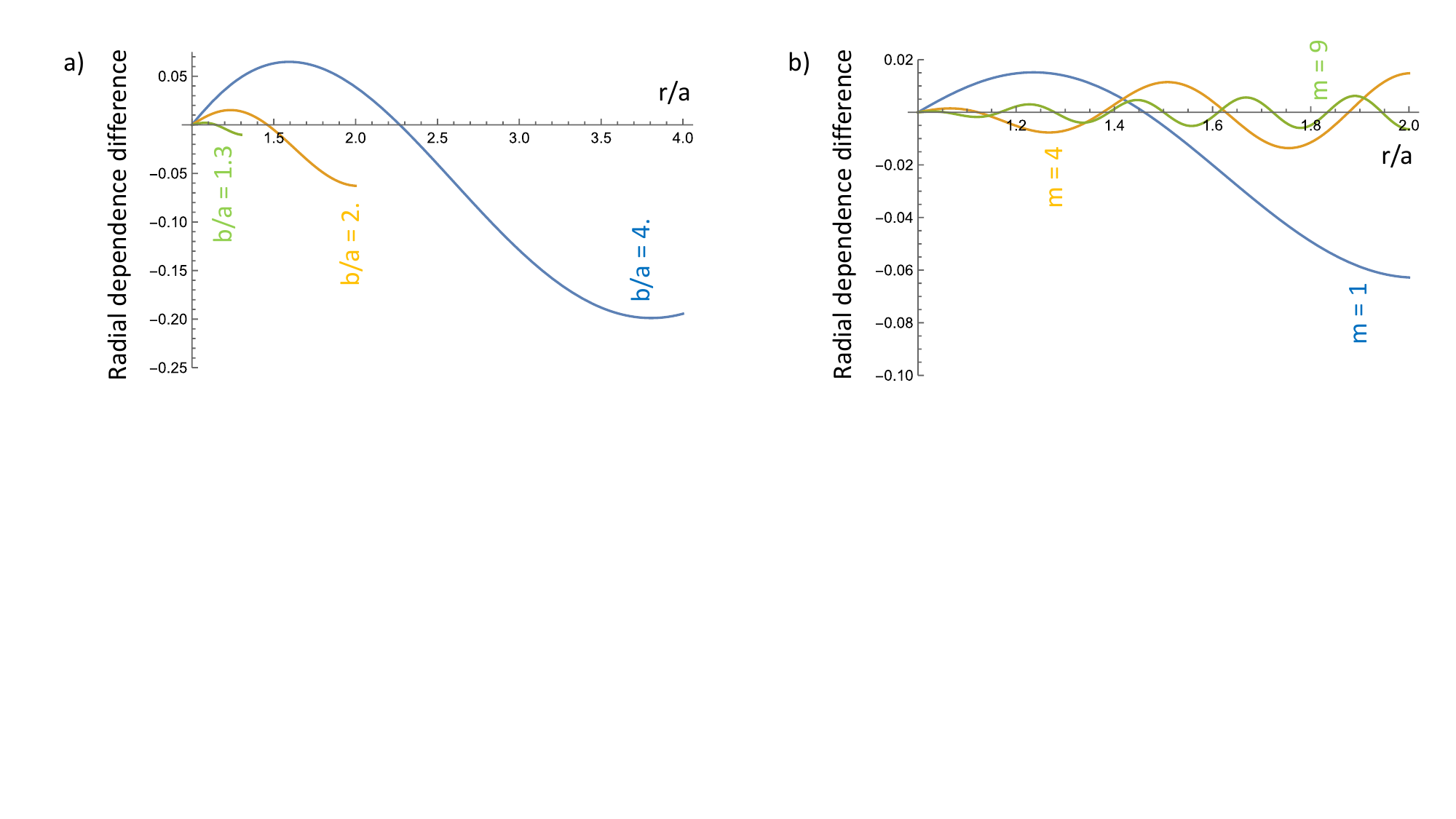} %{TM11_zpf_curve.png}
\vspace*{-6cm}
\caption{Difference between the gauge profile Eq. (\ref{eqgaugerTEcoax}) and the $g_{vir}(r)$ function (no units, to be compared to 1). 
 a) For mode $m=1$, different $b/a$ ratios. b) For $b/a=2.$, different mode numbers $m$.
 See text for details.
%
%%%% assez clair??????
}
\label{fig_3}
\end{figure}
%%%%%%

%%%%%%
% --- TE waves (two-conductor cylindrical geometry) ---
\begin{table}[H]
\centering
\caption{Gauge description for $\mathrm{TE}_{n=0,m}$ modes in a coaxial waveguide ($m>0$). % ($a<r<b$). 
%Gauge fixing and normalization identical to the Cartesian case. 
%The normalization constant $A_{n,m}$ is the same as in Table~\ref{tab:TE_mn_coax_profiles}, 
%and the dispersion relation is $J_n'(k_c a)Y_n'(k_c b)-Y_n'(k_c a)J_n'(k_c b)=0$.
%%%%%%%%%%%%%% c'est TOUT!!!!
}
%\label{tab:gauge_TE_mn_coax}
\setlength{\tabcolsep}{5pt}
\renewcommand{\arraystretch}{1.}
\resizebox{\textwidth}{!}{%
\begin{tabular}{@{}l l@{}}
\toprule
\textbf{Function} & \textbf{Expression} \\
\midrule
$A_r(r,\theta,z,t)$
& $= \phi_m \,k_c \frac{ \left( J_0[k_c a]\, Y_1[k_c r]-J_1[k_c r]\, Y_0[k_c a]\right) }{A_m' \, J_0(k_c a) } \tilde{f}(z,t)$  \\[10pt]

$A_\theta(r,\theta,z,t)$
& $=\phi_m \,\frac{-1}{h_{\rm eff}}\left( \frac{ 2 \left( J_1[k_c a]\, Y_1[k_c r]-J_1[k_c r]\, Y_1[k_c a]\right) }{A_m \, (k_c a) } \right) \tilde{f}(z,t)$ \\[10pt]

$A_z(r,\theta,z,t)$
& $= \phi_m \,
 \beta \left( \frac{J_0[k_c a]\, Y_0[k_c r]-J_0[k_c r]\, Y_0[k_c a]}{A_m' \, J_0(k_c a) } \right)
f(z,t)$ \\[10pt]

$V(r,\theta,z,t)$
& $=\phi_m \,
 \omega \left( \frac{J_0[k_c a]\, Y_0[k_c r]-J_0[k_c r]\, Y_0[k_c a]}{A_m' \, J_0(k_c a) } \right)
f(z,t)\,\,\,\,\,\,\,\,\,\,\,\,\,\,\,\,\,\, $
\\[10pt]  

\midrule
\textbf{Gauge fixing} & $a_\pi =0 ,\ \ \tilde a_\pi=0 , \  \   b_\pi=0,\ \ \tilde b_\pi=0 $ \ \ ($\Delta V, \Delta A_z$ approximate)   \\
\textbf{Gauge invariance} & None  \\
\bottomrule
\end{tabular}
} % end resizebox  
\label{tabTEpotcylSPECn}
\end{table} 

For the $n=0$ waves, we must consider  virtual electrodes crossing the guide with (arbitrary) azimuthal angle $\theta$.
The most natural way of extending the formalism of Ref. \cite{Delattre2024} would be to pose, for $a<r<b$:
\begin{eqnarray}
     \Delta V(r,z,t) & = & V(r,\theta,z,t)  +    V(r,\theta+\pi,z,t) , \label{eq:TEVcyl1} \\
     \Delta A_z(r,z,t) & = & A_z(r,\theta,z,t)  +   A_z(r,\theta+\pi,z,t) . \label{eq:TEVcyl2}
\end{eqnarray}
The potentials of top and bottom parts of the virtual electrode are added up, due to the symmetry.
In the gauge expressions Eqs. (\ref{petgauge1},\ref{petgauge2}), only the $\cos$ terms are relevant (with four gauge coefficients to be fixed).
But there is an issue here. The radial dependence created by the gauge choice writes:
\begin{equation}
A_z, V \propto \frac{J_0[k_c a]\, Y_0[k_c r]-J_0[k_c r]\, Y_0[k_c a]}{A_m' \, J_0(k_c a) } , \label{eqgaugerTEcoax}
\end{equation}
which is obtained by enforcing the $r=a$ boundary (at zero), and normalizing the amplitude following the same procedure as for the $E_\theta$ component (Subection \ref{sec:coax_TE}), defining thus numerically $A_m'$.
This function {\em is clearly different} from the $g_{vir}(r)$ one, Eq. (\ref{gvirTEcoax}).
%%%% CORR REF 2 3
This specificity, as compared to the Cartesian situation, arises mathematically from the fact that the Bessel functions {\it do not} share the same properties as the cosine and sine ones.
%%%%
We plot the difference between these two in Fig. \ref{fig_3}, for various ratios $b/a$ and mode numbers $m$.  
%%%%%%%%%%%
Only when $b/a \rightarrow 1$ or $m \gg 1$ do the two expressions match; but the difference is rather small for reasonable practical parameters (typically $< 15~\%$), and one can argue that:
\begin{eqnarray}
  \frac{\partial \varphi (r,z,t)}{\partial t}   & \approx &  \Delta V(r,z,t) , \label{eq:devorR1} \\
   \frac{\partial \varphi(r,z,t)}{\partial z}     & \approx & - \Delta A_z(r,z,t) ,  \label{eq:devorR2}
\end{eqnarray}
can be experimentally considered as good enough, replacing the previous (and exact) Devoret expressions  Eqs. (\ref{eq:phidot},\ref{eq:phiprime}).
The obtained gauge is given in Tab. \ref{tabTEpotcylSPECn}, reminding as well that Eqs. (\ref{eq:devorR1},\ref{eq:devorR2}) are only approximate.
This point shall be explicitly discussed in Subection \ref{gaugePLUS}. All coefficients are fixed, similarly to Ref. \cite{Delattre2024}.
%%%%%%%%%%%%%%%%%%%%%%%%%%%%%%%%%%%%%

\subsection{Hollow cylindrical waveguide}

The symmetries of the hollow cylinder are the same as for the coaxial line; so the same generic gauge expressions Eqs. (\ref{petgauge1},\ref{petgauge2}) for $p(r,\theta), \tilde{p}(r,\theta)$ apply.
%%%%%%% MAIS: il y a une subtilité, que tu n'as PAS vu... les potentiels DOIVENT ETRE EN Y_n, car les J_n sont systématiquement nuls sur les électrodes!!! C'est la condition au bord, qui du coup te donne des DElta V tous nuls!!!!
%%%%% On est OBLIGE d'accepter un terme divergent en r==0... subtil... J'explique donc....
Furthermore, there is no {\em a priori} reason to discard the Neumann functions $Y_n$, as opposed to what is discussed for the real fields $\mathbf{E},\mathbf{B}$.
Indeed, electric and magnetic fields must be finite everywhere, because they can be measured and directly  define (after volume integration) the constants of motion. On the other hand, the potentials in our gauge fixing view are relevant {\em only on the electrodes}.
% In this sense, a $r=0$ divergence in $\mathbf{A},V$ is nonphysical and tolerable.
%%%%%%%% OK?

\subsubsection{TM modes}

%For TM waves, $k^2-\beta^2=k_c^2$ with the transverse wavenumber fixed by the boundary
%condition. The gauge functions solving the transverse Helmholtz equations and sharing the TM symmetries write:
%\begin{equation}
%p(r,\theta)=a_\pi\,J_m(k_c r)\cos\!\big[m(\theta-\theta_0)\big]
%\;+\;b_\pi\,J_m(k_c r)\sin\!\big[m(\theta-\theta_0)\big],
%\label{eq:cylTMp}
%\end{equation}
%and the same expression with tilded coefficients for $\tilde p$. To define the generalized-flux relations exactly as in Eqs.~(87)–(88),
%
% on se répète déjà pas mal... quand on peut il faut faire sharp...
% du coup, j'ai mis un texte en facteur pour TM et TE ci-dessus.
Following the procedure presented in Ref. \cite{Delattre2024}, and taking into account the symmetries of the TM waves in a hollow cylinder (Subection \ref{subsec:hollow_charges_currents_enTM}), we define:
\begin{eqnarray}
     \Delta V(\theta,z,t) & = & V(r=a,\theta,z,t)  +  (-1)^{n }  \,V(r=a,\theta+\pi,z,t) , \label{eq:TMholl1} \\
     \Delta A_z(\theta,z,t) & = & A_z(r=a,\theta,z,t)  +   (-1)^{n } \, A_z(r=a,\theta+\pi,z,t) , \label{eq:TMholl2}
\end{eqnarray}
for any azimuthal position $\theta$ on the metallic guide.
%we use the
%pair of \emph{virtual angular electrodes} along the diametral planes
%$\theta=\theta_0$ and $\theta=\theta_0+\pi$.
%We then generalize the electrode differences by
%\begin{align}
%\Delta V^{[\theta]}(r,z,t)
%&= V(r,\theta_0,z,t)\;+\;(-1)^m\,V(r,\theta_0+\pi,z,t),
%\label{eq:cylTMdV}
%\\
%\Delta A^{[\theta]}(r,z,t)
%&= A_z(r,\theta_0,z,t)\;+\;(-1)^m\,A_z(r,\theta_0+\pi,z,t),
%\label{eq:cylTMdA}
%\end{align}
%so that for \textit{symmetric} angular profiles (even $m$) one \emph{adds} the two sides,
%whereas for \textit{antisymmetric} ones (odd $m$) one takes the \emph{difference}.
%
%  Mais enfin, NON! c'est l'onde TE n==0 qui est bizarre et a besoin d'electrode virtuelles!!! Ici, on a des électrodes réelles.... 
%%
%%%
%%
%With
%this convention the canonical relations keep the same form as in the cartesian paper:
%\begin{equation}
%\partial_t \phi^{[\theta]}(r,z,t)=\Delta V^{[\theta]}(r,z,t),\qquad
%\partial_z \phi^{[\theta]}(r,z,t)=-\,\Delta A^{[\theta]}(r,z,t),
%\end{equation}
%with a (radius-dependent) scalar profile:
%\begin{equation}
%    \phi^{[\theta]}(r,z,t)=\phi_m\,g_\phi(r)\,\tilde f(z,t),\qquad
%g_\phi(r)=2\,J_m(k_c r).
%\end{equation}
%
% putain, c'est franchement nimp'...
%%%%
%%%%
These "potential differences" verify Eqs. (\ref{eq:phidot},\ref{eq:phiprime}), and the corresponding gauge is given in Tab. \ref{tab:cyl_TMmn_gauge}.
%%%% expliquer ce qui est spécifique!!
As for the coaxial line, when $n=0$ only four gauge coefficients are relevant instead of eight.
%%%%%
It turns out that because of the boundary condition, the  functions $J_n[k_c r]$ present in the gauge expressions $p,\tilde{p}$ vanish on the electrode: the $a_\pi, \tilde{a}_\pi, c_\pi, \tilde{c}_\pi$ are therefore undefined, which makes the gauge invariance larger than for the rectangular guide \cite{Delattre2024} (which does even not support a $n=0$ mode, Subsection \ref{sec:TM_hollow}).
%%%
Likewise, the $Y_n[k_c r]$ gauge functions are mandatory for the Devoret rule to apply: all potential components diverge at $r=0$, which is acceptable since it is not part of an electrode.
%%%%
This is again a specificity of Bessel functions as compared to the regular sine and cosine ones, which do not lead to divergences in the Cartesian case \cite{Delattre2024}.
%%%%ADD REF 2 3

\begin{table}[H]
\centering
\caption{Gauge description for TM$_{n,m}$ waves in a cylindrical hollow waveguide, $n\geq 0, m>0$.}
\label{tab:cyl_TMmn_gauge}
\setlength{\tabcolsep}{5pt}
\renewcommand{\arraystretch}{1.1}
\resizebox{\textwidth}{!}{%
\begin{tabular}{@{}l l@{}}
\toprule
\textbf{Function} & \textbf{Expression} \\
\midrule

$A_r(r,\theta,z,t)$
& $=\;\phi_m \frac{k_c}{2} \left(  +\frac{1}{k_c \,a} \left[ \frac{ (J_{n+1}[k_c r]-J_{n-1}[k_c r])(J_{n+1}[k_c a]-J_{n-1}[k_c a]) }{J_{n-1}(k_c a)^2} \right] \right. $  \\[10pt]
& $  \left. - \frac{ Y_{n-1}[k_c r]- Y_{n+1}[k_c r]}{2\, Y_n[k_c a] }  \right)\cos[n(\theta-\theta_0)]\ \tilde{f}(z,t) $  \\[10pt]

$A_\theta(r,\theta,z,t)$
& $=\;\phi_m \frac{n}{r} \left(  +\frac{1}{k_c \,a} \left[ \frac{\, J_n[k_c r](J_{n+1}[k_c a]-J_{n-1}[k_c a]) }{J_{n-1}(k_c a)^2} \right]   + \frac{ Y_n[k_c r]}{2\, Y_n[k_c a] } \right)\sin[n(\theta-\theta_0)]\ \tilde{f}(z,t) $ \\[10pt]

$A_z(r,\theta,z,t)$
& $=\; \phi_m \, \beta \left( -\frac{k_c}{a\, \beta^2} \left[ \frac{\, J_n[k_c r](J_{n+1}[k_c a]-J_{n-1}[k_c a]) }{J_{n-1}(k_c a)^2} \right]  + \frac{ Y_n[k_c r]}{2\, Y_n[k_c a] } \, \right)\cos[n(\theta-\theta_0)]\ f(z,t) $ \\[10pt]

$V(r,\theta,z,t)$
 & $=\; \phi_m \, \omega \left( \frac{ Y_n[k_c r]}{2\, Y_n[k_c a] } \right)\cos[n(\theta-\theta_0)]\ f(z,t)$
\\[10pt] 

\midrule
\textbf{Gauge fixing} & $b_\pi=0,\ \tilde b_\pi=0, d_\pi=0,\ \tilde d_\pi=0$ \\
\textbf{Gauge invariance} & $a_\pi,\  \ \tilde a_\pi,\ \ c_\pi,\  \ \tilde c_\pi$ \\
\bottomrule
\end{tabular}
} % end resizebox  
%\label{tabTEpotcyl}
\end{table}

\subsubsection{TE modes}

Consider first $n>0$. In this case, real electrodes are involved and the symmetries of the modes lead for the "potential differences" $\Delta V, \Delta A$ to the same expressions as for TM waves, 
Eqs. (\ref{eq:TMholl1},\ref{eq:TMholl2}).
The corresponding gauge choice is given in
Tab. \ref{tab:cyl_TEmn_gauge}, matching the Devoret relations Eqs. (\ref{eq:phidot},\ref{eq:phiprime}). 
As opposed to the previous case, both $J_n[k_c r]$ and $Y_n[k_c r]$
functions within the gauge expressions $p, \tilde{p}$ are relevant: we recover the degeneracy materialized by the angle $\alpha$ already discussed in the case of a coaxial line, see Subsection \ref{TMPotcoax}.
%%%
The gauge invariance condition writes this time:
\begin{eqnarray}
    a_\pi & = & - \frac{ Y_n[k_c a] }{  J_n[k_c a]} \ b_\pi ,  \label{eqTEGaugeholl1} \\
    c_\pi & = & - \frac{ Y_n[k_c a] }{ J_n[k_c a]} \ d_\pi ,  \label{eqTEGaugeholl2}
\end{eqnarray} 
and similarly with tilded parameters. 
%%%%%%%
One can arbitrarily chose $\alpha=0$ in order to remove the $Y_n$ functions from the potentials, but this is not required by the modeling ($r=0$ is not part of an electrode). We find out that the gauge invariance is again larger than for its Cartesian (rectangular guide) counterpart \cite{Delattre2024}. \\

\begin{table}[H]
\centering
\caption{Gauge description for TE$_{n,m}$ waves in a %circular 
hollow cylindrical waveguide. $n > 0, m>0$.}
\label{tab:cyl_TEmn_gauge}
\setlength{\tabcolsep}{5pt}
\renewcommand{\arraystretch}{1.1}
\resizebox{\textwidth}{!}{%
\begin{tabular}{@{}l l@{}}
\toprule
\textbf{Function} & \textbf{Expression} \\
\midrule

$A_r(r,\theta,z,t)$
& $=\;\phi_m \,\frac{k_c}{2} \left( \left[  \frac{(1-\alpha) \, [J_{n+1}(k_c r)-J_{n-1}(k_c r)]}{2\,J_n(k_c a) }+\frac{\alpha \, [Y_{n+1}(k_c r)-Y_{n-1}(k_c r)]}{2\, Y_n(k_c a) } \right] \right. $   \\[10pt]
& $  \left. + \frac{ 4n^2  }{ r \,k_c^2 a} 
\left[ \frac{J_{n}[k_c a]\,J_{n }[k_c r]}{(k_c a)\,J_{n}[k_c a]^2-2n\,J_{n}[k_c a]\,J_{n+1}[k_c a]+(k_c a)\,J_{n+1}[k_c a]^2  }  \right]  \right)\sin[n(\theta-\theta_0)]\ \tilde{f}(z,t) $  \\[10pt]

$A_\theta(r,\theta,z,t)$
& $=\;\phi_m  \left( - \frac{n}{r} \left[ \frac{(1-\alpha) \, J_n(k_c r)}{2 \,J_n(k_c a) }+\frac{\alpha \, Y_n(k_c r)}{2\, Y_n(k_c a) } \right] \right. $  \\[10pt]
& $\left.   -\frac{n}{a} \left[\frac{J_{n}[k_c a]\,(J_{n+1}[k_c r]-J_{n-1}[k_c r]\,) }{(k_c a)\,J_{n}[k_c a]^2-2n\,J_{n}[k_c a]\,J_{n+1}[k_c a]+(k_c a)\,J_{n+1}[k_c a]^2  } \right]  \right)\cos[n(\theta-\theta_0)]\ \tilde{f}(z,t) $ \\[10pt]

$A_z(r,\theta,z,t)$
& $=\; \phi_m \, \beta \left(  \frac{(1-\alpha) \, J_n(k_c r) }{2\,J_n(k_c a) }+\frac{\alpha \, Y_n(k_c r) }{2\,  Y_n(k_c a)}   \right)\sin[n(\theta-\theta_0)]\ f(z,t) $ \\[10pt]

$V(r,\theta,z,t)$
 & $=\; \phi_m \, \omega \left( \frac{(1-\alpha) \, J_n(k_c r) }{2\,J_n(k_c a) }+\frac{\alpha \, Y_n(k_c r) }{2\,  Y_n(k_c a)} \right)\sin[n(\theta-\theta_0)]\ f(z,t)$
\\[10pt] 

\midrule
\textbf{Gauge fixing} & $ a_\pi  ,\ \ \tilde a_\pi   , \  \  c_\pi  ,\ \ \tilde c_\pi $  from Eqs. (\ref{eqTEGaugeholl1},\ref{eqTEGaugeholl2}) \\
\textbf{Gauge invariance} & $ b_\pi,\ \ \tilde b_\pi, \ \  d_\pi,\ \ \tilde d_\pi$ and $\alpha$  \\
\bottomrule
\end{tabular}
} % end resizebox  
\end{table}

As for the coaxial guide, the TE $n=0$ modes deal with virtual electrodes. 
%%%% CORR REFS 2 3
In the former case, we could define approximate 
Eqs. (\ref{eq:devorR1},\ref{eq:devorR2})
restoring somehow our intuitive expectations for  "Devoret-like" relationships.
For the hollow cylinder, {\em this is definitely impossible}.
In our gauge choice, we must discard the $Y_0$ function because the $r=0$ position is part of our virtual electrode. We are left with a single gauge coefficient $\tilde{a}_\pi$ which cannot be fixed in a convenient way if we follow strictly the previous method, as can be seen in Tab. \ref{tabTEpothollowSPECn}.
We must conclude that there is something very peculiar about these TE$_{0,m}$ modes, which have no equivalent in the Cartesian geometry (see Subsection \ref{sec:TE_hollow}).
This deserves to be analyzed, and it is the purpose of the following Subsection.

%We use the same "potential difference" definition, for $0<r<a$:
%\begin{eqnarray}
 %    \Delta V(r,z,t) & = & V(r,\theta,z,t)  +    V(r,\theta+\pi,z,t) , \label{eq:TEVholl1} \\
 %    \Delta A(r,z,t) & = & A_z(r,\theta,z,t)  +   A_z(r,\theta+\pi,z,t) . \label{eq:TEVholl2}
%\end{eqnarray}
%These expressions verify then Devoret relationships Eqs. (\ref{eq:devorR1},\ref{eq:devorR2}), and lead to the gauge choice given in Tab. \ref{tabTEpothollowSPECn}.
%We now apply the same procedure to TE$_{m,n}$ waves. Here $k^2-\beta^2=k_c^2$
%with the transverse eigenvalue fixed by the TE boundary condition
%$J_m'(k_cR)=0$. We keep the cylindrical gauge profiles identical to the TM case :
%\begin{equation}
%    p(r,\theta)=a_\pi J_m(k_c r)\cos[m(\theta-\theta_0)]
%\;+\;b_\pi J_m(k_c r)\sin[m(\theta-\theta_0)],
%\end{equation}
%
%(and the same expression with tilded coefficients), but \emph{lateral} (radial)
%virtual electrodes must be used to define the canonical differences:
%\(
%\Delta V^{[r]}(\theta_0)=V(0,\theta_0)-V(R,\theta_0),
%\;
%\Delta A^{[r]}(\theta_0)=A_z(0,\theta_0)-A_z(R,\theta_0).
%\)
%With this choice, the canonical relations keep their form,
%$\partial_t\phi^{[r]}=\Delta V^{[r]}$ and $\partial_z\phi^{[r]}=-\Delta A^{[r]}$,
%while TE symmetry ($E_z\equiv 0$, $B_z\neq 0$) is enforced by a constrained pair
%$(A_z,V)$ such that $-\partial_z V-\partial_t A_z\equiv 0$.
%
% Bon... no more comment, j'ai réécrit....

\begin{table}[H]
\centering
\caption{Gauge description for $\mathrm{TE}_{n=0,m}$ modes in a hollow cylindrical waveguide ($m>0$). % ($a<r<b$). 
%Gauge fixing and normalization identical to the Cartesian case. 
%The normalization constant $A_{n,m}$ is the same as in Table~\ref{tab:TE_mn_coax_profiles}, 
%and the dispersion relation is $J_n'(k_c a)Y_n'(k_c b)-Y_n'(k_c a)J_n'(k_c b)=0$.
%%%%%%%%%%%%%% c'est TOUT!!!!
}
%\label{tab:gauge_TE_mn_coax}
\setlength{\tabcolsep}{5pt}
\renewcommand{\arraystretch}{1.1}
\resizebox{\textwidth}{!}{%
\begin{tabular}{@{}l l@{}}
\toprule
\textbf{Function} & \textbf{Expression} \\
\midrule
$A_r(r,\theta,z,t)$
& $= -\tilde{a}_\pi \,k_c  \, J_1[k_c r] \, \tilde{f}(z,t)$  \\[10pt]

$A_\theta(r,\theta,z,t)$
& $=\phi_m \,\frac{-1}{h_{\rm eff}}\left( \frac{ 2   J_1[k_c r]  }{A_m   } \right) \tilde{f}(z,t)$ \\[10pt]

$A_z(r,\theta,z,t)$
& $=  -\tilde{a}_\pi \, \beta \, J_0[k_c r] \, 
f(z,t)$ \\[10pt]

$V(r,\theta,z,t)$
& $=-\tilde{a}_\pi \, \omega \, J_0[k_c r] \, 
f(z,t)\,\,\,\,\,\,\,\,\,\,\,\,\,\,\,\,\,\, $
\\[10pt]  

\midrule
\textbf{Gauge fixing} & $  a_\pi=0  , \ \  b_\pi=0,\ \ \tilde b_\pi=0 $  \\
\textbf{Gauge invariance} & $  \tilde{a}_\pi $\ \ ($\Delta V, \Delta A_z$ incompatible, see text) \  \  \  \   \  \  \  \ \  \  \  \ \  \  \  \  \  \  \  \ \  \  \  \  \  \  \  \ \  \  \  \ \\
\bottomrule
\end{tabular}
} % end resizebox  
\label{tabTEpothollowSPECn}
\end{table}

\subsection{Gauge fixing strategies: Cartesian vs cylindrical geometries}
\label{gaugePLUS}

Section \ref{sec:charges_currents_constants_cyl}
presented the procedure recasting all physical quantities in terms of the variable $\varphi$. 
Choosing a reference electrode when the transmission line has non-equivalent surfaces (coaxial case), a lengthscale $h_{\rm eff}$ is introduced. The formalization is then {\em unique}.
However, from the preceding Section it appears that linking this $\varphi$ to the potentials $\mathbf A, V$ {\em is not straightforward}. 
The strategy of Ref. \cite{Delattre2024} 
had been constructed from the generalized flux concept first proposed by Devoret in Ref. \cite{DevoretQED} for quantum circuits:
\begin{equation}
\varphi(t) = \int_{-\infty}^{\,t} \! \Delta V(t')\, dt' , \label{DevoretOrigins}
\end{equation}
where $\Delta V$ is the voltage drop between two nodes.
The fluxes $\varphi$ and their conjugate variables, the branch charges $Q$, %= \int_{-\infty}^{\,t} \! I(t')\, dt' $,
 are all we need to describe the quantum circuit. Surplus variables are eliminated using Kirchhoff's laws, and the remaining $\varphi,Q$ pairs enable to write an Hamiltonian for the whole circuit.   These constitute the system's degrees of freedom which must be quantized, following conventional recipes \cite{DevoretQED}.

The basic assumption of circuit theory is that each part defined by two nodes can be considered as a {\em lumped element} (it is much smaller than the relevant signal's wavelength). This enables to write $\Delta V = \int \mathbf{E}.d\mathbf{s}$ across it, independently of the integration path
\cite{DevoretQED,PozarMW}.
This is extremely convenient from an engineering point of view: 
the quasi-static type solutions of Maxwell's equations 
lead to the well-known circuit analysis toolbox \cite{PozarMW}. The voltage $\Delta V$ is a well-defined measurable quantity, just like the current $I$ flowing through the branch; and they characterize perfectly its state \cite{DevoretQED}.

By definition, a transmission line {\em is not} a lumped element. But if one considers only TEM transport confined within two conductors (say {\em elec. 1} and {\em elec. 2}), the situation remains fairly simple:  
the transverse fields verify Laplace's equation, which means that exactly like in the static case we can derive them form a scalar potential $\Phi$.
The quantity $\Delta \Phi = \Phi(\mbox{elec. 1})-\Phi(\mbox{elec. 2}) = \int \mathbf{E}.d\mathbf{s}$ is 
again independent of the path: this is what engineers {\em define} as being the voltage difference \cite{PozarMW}.
Very conveniently, we end up with %the same formalism than in 
a formalism compatible with circuit theory for $\Delta \Phi$, which can be easily quantized, and travels in the guide according to a d'Alembert equation  \cite{Clerk2010,PozarMW}.

But strictly speaking, this quantity $\Delta \Phi$ can be identified to $\Delta V$ {\em only} when imposing a longitudinal gauge.
%%% CORR REF 3 and 2
To our knowledge, this fact had never been pointed out before, and the quatization of the field is simply performed in an "ad hoc fashion" without further justification \cite{Clerk2010}. The aim of the paper is to present the properties and consequences of a gauge approach to the field quantization. 
%%%%
Eq. (\ref{DevoretOrigins}) can then be recast in:
\begin{equation}
\begin{cases}
&  \partial \varphi/\partial t   =  \Delta V , \\
&  \partial \varphi/\partial z   =  -\Delta A_z , \label{rule1}
\end{cases} 
\end{equation}
with the second equivalent equation being a consequence of the Lorenz gauge, Eq. (\ref{eq:lorenz}).
$\Delta A_z$ is defined as $A_z(\mbox{elec. 1})-A_z(\mbox{elec. 2})$, exactly like we have $\Delta V = V(\mbox{elec. 1})-V(\mbox{elec. 2})$.
%%%%%%% on va faire SHARP...
But what about non-TEM waves? What has been demonstrated in Ref. \cite{Delattre2024} is that for a Cartesian geometry, one can {\em always} find a gauge that fulfills these relationships, for each TEM, TM and TE traveling solution. The "potential differences" must be generalized, taking into account the transverse symmetry of the $\mathbf E, \mathbf B$ fields:
\begin{eqnarray}
    \Delta V (\eta,z,t) & = & V(\{ \mathbf{\eta} \} \in  \mbox{elec. 1},z,t) + \sigma_z  \,   V(\{ \mathbf{\eta} \} \in  \mbox{elec. 2},z,t) ,   \\
  \Delta A_z (\eta,z,t) & = & A_z(\{ \mathbf{\eta} \} \in  \mbox{elec. 1},z,t)  + \sigma_z  \,   A_z(\{ \mathbf{\eta} \} \in  \mbox{elec. 2},z,t) , 
\end{eqnarray}
with $\sigma_z  \in\{\pm1\}$ a "parity coefficient" \cite{Delattre2024}. 
$\mathbf \eta$ represents the transverse coordinate running along an electrode (and symmetrically along the other one), parametrized from $\{ x ,y  \}$ for Cartesian.
Therefore in general, the voltage drop $\Delta V$ depends on $\mathbf \eta$, which was not the case for TEM waves.
%%%
This approach has been our starting point in the preceeding Sections.
%%%

It turns out that another gauge equation can be matched, for  {\em any type} of traveling solution in the Cartesian waveguides:
\begin{equation}
\frac{1+ \mathcal{K}}{h_{\rm eff}} \, \varphi = - \Delta A_{\mathbf{n}}  ,  \label{rule2}
\end{equation}
with $\mathcal{K}$ a coefficient related to the electrodes' geometry. We obtain $\mathcal{K}=1$
for both parallel plates and rectangular guides. The quantity $\Delta A_{\mathbf{n}}$ is then defined as:
\begin{eqnarray}
  \Delta A_{\mathbf{n}} (\eta,z,t) & = & \mathbf{n}. \mathbf{A}(\{ \mathbf{\eta} \} \in  \mbox{elec. 1},z,t)  + \sigma_{\mathbf{n}}  \, \, \mathbf{n}. \mathbf{A}(\{ \mathbf{\eta} \} \in  \mbox{elec. 2},z,t) , 
\end{eqnarray}
with $\mathbf{n}$ the normal to the electrode's surface, and $\sigma_{\mathbf{n}}  \in\{\pm1\}$. 
For this quantity, the symmetry coefficient $\sigma_{\mathbf{n}}$ is actually the same as for the charges and currents. 
This "transverse" gauge feature had not been pointed out in Ref. \cite{Delattre2024}.\\

Remarkably, each TEM, TM and TE mode of  
the cylindrical geometry supports as well  
  a gauge   that verifies the rule  Eq. (\ref{rule2}). 
%%%%%%%%%%%%%%%%%%
For the fully symmetric hollow pipe, we again have  $\mathcal{K}=1$ (for all wave types, including the TE $n=0$ one). In the case of the coaxial guide, we obtain $0<\mathcal{K}<1$:
\begin{eqnarray}
\mbox{TEM:}\,\,\,\,\, \mathcal{K} & = & \frac{a}{b}  , \\
\mbox{TM:}\,\,\,\,\,\mathcal{K} & = & \frac{a}{b} \, \frac{\pi}{2} \, \Big| \, Y_n(k_c a) \, [\,(k_c b) J_{n-1}(k_c b) -n J_n(k_c b)]   \nonumber \\  
&&   \,\,\,\,\,\,\,\,\,  - J_n(k_c a) \, [\,(k_c b)Y_{n-1}(k_c b)- n Y_n(k_c b) ] \, \Big|, \\
\mbox{TE ($n \neq 0$):}\,\,\,\,\, \mathcal{K} & = & \frac{a}{b} \, \frac{\pi}{2} \, \Big| \, J_n(k_c b) \, [\,n Y_n(k_c a) -(k_c a)Y_{n-1}(k_c a)]   \nonumber \\  
&&   \,\,\,\,\,\,\,\,\,  - Y_n(k_c b) \, [\,n J_n(k_c a) -(k_c a)J_{n-1}(k_c a)] \, \Big| .
\end{eqnarray}
For the TE $n = 0$ coaxial waves, we get $\mathcal{K}=1$; this intuitively matches the fact that the virtual electrodes are perfectly symmetric. Taking the limit $b/a \rightarrow 1$, we recover the parallel plate results with $\mathcal{K} \rightarrow 1$ in the above equations.

\begin{table}%[H]
\centering
\caption{Definition of the "potential differences" for each configuration encountered in a cylindrical geometry. The gauge property is reminded (last column), and $\sigma_{\mathbf{n}}$ is also the symmetry coefficient of charges/currents, see text for details.
}
\label{tab:gauge_plus}
\begin{tabular}{@{}c c c@{}}
\toprule
\textbf{Wave type} & \textbf{ "Potential differences" } & \textbf{ Symmetry properties } \\
\midrule
\midrule
$\begin{aligned}
\mbox{\textbf{TEM}} \\ 
\mbox{coaxial} \end{aligned}$ &
$\begin{aligned}
\Delta V(z,t) &=   V(r=a,z,t)  +\sigma_z \,  V(r=b,z,t)  \\
\Delta A_z(z,t) &= A_z(r=a,z,t) +\sigma_z \, A_z(r=b,z,t) \\
\Delta A_{\mathbf{n}}(z,t) &= A_r(r=a,z,t) -\sigma_{\mathbf{n}} \, A_r(r=b,z,t) \\
\end{aligned}$ &
$\begin{aligned}
& \!\!\!\!\!\!\! \mbox{Transform  trans./long.} \\
\sigma_z &=-1 \\
\sigma_{\mathbf{n}}  &=-1
\end{aligned}$\\

\midrule
$\begin{aligned}
\mbox{\textbf{TM$_{n,m}$}} \\
\mbox{coaxial}  \end{aligned}$ &
$\begin{aligned}
\Delta V(\theta,z,t) &= V(r=a,\theta,z,t) +\sigma_z \, \,  V(r=b,\theta,z,t) \\
\Delta A_z(\theta,z,t) &= A_z(r=a,\theta,z,t)  +\sigma_z \,  A_z(r=b,\theta,z,t)  \\
\Delta A_{\mathbf{n}}(\theta,z,t) &= A_r(r=a,\theta,z,t) -\sigma_{\mathbf{n}} \, A_r(r=b,\theta,z,t)\\
\end{aligned}$ &
$\begin{aligned}
& \!\!\!\!\!\!\! \mbox{Transform  trans./long.} \\
\sigma_z & = +(-1)^m \\
\sigma_{\mathbf{n}}  &=-(-1)^m
\end{aligned}$\\

\midrule
$\begin{aligned}
\mbox{\textbf{TE$_{n\neq0,m}$}} \\ 
\mbox{coaxial} \end{aligned}$ &
$\begin{aligned}
\Delta V(\theta,z,t) &= V(r=a,\theta,z,t) +\sigma_z \, \,  V(r=b,\theta,z,t) \\
\Delta A_z(\theta,z,t) &= A_z(r=a,\theta,z,t)  +\sigma_z \,  A_z(r=b,\theta,z,t)  \\
\Delta A_{\mathbf{n}}(\theta,z,t) &= A_r(r=a,\theta,z,t) -\sigma_{\mathbf{n}} \, A_r(r=b,\theta,z,t) \\
\end{aligned}$ &
$\begin{aligned}
& \!\!\!\!\!\!\! \mbox{Transform  trans./long.} \\
\sigma_z & = +(-1)^{m+1} \\
\sigma_{\mathbf{n}}  &= -(-1)^{m+1}
\end{aligned}$\\

\midrule
$\begin{aligned}
\mbox{\textbf{TE$_{n=0,m}$}}\\
 \mbox{coaxial} \end{aligned}$ &
$\begin{aligned}
\Delta V_{\rm eff}(r,z,t) &= \phi_m \, \omega \, g_{vir}(r) \, f(z,t) \\
\Delta A_{z\, \rm eff}(r,z,t) &= \phi_m \, \beta \, g_{vir}(r) \, f(z,t) \\
\Delta A_{\mathbf{n}}(r,z,t) &=  A_\theta(r,\theta,z,t) -\sigma_{\mathbf{n}} \, A_\theta(r,\theta+\pi,z,t) \\
\end{aligned}$ &
$\begin{aligned}
& \!\!\!\!\!\!\! \mbox{Only transverse} \\
\sigma_z & = +1 \\
\sigma_{\mathbf{n}}  &=-1
\end{aligned}$\\

\midrule
\midrule
$\begin{aligned}
\mbox{\textbf{TM$_{n,m}$}} \\
\mbox{cylinder} \end{aligned}$ &
$\begin{aligned}
\Delta V(\theta,z,t) &= V(r=a,\theta,z,t)  +  \sigma_z \, V(r=a,\theta+\pi,z,t) \\
\Delta A_z(\theta,z,t) &= A_z(r=a,\theta,z,t)  +   \sigma_z  \, A_z(r=a,\theta+\pi,z,t) \\
\Delta A_{\mathbf{n}}(\theta,z,t) &= -A_r(r=a,\theta,z,t) -\sigma_{\mathbf{n}} \, A_r(r=a,\theta+\pi,z,t) \\
\end{aligned}$ &
$\begin{aligned}
& \!\!\!\!\!\!\! \mbox{Transform trans./long.} \\
\sigma_z & = +(-1)^{n} \\
\sigma_{\mathbf{n}}  &=-(-1)^{n+1}
\end{aligned}$\\

\midrule
$\begin{aligned}
\mbox{\textbf{TE$_{n\neq0,m}$}}\\
\mbox{cylinder} \end{aligned}$ &
$\begin{aligned}
\Delta V(\theta,z,t) &= V(r=a,\theta,z,t)  +  \sigma_z \, V(r=a,\theta+\pi,z,t) \\
\Delta A_z(\theta,z,t) &= A_z(r=a,\theta,z,t)  +   \sigma_z  \, A_z(r=a,\theta+\pi,z,t) \\
\Delta A_{\mathbf{n}}(\theta,z,t) &= -A_r(r=a,\theta,z,t) -\sigma_{\mathbf{n}} \, A_r(r=a,\theta+\pi,z,t)\\
\end{aligned}$ &
$\begin{aligned}
& \!\!\!\!\!\!\! \mbox{Transform  trans./long.} \\
\sigma_z & = +(-1)^{n} \\
\sigma_{\mathbf{n}}  &=-(-1)^{n+1}
\end{aligned}$\\

\midrule
$\begin{aligned}
\mbox{\textbf{TE$_{n=0,m}$}}\\
 \mbox{cylinder} \end{aligned}$ &
$\begin{aligned}
\Delta V_{\rm eff}(r,z,t) &= \phi_m \, \omega \, g_{vir}(r) \, f(z,t)\\
\Delta A_{z\, \rm eff}(r,z,t) &= \phi_m \, \beta \, g_{vir}(r) \, f(z,t) \\
\Delta A_{\mathbf{n}}(r,z,t) &= A_\theta(r,\theta,z,t) -\sigma_{\mathbf{n}} \, A_\theta(r,\theta+\pi,z,t) \\
\end{aligned}$ &
$\begin{aligned}
& \!\!\!\!\!\!\! \mbox{Only transverse} \\
\sigma_z & = +1 \\
\sigma_{\mathbf{n}}   &=-1
\end{aligned}$\\

\bottomrule  %%%%%  AAARGH!!! TOUT PETEUX, a revoir... XXXXX
\end{tabular}
\end{table}

The key property is that for {\em all modes that do not involve virtual electrodes}, there exists a gauge transformation that swaps the potentials from matching Eq. (\ref{rule2}) to matching Eq. (\ref{rule1}).
It is then perfectly lawful to use Devoret's expressions Eq. (\ref{rule1}) to describe the dynamics of the fields, in agreement with conventional electronics (which is precisely what we did in the previous Subsections).
Interestingly, the gauge that verifies  Eq. (\ref{rule2}) also brings in:
\begin{eqnarray}
\Delta V & = & 0 , \\
\Delta A_z & = & 0 .
\end{eqnarray}
If this is a fundamental property, it is outside of our scope, and shall be addressed elsewhere.
%%%%
This symmetry {\em is broken} for the TE $n=0$ modes, the ones which require virtual electrodes to be defined.
Pleasantly enough, for the parallel plate guide the two gauge rules can be satisfied simultaneously \cite{Delattre2024}: Eq. (\ref{rule2}) is always true, and one can conveniently chose a gauge that also matches Eq. (\ref{rule1}).
The point is that {\em this is impossible} for the TE $n=0$ waves traveling in a cylindrical geometry, as demonstrated in Section \ref{sec:gauge_cyl}. If this contains some fundamental meaning is again outside of our scope. \\

How shall we deal with these pathological TE $n=0$ traveling waves?
There is a very simple, and pragmatic answer.
The Devoret relations Eq. (\ref{rule1}), together with the definitions of surface charges $\sigma_{s}$ and currents $\mathbf j_s$ ($s$ being any electrode) found in Section \ref{sec:charges_currents_constants_cyl} lead to:
\begin{eqnarray}
q  \propto \Delta V , \label{guard1} \\
I \propto \Delta A_z , \label{guard2}
\end{eqnarray}
with $q$ a charge per unit length and $I$ a longitudinal current, obtained by integration of $\sigma_{s}$ and $  \mathbf{j}_{s}.\mathbf{z}$ over a proper transverse dimension.
The point is that $q$ and $I$ are {\it physical measurable quantities} (actually the ones transported in the transmission line): this means that this gauge choice turns the somehow abstract quantities $\Delta V, \Delta A_z$ into sensible objects. And from an engineering point of view, this is all we need; when one deals with a voltage in a circuit, by no means do we directly refer to a gauge property. But what is essential, is that this potential {\it is an image} of the charges and currents flowing in the circuit, which is exactly what Eqs. (\ref{guard1},\ref{guard2}) guarantee. The natural solution to handling TE $n=0$ modes is therefore to {\em impose}  in the Devoret equations an effective voltage $\Delta V_{\rm eff}$  (and longitudinal potential $\Delta A_{z\,\rm eff}$) which is $\propto A_\theta$. This physically means that we actually "swap our gauge", but keep all the useful properties of the conventional formalism.
%%%%%%%%%
These effective potentials are defined on the virtual electrodes, so they correspond to charges/currents {\em that would be excited if one would use a non-invasive electrode} to detect them. 
%%%%
At the same time, the "real voltage" $\Delta V$ (and $\Delta A_z$) on the real electrodes {\em can be set to zero}, since it is a valid gauge choice (see e.g. Tab. \ref{tabTEpothollowSPECn}; this is true for any of the other TE $n=0$ situations).
%%%%
The outcomes of this Subsection are summarized in Tab. \ref{tab:gauge_plus}.

\section{Field quantization}
\label{sec:quantize}
\label{quantum}

%We index the relevant electrode by $[e]\in\{(a),(b),[\theta]\}$. Following Table~IX,
%the \emph{distributed} coefficients along $z$ are:
%\begin{equation}
%    C_d^{[e]}=\frac{\varepsilon}{h_{\mathrm{eff}}^{[e]}},\qquad
%\big(L_d^{[e]}\big)^{-1}=\frac{1}{\mu\,h_{\mathrm{eff}}^{[e]}}\!,
%\end{equation}
%with $h_{\mathrm{eff}}^{[e]}$ fixed by the geometry and the electrode choice
%(real or virtual). We use a single longitudinal degree of freedom carried by: 
%%
%\begin{equation}
%    \varphi_{\max}^{[e]}(z,t)=\phi_m^{[e]}\,\tilde f(z,t),
%\end{equation}
%%
%%% pourquoi in ^[e]???? Là tu alourdis pour rien...
%%% tes définitions de Cd et Ld arrivent un peu tard, ou alors elles se répètent...
We shall now quantize the $X$ and $Y$ quadratures, following strictly the method of Ref. \cite{Delattre2024}. To start with, let us define:
\begin{equation}
    \varphi_{max}(z,t) = \phi_m \, \tilde{f}(z,t) , 
\end{equation}
which is nothing but the amplitude of the $\varphi$ functions without the $g_{real}, g_{vir}$ transverse dependencies, Eqs. (\ref{varphi1},\ref{varphi2}). Integrating $H$ (energy) and $\mathbf P$ (momentum) over the transverse direction (the $a\,d\theta$ periphery or the virtual plane) leads to:
\begin{eqnarray}
H & = &\!\!\! \int_{z=0}^L  \left[ \frac{1}{2}\, C_H^{-1} \left( C_H \frac{\partial \varphi_{max}(z,t)}{\partial t}\right)^{\!\!2} + \frac{1}{2} \, L_H^{-1} \left(  \frac{\partial \varphi_{max}(z,t)}{\partial z}\right)^{\!\!2} \right. \nonumber \\
&& \left. + \frac{1}{2}\, C_P \, (c\, k_c)^2 \,  \varphi_{max}(z,t)^{2}  \right]   dz/L , \label{finalHz} \\
\mathbf{P} & = &\!\!\!  \int_{z=0}^L     \left[\left( C_P \frac{\partial \varphi_{max}(z,t)}{\partial t} \right)\, \left( -  \frac{\partial \varphi_{max}(z,t)}{\partial z} \right) \right]  dz/L \, \mathbf{z} , \label{finalPz}
\end{eqnarray}
for {\em all configurations} discussed here. This writing is strictly the same as the one of Cartesian guides \cite{Delattre2024}, and the obtained coefficients are listed in Tab. \ref{tab:cyl_modal_coeffs_fixed} (together with a reminder of other mode-dependent quantities).
%% petit comment pour corriger le précédent papier...
The velocity appearing in this tabular is the one that enters the $\varphi$ propagation equation, together with $k_c$. Remember that only for TEM waves do we have $v_\phi=c$; for both TM and TE solutions, the phase velocity verifies $v_\phi=c \, k/|\beta| \neq c$.
$C_H$ and $C_P$ are total capacitances (in F) while $L_H$ is a "total inductance" (actually in H$/$m$^2$), characteristic of each mode.
The proper mode inductance writes $L_H/\beta^2$, and arises from the $\partial \varphi_{max}/\partial z$ that appears in Eq. (\ref{finalHz}).
Finally, we introduce:
\begin{equation}
    Q_{max}(z,t) = + C_P \, \frac{\partial \varphi_{max}(z,t)}{\partial t}, 
\end{equation}
the charge amplitude (in Coulomb) that propagates along with $\varphi_{max} \propto \tilde{f}[z,t]$ (in quadrature, with a $f[z,t]$ dependence). \\
We now promote the two quadratures to operators ($X\to\hat X$, $Y\to\hat Y$), as is customary \cite{Clerk2010}. 
%%% ADDED REF 3
Performing the last integration over $z$ %and using the total coefficients.
we obtain:
\begin{align}
\hat H&=(2\,C_P \,\omega\,\phi_m^{2})\,
\omega\, \left(\frac{\hat X^{2}+\hat Y^{2}}{4} \right),\\
\hat {\mathbf P}&=(2\,C_P \,\omega\,\phi_m^{2})\,
\beta\,\left(\frac{\hat X^{2}+\hat Y^{2}}{4} \right)\, \mathbf z.
\end{align}
The generalized flux and charge operators %, 
therefore obey the commutation relation:
\begin{equation}
\big[\hat\varphi_{\max} (z,t),\hat Q_{\max} (z,t)\big]
=\big(2\,C_P \,\omega\,\phi_m^{2} \big)\,\frac{[\hat X,\hat Y]}{2}.
\end{equation}
%%% il faut le mettre là de suite!!!
%  Such as usual, introducing 
These can be conveniently re-expressed in terms of the creation/annihilation operators $\hat b^\dagger,\hat b$:
\begin{equation}
    \hat X=\hat b^\dagger+\hat b,\qquad \hat Y=i \, (\hat b^\dagger-\hat b),
\end{equation}
which straightforwardly lead to:
%All formulas above hold verbatim with the \emph{angular} total coefficients
%$C_P^{[\theta]}$, $C_H^{[\theta]}$, $L_H^{[\theta]}$ when the symmetry dictates
%the use of virtual angular electrodes; for TEM and TM$_{0,n}$, the relevant
%electrodes are radial, i.e. $(a)$ or $(b)$ following our convention. From Eqs.~(115)–(118) we obtain:
%%%%%%%%%%%%%%%%%%%%%
\begin{align}
\big[\hat X,\hat Y\big] \;&=\;  2i\,\big[\hat b,\hat b^\dagger\big],\\
\hat X^{2}+\hat Y^{2} \;&=\; 4\!\left(\hat b^\dagger\hat b \;+\; \frac{\big[\hat b,\hat b^\dagger\big]}{2}\right).
\label{eq:bdagbfinal}
\end{align}
%%% XXX
%%%
 Technically,  $\hat b, \hat b^\dagger$ are implicitly $t=0$ operators, in a Heisenberg picture \cite{DevoretQED}. Their time-evolved expressions are obviously $\hat b \,e^{-i \omega t}$ and $\hat b^\dagger \,e^{+i \omega t}$ respectively, which lead to the time-dependence of the $f, \tilde f$ functions 
 %%%% ADD REF 2 3
 through a trivial integration of the dynamics equation: $d\,\hat{b}/dt = -i/\hbar \, [\hat{b},\hat{H}]$ (and similarly for $\hat{b}^\dag$) \cite{CohenQED}.
%%%
%%%
We recover textbook formulations by imposing:
%Imposing the canonical quantization $[\,\hat\varphi_{\max}^{[e]},\hat Q_{\max}^{[e]}\,]=i\hbar$
%fixes the normalization:
%
\begin{equation}
   2\,C_P \, \omega \,  \phi_m^{2}\,=\,\hbar ,
\end{equation}
%Finally, to recover the textbook form, we simply impose:
\begin{equation}
\big[\hat b,\hat b^\dagger\big]=1,
\label{eq:bosons}
\end{equation}
which {\em fundamentally} link our field amplitude $E_m$ (hidden in $\phi_m$) to Planck's reduced constant $\hbar$, and defines our modes as being {\em bosons}. 
More details and mathematical properties can be found in Ref. \cite{Delattre2024}, but let us discuss the physical meaning of this result:
%%%% CORR REF 3:
the $\hat{\varphi}$ variable which produces the quantum properties (especially $\phi_m \propto \sqrt{\hbar}$) is {\it intrinsically a gauge property}. This point, to our knowledge, had never been commented so far.
In particular, quantum mechanics enables to understand the cutoff energy of TM waves:
\begin{equation}
    \Delta_c = \hbar \, \omega_c ,
\end{equation}
as being an energy gap per photon, needed to launch the wave. Similarly for TE waves: 
%%%
%In the TE case, the addendum can be
% recast as a Klein–Gordon “rest energy”:
\begin{equation}
 m\,c^{2}  \;=\;\hbar\,\omega_c ,
\end{equation}
where $ m $ is an effective %(waveguide)
photon mass, appearing in the Klein-Gordon equation.
%   et c'est TOUT !!!!!!!!!!!!!!!!!!!!!!!!!!!!
%
%With the quantization condition $2\,C_P^{[e]}\,\omega\,(\phi_m^{[e]})^2=\hbar$
%for the relevant electrode pair $[e]\in\{(a),(b),[\theta],[r]\}$,
%the electric zero-point field referenced to that pair reads:
%%
%\begin{equation}
%E_{\mathrm{zpf}}^{[e]}
%=\frac{1}{h_{\mathrm{eff}}^{[e]}}\,
%\sqrt{\frac{\hbar\,\omega}{2\,C_P^{[e]}}}, \qquad
%\text{with}\quad
%C_P^{[e]}=
%\begin{cases}
%C_H^{[e]} & \text{(TEM, TE)},\\[4pt]
%\big(\dfrac{k}{\beta}\big)^{\!2} C_H^{[e]} & \text{(TM)}.
%\end{cases}
%\label{eq:cyl:zpf_master}
%\end{equation}
%

\begin{table}[H]
\centering
\caption{Modal coefficients $C_H$, $C_P$, $L_H$ %(et versions primées) et paramètres associés, en géométrie cylindrique.
(together with $k_c$ and the velocity appearing in the Klein-Gordon equation). We remind that $C_d=\epsilon/h_{\mathrm{eff}}$ and $Ld^{-1}=1/(\mu\, h_{\mathrm{eff}})$. $k_c \neq 0$ and $h_{\mathrm{eff}}$ are given in the core of the text. We define $\gamma[n]=1$ for $n=0$ and $\gamma[n]=2$ for $n\neq0$.
Note that non-TEM waves have the same writing in coaxial and hollow cylinder guides. }
\label{tab:cyl_modal_coeffs_fixed}
\begin{tabular}{@{}c c c@{}}
\toprule
\textbf{Wave type} & \textbf{   Modal coefficients   } & \textbf{   Potential–related parameters   } \\
\midrule
\midrule
\textbf{TEM}, coaxial &
$\begin{aligned}
C_H &= C_d \,(2\pi a) \, L\\
C_P &= C_H\\
L_H^{-1} &= L_d^{-1}\,(2\pi a) \,  L\\
\end{aligned}$ &
$\begin{aligned}
k_c&=0, \mbox{K-G velocity } c\\
%\sigma  &=-1
\end{aligned}$\\

\midrule
\textbf{TM$_{n,m}$}, coaxial &
$\begin{aligned}
C_H &= C_d \,(2\pi a/\gamma[n]) \, L\\
C_P &= C_H\,(k/\beta)^2 \\
L_H^{-1} &= L_d^{-1}\,(k/\beta)^4\,(2\pi a/\gamma[n]) \,  L\\
\end{aligned}$ &
$\begin{aligned}
k_c& \neq 0, \mbox{K-G velocity } v_\phi\\
%\sigma  &=+(-1)^m
\end{aligned}$\\

\midrule
\textbf{TE$_{n,m}$}, coaxial &
$\begin{aligned}
C_H &= C_d \,(2\pi a/\gamma[n]) \, L\\
C_P &= C_H\\
L_H^{-1} &= L_d^{-1}\,(2\pi a/\gamma[n]) \,  L\\
\end{aligned}$ &
$\begin{aligned}
k_c& \neq 0, \mbox{K-G velocity } c\\
%\sigma  &=+(-1)^{m+1}
\end{aligned}$\\

%\midrule
%\textbf{TE$_{n=0,m}$}, coaxial &
%$\begin{aligned}
%C_H &= C_d \,(2\pi a) \, L\\
%C_P &= C_H\\
%L_H^{-1} &= L_d^{-1}\,(2\pi a) \,  L\\
%\end{aligned}$ &
%$\begin{aligned}
%k_c& \neq 0, \mbox{K-G velocity } c\\
%%\sigma  &=+1
%\end{aligned}$\\

\midrule
\midrule
\textbf{TM$_{n,m}$}, cylinder &
$\begin{aligned}
C_H &= C_d \,(2\pi a/\gamma[n]) \, L\\
C_P &= C_H\,(k/\beta)^2\\
L_H^{-1} &= L_d^{-1}\,(k/\beta)^4\,(2\pi a/\gamma[n]) \,  L\\
\end{aligned}$ &
$\begin{aligned}
k_c& \neq 0, \mbox{K-G velocity } v_\phi\\
%\sigma  &=+(-1)^n
\end{aligned}$\\

\midrule
\textbf{TE$_{n,m}$}, cylinder &
$\begin{aligned}
C_H &= C_d \,(2\pi a/\gamma[n]) \, L\\
C_P &= C_H\\
L_H^{-1} &= L_d^{-1}\,(2\pi a/\gamma[n]) \,  L\\
\end{aligned}$ &
$\begin{aligned}
k_c& \neq 0, \mbox{K-G velocity } c\\
%\sigma  &=+(-1)^{n}
\end{aligned}$\\

%\midrule
%\textbf{TE$_{n=0,m}$}, cylinder &
%$\begin{aligned}
%C_H &= C_d \,(2\pi a) \, L\\
%C_P &= C_H\\
%L_H^{-1} &= L_d^{-1}\,(2\pi a) \,  L\\
%\end{aligned}$ &
%$\begin{aligned}
%k_c& \neq 0, \mbox{K-G velocity } c\\
%%\sigma  &=+1
%\end{aligned}$\\

\bottomrule  
\end{tabular}
\end{table}

\section{Conclusion}
%%%%%%%%%% ouuuuf, on y est....

We %have developed a 
report on the canonical quantization of electromagnetic fields in cylindrical waveguides, %extending the Devoret formalism to a geometry of fundamental importance in both classical and quantum technologies. 
adapting the formalism already developed for Cartesian geometries \cite{Delattre2024}.
%
%
%By deriving the explicit modal structure of TE and TM families in terms of Bessel functions, 
%%
%%
We identify the %associated 
generalized flux $\varphi$ and its conjugate charge $Q$ % variables
that enable to recast all the field's properties. 
%%%
% and constructed the corresponding Hamiltonian. The resulting framework mirrors the Cartesian treatment while uncovering specific features of cylindrical confinement, such as azimuthal degeneracy, radial mode scaling, and the existence of universal TE$_{0n}$ families.
The theory invokes a gauge fixing required to deduce $\varphi$ from the potentials   $\mathbf A,   V$, which %we take as being 
must be 
a fundamental property {\em for all types of waves}, namely TEM, TM and TE.
%%% c'est finalement ca le plus important dans le modèle. Le reste n'est que du calcul a la con...
%%%%
We discuss how this can be recast in 
"potential differences" $\Delta V, \Delta A_z$, producing  "Devoret relationships" particularly relevant for quantum engineering \cite{DevoretQED,Clerk2010}.
The conclusions reached in Ref. \cite{Delattre2024} also apply here, especially the introduction of "virtual electrodes", but with some specificities linked to the higher symmetry of the problem at hand. %%% je le dis, et en plus j'insiste sur le pourquoi: plus de symmétrie!
%%%% maintenant: uniquement les trucs profonds... On répète pas tout la bazar

Exactly like in the Cartesian case, the cutoff frequency $\omega_c$ appears as a consequence of an energy gap for TM waves, and of a photon mass for TE. %%%
Modal parameters $C_H, C_P$ and $L_H$ are introduced,  leading to  expressions for energy $H$ and momentum $\mathbf P$ identical to those of Ref. \cite{Delattre2024}. These results are robust, %, and certainly %%% juste certainly parce qu'on a pas fait de topo ici!!!
% linked to the topological equivalence between the two geometries. 
%%%%%%%%%%%
% More broadly, the approach developed here illustrates how geometry, topology, and symmetry constrain the quantization procedure, 
%
%% un petit mot sur les différences!!
even though extra modes are allowed in the coaxial configuration as compared to the parallel plate guide (all $n \neq 0 $ modes), and  TM $n=0$ waves do exist in the hollow cylinder (but not in a rectangular guide).
%; the TE $n=0$ modes hosted by the hollow tube are peculiar, with no analogue in the square geometry. As well, the remaining gauge invariance is higher here for some configurations than in the Cartesian case. All of this because of the higher symmetry.
%%%%%
%Beyond its theoretical consistency, this work provides practical tools for analyzing quantum devices based on cylindrical and coaxial structures, from high-Q cavities to traveling-wave amplifiers and hybrid optomechanical systems. By keeping the dependence on physical constants and geometrical parameters explicit, 
%%%
%%%% OK? maintenant, on peut parler des extensions possibles, et implications...
%%% A nouveau, on ne reviendra pas sur les fluctu quantiques etc.. parce que c'était pas un truc central, et je suis même pas sur que ce soit vraiment profond...

Beyond all similarities, the comparison cylinder/square is particularly enlightening {\em because of the peculiarities} encountered here.
The potentials considered in this work can diverge at $r=0$ with no physical relevance, while this is impossible in a Cartesian geometry.  For all waves supported by real electrodes, we pointed out the existence of a gauge transformation that links $\varphi$ to the {\em transverse} potential difference $\Delta A_{\mathbf n}$, and at the same time generates $\Delta A_z = \Delta V=0$.
On the other hand, the TE $n=0$ modes which require virtual electrodes to be described,  
{\em do not support} this transformation and seem in this sense quite special.
%%%%% ADD REF 3!!
This peculiar point had been missed in our previous work \cite{Delattre2024}, and {\it only the direct comparison} Cartesian/Cylindrical could reveal it in the present paper.
%%%%%
If these facts 
%%%% CORR REF 2 3
%bear a fundamental meaning 
produce measurable phenomena 
remains an open question, and might be investigated in the future. 
%%%%

An important outcome of the model is the capability to define capacitance and inductance densities {\em for all types of modes} TEM, TM and TE, which makes the link with conventional wave propagation: the well known {\em telegrapher's equation}. 
%%%% ADD REF 2
As well, considering a shorter guide and specific non-periodic boundary conditions, the concepts addressed here shall apply equally well to the case of a {\it microwave cavity}, especially when light interacts with a specific object localized inside it (e.g. a quantum bit, a spin wave, a mechanical element).
Formalizing it properly is a task on its own that involves the proper introduction of what a "quantum load" is, enabling the introduction of the quantum  Scattering Matrix. We leave it for future works, making the link to the well-known input-output theory \cite{GardinerZoller}. 
%%%%
%How to formalize this, in particular in the  shall be left for future works. % (en racine de Hz)...
%%% avec les quantum pulses... casse gueule...
% et quantum version... du load???  casse gueule...
% un mot sur spin et hélicité??? casse gueule...

The present formalism shall also be extended to other types of guides. Especially, the {\em bifilar} geometry (hosting only TEM modes), and the coplanar waveguide (hosting quasi-TEM) which is heavily used in quantum engineering. 
Both have their specificity which might require some adaptations of the reasoning: 
%%%%%%
the former corresponds to an open geometry, while the latter involves multiple electrodes which do not share a common symmetry. 
%%%%%%%%
As well, dielectric interfaces have not been addressed at all; 
%%%%%%%%
in particular, how shall we deal with {\em hybrid waves}, which have both an $E_z$ and $B_z$ component?
%%
%%% tu vois, je suis assez spécifique sur les extensions, et pas que des mots en l'air comme "ce serait cool de faire pareil avec un crystal photonique"...
%%% mais ca n'empeche pas de le dire subtilement en conclusion finale, bien sur!!!
%Our formulation bridges the gap between canonical quantization and experimental design, offering predictive power for emerging architectures in quantum engineering,
%%%%
This would 
%opening 
open the way to generalizations 
%%%
in even more complex media, 
%%%%
e.g. structured photonic environments~\cite{Ozawa2019TopoPhotonics} %, Lodahl2017QuantumPhotonics}   %%% est-ce vraiment la peine?????
and multimode quantum networks~\cite{Gely2017Multimode}. 
%%%%%
%%%%%

% Discuss mass in Quantif!! et conclu... of course...
%%
%%
%Future directions include the quantization of coaxial TEM modes, which complete the taxonomy of cylindrical families and connect directly to transmission-line theory.
%
%% nan, tu l'as fait! ;-)
%%%
%In this sense, canonical quantization in cylindrical waveguides is not only an extension of the Devoret framework, but also a stepping stone toward a unified treatment of quantum fields in engineered electromagnetic structures.
%%
%%% déjà dit... tu te répètes...

\section*{Acknowledgements}

This work is a follow-up of a previous article, Ref. \cite{Delattre2024}. We are thankful to the European Microkelvin Platform community (EMP), for strong support  over the years. We also wish to acknowledge useful interactions among the participants of the former "Groupement de Recherche MecaQ".

\section*{Data availability}

No data was used or created for this manuscript. A Mathematica\textsuperscript{\textregistered}$\,$ code is available at the following URL: 
\small{\underline{https://cloud.neel.cnrs.fr/index.php/s/CnnYPKn8XHYZgXa}}.
% It corresponds to all calculations presented here.

\section*{Conflict of interest}

The authors have no conflicts to disclose.

\appendix
\normalsize

%\section{Bifilar line}
%\label{bifilar}

%Might be added.
%%%% nan, j'ai la flemme j'en peux plus.

%\newpage

\bibliographystyle{apsrev4-2}
\bibliography{bibliography_COMPLETE}

\end{document}